\documentclass[11pt]{article}
\makeatletter \@addtoreset{equation}{section} \makeatother

\newcommand{\noi}{\vspace{12pt}\noindent}
\newcommand{\beq}{\begin{equation}}
\newcommand{\eeq}{\end{equation}}
\newcommand{\bea}{\begin{eqnarray}}
\newcommand{\eea}{\end{eqnarray}}

\newcommand{\e}[1]{{(\ref{#1})}}
\newcommand{\eq}[1]{{eq.\ (\ref{#1})}}
\newcommand{\es}[2]{{(\ref{#1}) and (\ref{#2})}}
\newcommand{\eqs}[2]{{eqs.\ (\ref{#1}) and (\ref{#2})}}
\newcommand{\Ref}[1]{{Ref.~\cite{#1}}}

\newcommand{\equi}[1]{\stackrel{{#1}}{=}}

\newcommand{\ie}{{${ i.e., \ }$}}
\newcommand{\eg}{{${ e.g., \ }$}}

\newcommand{\cf}{{cf.\ }}

\newcommand{\wrt}{{with respect to }}
\newcommand{\wrtt}{{with respect to the }}

\newcommand{\rhs}{{right--hand side }}

\renewcommand{\~}{ \ }
\renewcommand{\=}{ \ = \ }
\newcommand{\eps}{\varepsilon^{}}
\renewcommand{\tilde}{\widetilde}
\renewcommand{\bar}{\overline}

\newcommand{\gh}{{\rm gh}}

\newcommand{\rank}{{\rm rank}}

\newcommand{\cl}{{\rm cl}}
\newcommand{\for}{{\rm for}}

\newcommand{\Hf}{{1 \over 2}}

\newcommand{\afn}{{\rm afn}}
\newcommand{\rafn}{{\rm rafn}}
\newcommand{\safn}{{\rm safn}}
\newcommand{\puregh}{{\rm puregh}}

\newcommand{\aaa}{\alpha^{}}
\newcommand{\AAA}{A^{}}
\newcommand{\bbbb}{\beta^{}}
\newcommand{\BBBB}{B^{}}
\newcommand{\cccc}{\gamma^{}}
\newcommand{\MMM}{M^{}}
\newcommand{\mmm}{m^{}}
\newcommand{\brst}{{\bf \rm s}^{}}
\newcommand{\bbrst}{\overline{{\bf \rm s}}^{}}
\newcommand{\CME}{{\rm CME}^{}}
\newcommand{\bCME}{\overline{\rm CME}^{}}
\newcommand{\EOM}{T^{}}
\newcommand{\JI}{J^{}}
\newcommand{\bJI}{\overline{J}^{}}

\newcommand{\bNI}{\overline{\cal N}^{}}
\newcommand{\Squad}{S^{\rm quad}}

\newcommand{\Vquad}{V^{\rm quad}}
\newcommand{\bVquad}{\overline{V}^{\rm quad}}
\newcommand{\Sfixed}{S^{\rm fixed}}
\newcommand{\bSfixed}{\overline{S}^{\rm fixed}}
\newcommand{\Vfixed}{V^{\rm fixed}}
\newcommand{\bVfixed}{\overline{V}^{\rm fixed}}
\newcommand{\Snoeth}{S^{\rm Noether}}
\newcommand{\bSnoeth}{\overline{S}^{\rm Noether}}
\newcommand{\Vnoeth}{V^{\rm Noether}}
\newcommand{\bVnoeth}{\overline{V}^{\rm Noether}}
\newcommand{\cO}{{\cal O}}
\newcommand{\cB}{{\cal B}}

\newcommand{\twobyone}[2]{\left[\begin{array}{c}{#1} \cr
                                {#2} \end{array} \right]}
\newcommand{\onebytwo}[2]{\left[\begin{array}{ccc}{#1}&{#2}
                                        \end{array} \right]}
\newcommand{\twobytwo}[4]{\left[\begin{array}{ccc}{#1}&{#2} \cr
                                  {#3} & {#4} \end{array} \right]}

\newcommand{\twostack}[2]{\begin{array}{c} \lower.8ex\hbox{${#1}$}
                     \cr \raise.8ex\hbox{${#2}$} \end{array}}
\newcommand{\deder}[1]{{ 
 {\stackrel{\raise.2ex\hbox{$\leftarrow$}}{\delta^{r}}   } 
\over {   \delta {#1}}  }}
\newcommand{\dedel}[1]{{ 
 {\stackrel{\lower.3ex \hbox{$\rightarrow$}}{\delta^{\ell}}   }
 \over {   \delta {#1}}  }}
\newcommand{\papar}[1]{{ 
 {\stackrel{\raise.2ex\hbox{$\leftarrow$}}{\partial^{r}}   } 
\over {   \partial {#1}}  }}
\newcommand{\papal}[1]{{ 
 {\stackrel{\lower.3ex \hbox{$\rightarrow$}}{\partial^{\ell}}   }
 \over {   \partial {#1}}  }}
\newcommand{\rpa}[1]{{ 
 \stackrel{\raise.2ex\hbox{$\leftarrow$}}{\partial^{r}_{#1}}   }}
\newcommand{\lpa}[1]{{ 
 \stackrel{\lower.3ex\hbox{$\rightarrow$}}{\partial^{\ell}_{#1}}  }}
\newcommand{\proofbox}{\begin{flushright}
${\,\lower0.9pt\vbox{\hrule \hbox{\vrule
height 0.2 cm \hskip 0.2 cm \vrule height 0.2 cm}\hrule}\,}$
\end{flushright}}

\newtheorem{theorem}{Theorem}[section]

\newtheorem{corollary}[theorem]{Corollary}

\newtheorem{lemma}[theorem]{Lemma}
\newtheorem{principle}[theorem]{Principle}
\newtheorem{proposition}[theorem]{Proposition}

\begin{document}
\thispagestyle{empty}
\title{\Large{\bf Remarks on Existence of Proper Action\\
for Reducible Gauge Theories}}
\author{{\sc Igor~A.~Batalin}$^{a,b}$ and {\sc Klaus~Bering}$^{a,c}$ \\~\\
$^{a}$The Niels Bohr Institute\\The Niels Bohr International Academy\\
Blegdamsvej 17\\DK--2100 Copenhagen\\Denmark \\~\\
$^{b}$I.E.~Tamm Theory Division\\
P.N.~Lebedev Physics Institute\\Russian Academy of Sciences\\
53 Leninsky Prospect\\Moscow 119991\\Russia\\~\\
$^{c}$Institute for Theoretical Physics \& Astrophysics\\
Masaryk University\\Kotl\'a\v{r}sk\'a 2\\CZ--611 37 Brno\\Czech Republic}
\maketitle
\vfill
\begin{abstract}
In the field--antifield formalism, we review existence and uniqueness proofs 
for the proper action in the reducible case. We give two new existence proofs 
based on two resolution degrees called ``reduced antifield number'' and 
``shifted antifield number'', respectively. In particular, we show that for
every choice of gauge generators and their higher stage counterparts, there
exists a proper action that implements them at the quadratic order in the
auxiliary variables.
\end{abstract}
\vfill
\begin{quote}
PACS number(s): 11.10.-z; 11.10.Ef; 11.15.-q; 11.15.Bt. 
\hrule width 5.cm \vskip 2.mm \noindent 
$^{b}${\small E--mail:~{\tt batalin@lpi.ru}} \hspace{10mm}
$^{c}${\small E--mail:~{\tt bering@physics.muni.cz}} \\
\end{quote}

\newpage
%\tableofcontents

\section{Introduction}
\label{secintro}

\noi
This paper considers existence and uniqueness theorems for a proper action $S$
of a general open reducible gauge theory in the field--antifield formalism
\cite{bv81,bv83}. We should stress that the term {\em uniqueness theorem} 
here should be understood in a generalized sense, \ie as a theorem
that specifies the natural arbitrariness/ambiguity of solutions (see
Theorem~\ref{theorembrst3}). In our case, the uniqueness theorem states that
all proper solutions to the classical master equation $(S,S)\!=\!0$ can locally
be reached from any other proper solution via a finite anticanonical 
transformation.

\noi
It is always possible to locally close or Abelianize the gauge algebra of a
physical system by rotating and by adding off--shell contributions to the 
gauge generators \cite{b81,bv84}. Moreover, closed and Abelian theories
have well--known proper actions, so why do we need another existence proof?
The answer is that although closure and Abelianization are great theoretical 
tools, they have only limited practical use in field theory, where they
often destroy space--time locality. Thus, ideally, one would like to establish 
the existence of the proper action $S$ without tampering in any way with the
original gauge algebra, and its higher--stage counterparts. This raises the 
question: How many terms in the action can one preserve while constructing the
proper action? Specifically, we shall prove that the most general gauge 
generators $R^{i}{}_{\aaa_{0}}$ and higher--stage counterparts 
$Z^{\aaa_{s-1}}{}_{\aaa_{s}}$ locally fit in unaltered form into a proper solution
\beq
S\=\Squad+\cO\left((\Phi^{*})^{2},c^{2}\right)\~,\~\~\~\~\~\~\~\~
\Squad\~:=\~S^{}_{0}+\varphi^{*}_{i}R^{i}{}_{\aaa_{0}}\~c^{\aaa_{0}}
+\sum_{s=1}^{L}c^{*}_{\aaa_{s-1}}Z^{\aaa_{s-1}}{}_{\aaa_{s}}\~c^{\aaa_{s}}\~,
\eeq
to the classical master equation, see Theorem~\ref{theoremss2}. In other words,
the Existence Theorem~\ref{theoremss2} states that any action $\Squad$, which
is quadratic in auxiliary variables, and which satisfies maximal rank
conditions and Noether identities, can locally be completed into a proper
action $S\!=\!\Squad\!+\!\cO\left((\Phi^{*})^{2},c^{2}\right)$.

\noi
With this said, we must admit, that our current treatment will still make use
of space--time non--local items, at least behind the screen. In particular, we
will use the existence of a set of transversal and longitudinal fields
$\bar{\varphi}^{i}\!\equiv\!\{\xi^{I};\theta^{\AAA_{0}}\}$, which are often
non--local, see Section~\ref{secltfields}. Also we should say that we 
will for simplicity use DeWitt's condensed index notation, where space--time
locality is suppressed. So we work, strictly speaking, only with a finite
number $2N$ of variables
$\Gamma^{A}\!\equiv\!\{\Phi^{\alpha};\Phi^{*}_{\alpha}\}$.

\noi
There is of course an analogous story for the Hamiltonian/canonical formalism,
which we omit for brevity. Also we do not discuss quantum corrections in this
paper. Even today, there exists only a relatively limited number of general
results in manifestly space--time local field--antifield formalism
\cite{hen91,barbrahen95}. See \Ref{pigsor95} and \Ref{barbrahen00} for a 
treatment of Yang--Mills type theories. 

\noi
The standard existence proof is bases on two key elements. {}Firstly, the use
of {\em antifield number} 
\beq
\afn(\Phi^{\alpha})\=0\~,\~\~\~\~\~\~\~\~
\afn(\Phi^{*}_{\alpha})\=-\gh(\Phi^{*}_{\alpha})\~,
\eeq
as resolution degree see Table~\ref{multtable1}. (The antifield number is
sometimes called {\em antighost number}, because it is just the negative part
of the ghost number.) Secondly, the use of a nilpotent acyclic Koszul--Tate 
operator $\brst_{-1}$ of antifield number minus one, 
\bea
\brst_{-1}&=&V^{}_{\alpha}\papal{\Phi^{*}_{\alpha}}~,
\~\~\~\~\~\~\~\~\afn(\brst_{-1})\=-1\~, \label{ktm2} \\
\brst_{-1}\varphi^{*}_{i}&=&V^{}_{i}
\=(S^{}_{0} \papar{\varphi^{i}})\~,\label{ktm1} \\
\brst_{-1}c^{*}_{\aaa_{0}}&=&V^{}_{\aaa_{0}}
\=\varphi^{*}_{i}R^{i}{}_{\aaa_{0}}\~,\label{ktp0}\\
\brst_{-1}c^{*}_{\aaa_{1}}&=&V^{}_{\aaa_{1}}
\=c^{*}_{\aaa_{0}}Z^{\aaa_{0}}{}_{\aaa_{1}}
+\frac{(-1)^{\eps_{j}}}{2}\varphi^{*}_{j}\varphi^{*}_{i}B^{ij}_{\aaa_{1}}
\~,\~\~\~\~\~\~\~\~B^{ji}_{\aaa_{1}}\=-(-1)^{\eps_{i}\eps_{j}}B^{ij}_{\aaa_{1}}
\~,\label{ktp1}\\
\brst_{-1}c^{*}_{\aaa_{2}}&=&V^{}_{\aaa_{2}}
\=c^{*}_{\aaa_{1}}Z^{\aaa_{1}}{}_{\aaa_{2}}
-(-1)^{\eps_{\aaa_{0}}}c^{*}_{\aaa_{0}}\varphi^{*}_{i}B^{i\aaa_{0}}_{\aaa_{2}}
+\cO\left((\Phi^{*})^{3}\right)\~.\label{ktp2} \\
\brst_{-1}c^{*}_{\aaa_{3}}&=&V^{}_{\aaa_{3}}
\=c^{*}_{\aaa_{2}}Z^{\aaa_{2}}{}_{\aaa_{3}}
+(-1)^{\eps_{\aaa_{1}}}c^{*}_{\aaa_{1}}\varphi^{*}_{i}B^{i\aaa_{1}}_{\aaa_{3}}
+\Hf c^{*}_{\bbbb_{0}}c^{*}_{\aaa_{0}}B^{\aaa_{0}\bbbb_{0}}_{\aaa_{3}}
+\cO\left((\Phi^{*})^{3}\right)\~,\label{ktp3} \\
&& \~\~\~\~\~\~\~\~\~\~\~\~\~\~\~\~B^{\bbbb_{0}\aaa_{0}}_{\aaa_{3}}
\=(-1)^{\eps_{\aaa_{0}}\eps_{\bbbb_{0}}}B^{\aaa_{0}b^{}_{0}}_{\aaa_{3}}\~,
\label{ktp3b} \\
\brst_{-1}c^{*}_{\aaa_{4}}&=&V^{}_{\aaa_{4}}
\=c^{*}_{\aaa_{3}}Z^{\aaa_{3}}{}_{\aaa_{4}}
-(-1)^{\eps_{\aaa_{2}}}c^{*}_{\aaa_{2}}\varphi^{*}_{i}B^{i\aaa_{2}}_{\aaa_{4}}
+c^{*}_{\aaa_{1}}c^{*}_{\aaa_{0}}B^{\aaa_{0}\aaa_{1}}_{\aaa_{4}}
+\cO\left((\Phi^{*})^{3}\right)\~,\label{ktp4} \\
&\vdots& \nonumber
\eea
Historically, the existence and uniqueness theorems for an arbitrary
irreducible gauge theory in an arbitrary basis were established in \Ref{vt82} 
and \Ref{bv85}. In the reducible case, an $Sp(2)$ covariant proof (which at one
point uses rotations of gauge generators $R^{i}{}_{\aaa_{0}}$ and higher--stage 
counterparts $Z^{\aaa_{s-1}}{}_{\aaa_{s}}$) was given in \Ref{blt90}. A proof that 
does {\em not} change the gauge generators $R^{i}{}_{\aaa_{0}}$ and higher--stage
counterparts $Z^{\aaa_{s-1}}{}_{\aaa_{s}}$ was given in \Ref{fishen90}. The heart 
of the proof consists in showing the existence of a nilpotent and acyclic
Koszul--Tate operator. This was done in \Ref{fishen90} by referring to the 
analogous proof in the Hamiltonian setting \cite{fishen89,hentei92}. We should 
warn, that it is easy to get the false impression after a first reading of 
\Ref{fishen89}, that the nilpotency of the Koszul--Tate operator is merely a 
consequence of its Grassmann--odd nature. This is of course wrong, and this is 
{\em not} what these authors are really saying, as becomes evident when reading 
Theorem 3 in \Ref{fishen89}. (There are simple counterexamples of 
Grassmann--odd operators that satisfy a graded Leibniz rule, but are {\em not} 
nilpotent.) It should be stressed that the nilpotency of the Koszul--Tate 
operator is a non-trivial statement. It is equivalent to a $L$-stage tower of 
higher--stage Noether identities \e{noether1}--\e{noether2}, 
and their consistency relations \cite{basgiamansar05}. 
The first consistency relation appears at stage $2$. The
first few consistency relations read
\bea
0&\approx&B^{ij}_{\aaa_{1}}Z^{\aaa_{1}}{}_{\aaa_{2}}
+\left[ R^{i}{}_{\aaa_{0}} B^{j\aaa_{0}}_{\aaa_{2}}
(-1)^{\eps_{j}\eps_{\aaa_{0}}}
-(-1)^{\eps_{i}\eps_{j}}(i \leftrightarrow j)\right]\~, \label{bconrelb2} \\
B^{i\aaa_{0}}_{\aaa_{2}}Z^{\aaa_{2}}{}_{\aaa_{3}} 
&\approx&R^{i}{}_{\bbbb_{0}}B^{\bbbb_{0}\aaa_{0}}_{\aaa_{3}}
+Z^{\aaa_{0}}{}_{\aaa_{1}}B^{i\aaa_{1}}_{\aaa_{3}}
(-1)^{\eps_{i}(\eps_{\aaa_{0}}+\eps_{\aaa_{1}})}\~, \label{bconrelb3} \\
0&\approx&B^{\aaa_{0}b^{}_{0}}_{\aaa_{3}}Z^{\aaa_{3}}{}_{\aaa_{4}}
+\left[ Z^{\aaa_{0}}{}_{\aaa_{1}} B^{\bbbb_{0}\aaa_{1}}_{\aaa_{4}}
(-1)^{\eps_{\bbbb_{0}}\eps_{\aaa_{1}}}
+(-1)^{\aaa_{0}\bbbb_{0}}(\aaa_{0} \leftrightarrow b^{}_{0})\right]
\~, \label{bconrelb4i} \\
B^{i\aaa_{1}}_{\aaa_{3}}Z^{\aaa_{3}}{}_{\aaa_{4}} 
&\approx&R^{i}{}_{\aaa_{0}}B^{\aaa_{0}\aaa_{1}}_{\aaa_{4}}
+Z^{\aaa_{1}}{}_{\aaa_{2}}B^{i\aaa_{2}}_{\aaa_{4}}
(-1)^{\eps_{i}(\eps_{\aaa_{1}}+\eps_{\aaa_{2}})}\~, \label{bconrelb4ii} \\
&\vdots& \nonumber
\eea
where ``$\approx$'' means equality modulo equations of motion for
$\varphi^{i}$. The number of consistency relations grows with the reducibility 
stage. It is natural to wonder if these consistency relations can be 
satisfied without changing the original gauge generators $R^{i}{}_{\aaa_{0}}$
and their higher--stage $Z^{\aaa_{s-1}}{}_{\aaa_{s}}$ counterparts 
\cite{vanvan94,vandorenthesis}? This turns out to be possible, see 
Theorem 3 in \Ref{fishen89} (or Theorem 10.2 in 
\Ref{hentei92}, or Lemma~\ref{lemmaacycnilp} in this paper). The proof uses the
acyclicity property of previous stages to prove the existence of a nilpotent 
extension of the Koszul--Tate operator up to a certain stage without changing 
$R^{i}{}_{\aaa_{0}}$ and $Z^{\aaa_{s-1}}{}_{\aaa_{s}}$. 

\noi
Besides the gauge generators $R^{i}{}_{\aaa_{0}}$ and $Z^{\aaa_{s-1}}{}_{\aaa_{s}}$, 
$s\!\in\!\{1,\ldots,L\}$, it turns out as an added bonus, that the 
antisymmetric first--stage structure function $B^{ij}_{\aaa_{1}}$ in the
first--stage Noether identity \e{noether15} can also be preserved. (This is 
because there is no consistency relations at first stage.)

\noi
On the other hand, the proof does reveal that it will in general be necessary 
to change the given $B^{i\aaa_{s-2}}_{\aaa_{s}}$ structure functions in the
higher--stage Noether identities \e{noether2} as 
\beq
B^{i\aaa_{s-2}}_{\aaa_{s}}\~\longrightarrow\~B^{i\aaa_{s-2}}_{\aaa_{s}}
+R^{i}{}_{\aaa_{0}}X^{\aaa_{0}\aaa_{s-2}}_{\aaa_{s}}
+(S^{}_{0}\papar{\varphi^{j}})Y^{ji\aaa_{s-2}}_{\aaa_{s}}\~,\~\~\~\~\~\~\~\~
Y^{ji\aaa_{s-2}}_{\aaa_{s}}
\=-(-1)^{\eps_{i}\eps_{j}}Y^{ij\aaa_{s-2}}_{\aaa_{s}}\~. \label{bshift}
\eeq
The higher--stage Noether identities \e{noether2} are unaffected by such a
change \e{bshift}, due to the zero--stage Noether identity \e{noether1}.

\noi
So what remains is to prove the acyclicity. Unfortunately, the treatments in
\Ref{fishen89} and \Ref{hentei92} of acyclicity at higher stages are very
brief. One of the main purposes to introduce reduced and shifted antifield
number, is to properly spell out, in great detail and in a systematic way, an
acyclicity proof for all stages.  

\noi
It is very simple to motivate the construction of shifted antifield number
``$\safn$''. Consider what happens if one raises the resolution degree of all 
the antifields $\Phi^{*}_{\alpha}$ by $1$ unit. Then the $R^{i}{}_{\aaa_{0}}$ 
and the $Z^{\aaa_{s-1}}{}_{\aaa_{s}}$ terms, which are linear in the
antifields, will have their resolution degree raised by $1$, while the
$B^{i\aaa_{s-2}}_{\aaa_{s}}$ terms and higher terms, which are at least
quadratic in the antifields, will have their resolution degree raised by at 
least $2$, and hence they become subleading, and can be dropped from the new
``shifted'' Koszul--Tate operator $\brst_{(-1)}$. That is the good news! The
bad news is that the very first term
\beq
V^{}_{i}\papal{\varphi^{*}_{i}}
\~\equiv\~ (S^{}_{0} \papar{\varphi^{i}})\papal{\varphi^{*}_{i}}
\label{afnm2}
\eeq
in the Koszul--Tate operator now has resolution degree $-2$, which would be
devastating. But there is a remedy. The term \e{afnm2} is proportional to the
original equations of motion. The remainder of our constructions is concerned
with somehow assigning shifted antifield number $\geq 1$ to the original
equations of motion, so that the resolution degree of the term \e{afnm2}
becomes $\geq -1$. In practice, we implement this by assigning shifted
antifield number, $\safn(\xi^{I})\!=\!1$, to a set of so--called transversal
fields $\xi^{I}$, see Section~\ref{secltfields}.

\noi
Another idea is to democratically assign resolution degree $1$ to {\em all} 
antifields $\Phi^{*}_{\alpha}$. We call this degree for reduced antifield
number. Then the terms
\beq
\varphi^{*}_{i} R^{i}{}_{\aaa_{0}} \papal{c^{*}_{\aaa_{0}}}
+\sum_{s=1}^{L}c^{*}_{\aaa_{s-1}} Z^{\aaa_{s-1}}{}_{\aaa_{s}}
\papal{c^{*}_{\aaa_{s}}}
\label{afnm3}
\eeq
will have resolution degree $0$. Thus one would like the degree $0$ sector to 
become the leading resolution sector. Again the
$B^{i\aaa_{s-2}}_{\aaa_{s}}$ terms and higher terms, which are at least
quadratic in the antifields, will become subleading in resolution degree, and 
can be dropped from the new ``reduced'' Koszul--Tate operator $\brst_{-1[0]}$.
However, the very first term \e{afnm2} now has resolution degree $-1$, which 
would be devastating. Again we cure this by assigning reduced antifield number,
$\rafn(\xi^{I})\!=\!1$, to a set of transversal fields $\xi^{I}$, see 
Section~\ref{secltfields}.

\noi
The paper is organized as follows. In Section~\ref{secgt} we review the
starting point of an arbitrary gauge theory, and introduce the field--antifield
formalism. In Section~\ref{secfakeexist} we consider an existence and
uniqueness proof via the standard methods of antifield number and Koszul--Tate
operator. And finally, in Section~\ref{secrealexist} we consider a complete 
existence proof via the new methods of reduced (shifted) antifield number, 
and reduced (shifted) Koszul--Tate operator, respectively. Note that the proofs
in Section~\ref{secfakeexist} are incomplete in the sense that one needs the 
new technology of reduced or shifted antifield number (which is developed in 
Section~\ref{secrealexist}) to prove the acyclicity of the Koszul--Tate 
operator.  

\noi
In an appendix~\ref{secrot}, we show for completeness how the
$B^{i\aaa_{s-2}}_{\aaa_{s}}$ terms can be removed by (space--time non--local) 
change of the gauge generators $R^{i}{}_{\aaa_{0}}$ and
$Z^{\aaa_{s-1}}{}_{\aaa_{s}}$. 

\noi
{\sc General remarks about notation:}~~An integer subindex without parenthesis 
refers to ordinary antifield number, while an integer subindex in round or 
square parenthesis refers to shifted or reduced antifield number, respectively.
{}For instance, $\brst_{-1}$ denotes the ordinary Koszul--Tate operator, which 
carries antifield number $\afn(\brst_{-1})\!=\!-1$, while $\brst_{(-1)}$ is the 
shifted Koszul--Tate operator, which on the other hand carries shifted 
antifield number $\safn(\brst_{(-1)})\!=\!-1$.
{\em Strong} equality ``$=$'' and {\em weak} equality ``$\approx$'' refer to 
off--shell and on--shell equality \wrtt equations of motion for the original
fields $\varphi^{i}$, respectively.

\section{Gauge Theories}
\label{secgt}

\subsection{Starting Point}
\label{secstartingpt}

\noi
In its purest form, the starting point for quantization is just an action 
$S^{}_{0}\!=\!S^{}_{0}(\varphi)$, which depends on a set of fields
$\varphi^{i}$, $i\!\in\!\{1,\ldots,n\!\equiv\!\mmm_{-1}\}$. We shall hereafter
refer to $S^{}_{0}$ and $\varphi^{i}$ as the original action and the original
fields, respectively.

\noi
Under some mild regularity assumptions, it is possible to recast the starting
point into a form that we will use in this paper. ({}For a 
collection of starting points in the irreducible case, see Postulates 
2--2$'''$ in \Ref{bv85}.) Explicitly, we 
will assume that all the gauge--symmetries, and in the reducible case, all the 
gauge(--for--gauge)$^s$--symmetries, $s\!\in\!\{1,\ldots,L\}$, have been 
properly identified. This means, in terms of formulas, that there should be
given gauge--generators 
$R^{i}{}_{\aaa_{0}}\!\equiv\!Z^{\aaa_{-1}}{}_{\aaa_{0}}$, 
and gauge(--for--gauge)$^s$--generators $Z^{\aaa_{s-1}}{}_{\aaa_{s}}$, 
$s\!\in\!\{1,\ldots,L\}$, such that
\bea
(S^{}_{0}\papar{\varphi^{i}})R^{i}{}_{\aaa_{0}}&=&0\~,\label{noether1} \\
R^{i}{}_{\aaa_{0}}Z^{\aaa_{0}}{}_{\aaa_{1}}
&=&(S^{}_{0}\papar{\varphi^{j}})B^{ji}_{\aaa_{1}}\~,\label{noether15}\\
Z^{\aaa_{s-2}}{}_{\aaa_{s-1}}Z^{\aaa_{s-1}}{}_{\aaa_{s}}
&=&(S^{}_{0}\papar{\varphi^{i}})B^{i\aaa_{s-2}}_{\aaa_{s}}
\~,\~\~\~\~\~\~\~\~s\in\{1,\ldots,L\}\~,\label{noether2}
\eea
where the indices runs over the following sets
\beq
i\!\equiv\!\aaa_{-1}\in\{1,\ldots,n\!\equiv\!\mmm_{-1}\}\~,\~\~\~\~
\aaa_{0}\in\{1,\ldots,\mmm_{0}\}
\~,\~\~\~\~\ldots\~,\~\~\~\~
\aaa_{L}\in\{1,\ldots,\mmm_{L}\}\~.
\eeq
The letter $L$ denotes the number of reducibility stages of the theory. A
theory with reducibility stage $L\!=\!0$ equal to zero is by definition 
an {\em irreducible} theory. {}For each stage $s\!\in\!\{0,\ldots,L\}$,
the number of gauge(--for--gauge)$^s$--symmetries is denoted $\mmm_{s}$.
It is convenient to define multiplicities 
$\mmm_{s}:=0$ for $s>L$. (Or, the other way around, 
$L=\min\{s|\forall r\!>\! s:\mmm_{r}\!=\!0\}$.) The rank conditions read
\beq
0\~\leq\~\rank(\papal{\varphi^{i}}S^{}_{0}\papar{\varphi^{j}})\= \MMM_{-1}
\~,\~\~\~\~\~\~\~\~0\~\leq\~\rank(R^{i}{}_{\aaa_{0}})\= \MMM_{0}\~,
\label{rankcondition1}
\eeq
\beq
0\~\leq\~\rank(Z^{\aaa_{s-1}}{}_{\aaa_{s}})\= \MMM_{s}
\~,\~\~\~\~\~\~\~\~s\in\{0,\ldots,L\}\~, \label{rankcondition2}
\eeq
{\em near} the stationary $\varphi$-surface, \ie the $\varphi$-surface of 
extremals for $S^{}_{0}$. Here we have defined
\beq
\MMM_{s}\~:=\~\sum_{r=0}^{\infty}(-1)^{r} \mmm_{r+s}
\=\mmm_{s}-\MMM_{s+1}\~\leq\~\mmm_{s}\~. \label{mandm}
\eeq
The second equality in \eq{mandm} yields a recursion relation for $\MMM_{s}$.
Note that $\MMM_{s}\!=\!0$ for $s>L$. 
The word {\em near} means in some tubular $\varphi$-neighborhood of the
stationary $\varphi$-surface. Note that it is not enough to impose rank 
conditions only at the stationary $\varphi$-surface, because the rank could 
jump once one leaves the stationary $\varphi$-surface. This is not allowed. 
Hence one must impose the rank conditions in an {\em open} 
$\varphi$-neighborhood of the stationary $\varphi$-surface.

\noi
{\em A priori} (outside the field--antifield formalism) it could 
in principle happen that the given $B^{ji}_{\aaa_{1}}$ structure function in 
the first--stage Noether identity \e{noether15} does not have 
$i\!\leftrightarrow\!j$ skewsymmetry \e{ktp1}. Nevertheless, one may show
that there locally always exists a $i\!\leftrightarrow\!j$ skewsymmetric
$B^{ji}_{\aaa_{1}}$ structure function satisfying the first--stage Noether
identity \e{noether15}, \cf Appendix~\ref{secbanti}. On the other hand, within
the field--antifield formalism, the $B^{ji}_{\aaa_{1}}$ structure function is
manifestly $i\!\leftrightarrow\!j$ skewsymmetric, since it arises from an
action term
$\Hf(-1)^{\eps_{j}}\varphi^{*}_{j}\varphi^{*}_{i}B^{ij}_{\aaa_{1}}c^{\aaa_{1}}$,
\cf Section~\ref{secreduc}. We shall therefore for simplicity assume from now
on, that the $B^{ji}_{\aaa_{1}}$ structure function, 
that appears in the first--stage Noether identity \e{noether15}, 
is a $i\!\leftrightarrow\!j$ skewsymmetric tensor, that is defined in at least 
some tubular $\varphi$-neighborhood of the stationary $\varphi$-surface.

\noi
The choices of gauge(--for--gauge)$^s$--generators $Z^{\aaa_{s-1}}{}_{\aaa_{s}}$,
$s\!\in\!\{0,\ldots,L\}$, are not unique. An arbitrary other choice (with the
same multiplicities) is locally given as
\beq
\bar{R}^{i}{}_{\bbbb_{0}}\=R^{i}{}_{\aaa_{0}}
(\Lambda^{-1})^{\aaa_{0}}{}_{\bbbb_{0}}
+(S^{}_{0}\papar{\varphi^{j}})K^{ji}_{\bbbb_{0}}\~,
\~\~\~\~\~\~\~\~
K^{ji}_{\bbbb_{0}}
\=-(-1)^{\eps_{i}\eps_{j}}K^{ij}_{\bbbb_{0}}\~,\label{genambi1}
\eeq
\beq
\bar{Z}^{\bbbb_{s-1}}{}_{\bbbb_{s}}
\~\approx\~\Lambda^{\bbbb_{s-1}}{}_{\aaa_{s-1}}Z^{\aaa_{s-1}}{}_{\aaa_{s}}
(\Lambda^{-1})^{\aaa_{s}}{}_{\bbbb_{s}}
\~,\~\~\~\~\~\~\~\~s\!\in\!\{1,\ldots,L\}\~.\label{genambi2}
\eeq
To summarize, we shall assume that some gauge(--for--gauge)$^s$--generators, 
$Z^{\aaa_{s-1}}{}_{\aaa_{s}}$, $s\!\in\!\{0,\ldots,L\}$, have been given
to us (perhaps outside the field--antifield formalism), and they are hereafter 
considered as a part of the starting point. 

\begin{table}[t]
\caption{Multiplicity, Grassmann parity, ghost number, antifield number of 
the fundamental variables $\Gamma^{A}$.}

\label{multtable1}
\begin{center}
\begin{footnotesize}
\begin{tabular}{|l|c||c|c|c|c|c|c|}  \hline
Variable/Operator&Symbol&Multi-&Grass-&Ghost &Anti-&Shifted&Reduced \\ 
&&plicity&mann&number&field&antifield&antifield \\
&&&parity&&number&number&number \\ \hline\hline
Generic variable&$\Gamma^{A}
$&$2N$&$\eps_{A}$&$\gh_{A}$&$\afn_{A}$
&$\safn_{A}$&$\rafn_{A}$\rule[-1.5ex]{0ex}{4.5ex} \\ \hline\hline
{}Field&$\Phi^{\alpha}$&$N$&$\eps_{\alpha}$&$\gh_{\alpha}$&$0$&$\geq 0$&$\geq 0$
\rule[-1.5ex]{0ex}{4.5ex} \\ \hline
Original field&$\varphi^{i}\!\equiv\!\Phi^{\aaa_{-1}}$&$n\!\equiv\!\mmm_{-1}$
&$\eps_{i}\!\equiv\!\eps_{\aaa_{-1}}$&$0$&$0$&$\geq 0$&$\geq 0$
\rule[-1.5ex]{0ex}{4.5ex} \\ \hline
Transversal field&$\xi^{I}$&$\MMM_{-1}$&$\eps_{I}$
&$0$&$0$&$1$&$1$\rule[-1.5ex]{0ex}{4.5ex}\\ \hline
Longitudinal field&$\theta^{\AAA_{0}}$&$\MMM_{0}$&$\eps_{\AAA_{0}}$
&$0$&$0$&$0$&$0$\rule[-1.5ex]{0ex}{4.5ex}\\ \hline
Stage-$s$ ghost&$c^{\aaa_{s}}\!\equiv\!\Phi^{\aaa_{s}}$&$\mmm_{s}$
&$\eps_{\aaa_{s}}\!+\!s\!+\!1$&$s\!+\!1$&$0$&$0$&$0$
\rule[-1.5ex]{0ex}{4.5ex} \\ \hline\hline
Antifield&$\Phi^{*}_{\alpha}$&$N$&$\eps_{\alpha}\!+\!1$&$-\gh_{\alpha}\!-\!1$
&$\gh_{\alpha}\!+\!1$&$\gh_{\alpha}\!+\!2$&$1$ 
\rule[-1.5ex]{0ex}{4.5ex}  \\ \hline
Original antifield&$\varphi^{*}_{i}\!\equiv\!\Phi^{*}_{\aaa_{-1}}$
&$n\!\equiv\!\mmm_{-1}$&$\eps_{i}\!+\!1\!\equiv\!\eps_{\aaa_{-1}}\!+\!1$
&$-1$&$1$&$2$&$1$ \rule[-1.5ex]{0ex}{4.5ex} \\ \hline
Stage-$s$ ghost antifield&$c^{*}_{\aaa_{s}}\!\equiv\!\Phi^{*}_{\aaa_{s}}$
&$\mmm_{s}$&$\eps_{\aaa_{s}}\!+\!s$
&$-(s\!+\!2)$&$s\!+\!2$&$s\!+\!3$&$1$ \rule[-1.5ex]{0ex}{4.5ex} \\ \hline\hline
BRST operator&$\brst\!=\!(S,\cdot)$&$1$&$1$&$1$&$\geq-1$&$\geq-1$&$\geq 0$
\rule[-1.5ex]{0ex}{4.5ex} \\ \hline
Koszul-Tate operator&$\brst_{-1}$&$1$&$1$&$1$&$-1$&$\geq-1$&$\geq 0$
\rule[-1.5ex]{0ex}{4.5ex} \\ \hline
Shifted/Reduced KT op.&$\delta\!=\!\brst_{(-1)}\!=\!\brst_{-1[0]}$
&$1$&$1$&$1$&$-1$&$-1$&$0$\rule[-1.5ex]{0ex}{4.5ex} \\ \hline
Contracting homotopy op.&$\delta^{-1}$
&$1$&$1$&$-1$&$1$&$1$&$0$\rule[-1.5ex]{0ex}{4.5ex} \\ \hline
\end{tabular}
\end{footnotesize}
\end{center}
\end{table}

\subsection{Field--Antifield Formulation}
\label{secreduc}

\noi
Let us now reformulate the problem in the field--antifield language
\cite{bv81,bv83}. We shall for simplicity only consider the minimal sector.
(The non--minimal sector, which is needed for gauge--fixing, can be treated by
similar methods.) The minimal content of fields $\Phi^{\alpha}$, 
$\alpha\!\in\!\{1,\ldots,N\}$, for a gauge theory of reducible stage $L$, 
is
\beq
\Phi^{\alpha}
\=\{\varphi^{i}\!\equiv\!\Phi^{\aaa_{-1}};
c^{\aaa_{0}}\!\equiv\!\Phi^{\aaa_{0}};\ldots;
c^{\aaa_{L}}\!\equiv\!\Phi^{\aaa_{L}}\}
\~,\~\~\~\~\~\~\~\~\alpha\~\in\~\{1,\ldots,N\}\~.
\eeq
The $c^{\aaa_{0}}$ fields are the ghosts, or in a systematical terminology, 
the stage--zero ghosts. The $c^{\aaa_{s}}$ fields are the 
(ghost--for)$^{s}$--ghosts, or stage-$s$ ghost, $s\!\in\!\{1,\ldots,L\}$.
{}For Grassmann parity and ghost number assignments, see
Table~\ref{multtable1}. To simplify notation, the stage $s$ of a ghost
$c^{\aaa_{s}}$ can only be identified though its index--variable $\aaa_{s}$.
(Hopefully, this slight misuse of notation does not lead to confusion.) 
It is tempting to call the original fields $\varphi^{i}$ for stage--minus--one 
ghosts $c^{\aaa_{-1}}\!\equiv\!\varphi^{i}$, but we shall not do so. {}From now
on, the letter ``$c$'' without indices, and the word {\em ghost}, will always
refer to a ghost with strictly positive ghost number. In particular, the symbol
$c^{*}$ will {\em not} refer to original antifields $\varphi^{*}$, but only the
antifields for the ghost ``$c$'', \ie $\gh(c^{*})\leq -2$. We will collectively
refer to ghosts ``$c$'' and antifields ``$\Phi^{*}$'' as {\em auxiliary}
variables. Auxiliary variables are characterized by non--zero ghost number. 

\noi
{\sc Remark}:~~
Mathematically, the various stucture functions, such as, \eg 
$R^{i}{}_{\aaa_{0}}\!=\!R^{i}{}_{\aaa_{0}}(\varphi)$ and 
$B^{ji}_{\aaa_{1}}\!=\!B^{ji}_{\aaa_{1}}(\varphi)$, are tensors, 
or sections of appropriate vector bundles over the $\varphi$-basemanifold. 
They should be defined in at least some tubular $\varphi$-neighborhood of the
stationary $\varphi$-surface. The auxiliary variables can be thought of as a
local basis for the corresponding vector bundle. {}For instance, the antifield
\beq
\varphi^{*}_{i}\=(\papal{\varphi^{i}}\varphi^{\prime j})\varphi^{\prime *}_{j}
\=\varphi^{\prime *}_{j}(\varphi^{\prime j}\papar{\varphi^{i}})
\eeq
transforms as a co--vector under general coordinate transformations
$\varphi^{i}\to\varphi^{\prime j}$, and can thus be identified with a local 
basis $\partial_{i}$ for vector fields $X\!=\!X^{i}\partial_{i}$. On the other 
hand, the indices $\aaa_{s}$, $s\!\in\!\{0,\ldots,L\}$, do only transform 
under ridig $\varphi$-independent transformations, and the ghost $c^{\aaa_{s}}$
and their antifields $c^{*}_{\aaa_{s}}$ can often be taken as global coordinates.

\noi
The total number $N$ of fields is
\beq
N:=\sum_{s=-1}^{L}\mmm_{s}\= \sum_{s=-1}^{\infty}\mmm_{s}\~.
\eeq
{}For each field $\Phi^{\alpha}$, one introduces an antifield
$\Phi^{*}_{\alpha}$ of opposite Grassmann number 
$\eps(\Phi^{*}_{\alpha})\!=\!\eps(\Phi^{\alpha})\!+\!1$ and of ghost number
$\gh(\Phi^{*}_{\alpha})\!=\!-\gh(\Phi^{\alpha})\!-\!1$. The antibracket is 
defined as
\beq 
(f,g)\~:=\~f\left(\papar{\Phi^{\alpha}}\papal{\Phi^{*}_{\alpha}}
-\papar{\Phi^{*}_{\alpha}}\papal{\Phi^{\alpha}}\right)g\~.
\eeq
The problem that we address in this paper is the existence of a {\em proper}
solution $S$ with ghost number zero
\beq
\gh(S)\=0 \label{ghostcons}
\eeq
to the classical master equation
\beq
(S,S)\=0\~, \label{cmej}
\eeq 
such that $S$ has the correct original limit
\beq
S\= S^{}_{0}+\cO(\Phi^{*})\~\equi{\e{ghostcons}}\~S^{}_{0}+\cO(c)
\~,\label{origlimit}
\eeq
and satisfies the properness condition
\beq
\left. \rank(\papal{\Gamma^{A}}S\papar{\Gamma^{B}})\right|_{\Phi^{*}=0=c}\=N 
\label{properness1}
\eeq
at the stationary $\varphi^{i}$-surface, 
when all the auxiliary variables are put to zero,
$\Phi^{*}_{\alpha}\!=\!0$ and $c^{\aaa_{s}}\!=\!0$, $s\!\in\!\{0,\ldots,L\}$.
Here $\Gamma^{A}\!\equiv\!\{\Phi^{\alpha};\Phi^{*}_{\alpha}\}$, 
$A\!\in\!\{1,\ldots,2N\}$, is a collective notation for both fields 
$\Phi^{\alpha}$ and antifields $\Phi^{*}_{\alpha}$. 

\noi
The half rank $N$ is the maximal possible for a solution $S$ to the classical
master equation. The properness condition \e{properness1} is important, because
it implies (after the non--minimal sector has been included) that the
gauge--fixed proper action $S(\Phi,\Phi^{*}\!=\!\partial \Psi/\partial \Phi)$
is free of flat directions.

\noi
We will require one more condition besides eqs.\ 
\e{ghostcons}--\e{properness1}. It will encode the gauge(--for--gauge)$^s$
symmetries, $s\!\in\!\{0,\ldots,L\}$, into the proper action $S$. To explain
it, lets us divide an arbitrary proper solution $S$ into two parts,
\beq
S\=\Squad+S^{\rm non-quad}\~. \label{twosplit}
\eeq  
The first part $\Squad$ contains all terms that are at most quadratic in 
auxiliary variables, while the second part
$S^{}_{\rm non-quad}\!=\!\cO\left((\Phi^{*})^{2},c^{2}\right)$ contains all
terms that are at least cubic in auxiliary variables, which actually means all 
terms at least quadratic in ghosts or antifields, due to ghost number 
conservation \e{ghostcons}.
Because of ghost number conservation \e{ghostcons} and the original limit
requirement \e{origlimit}, the action $\Squad$ must be of the form 
\beq
\Squad\=S^{}_{0}+\sum_{s=0}^{L}\Phi^{*}_{\aaa_{s-1}}
Z^{\aaa_{s-1}}{}_{\aaa_{s}}\~c^{\aaa_{s}}
\=S^{}_{0}+\varphi^{*}_{i}R^{i}{}_{\aaa_{0}}\~c^{\aaa_{0}}
+\sum_{s=1}^{L}c^{*}_{\aaa_{s-1}}Z^{\aaa_{s-1}}{}_{\aaa_{s}}\~c^{\aaa_{s}}\~.
\label{sabel}
\eeq
In particular, there are no action terms that are linear in the auxiliary 
variables. {\em A priori} the structure functions 
$Z^{\aaa_{s-1}}{}_{\aaa_{s}}\!=\!Z^{\aaa_{s-1}}{}_{\aaa_{s}}(\varphi)$ in
the action \e{sabel} could be different from the given 
gauge(--for--gauge)$^s$--generators, $s\!\in\!\{0,\ldots,L\}$, specified in
\eqs{noether1}{noether2}. However, the classical master equation \e{cmej}
implies that the structure functions $Z^{\aaa_{s-1}}{}_{\aaa_{s}}$ in
\eq{sabel} also satisfy the Noether identities \e{noether1}--\e{noether2}, 
although perhaps with some other $B^{i\aaa_{s-2}}_{\aaa_{s}}$ structure functions. 
We will therefore demand that the structure functions $R^{i}{}_{\aaa_{0}}$ and 
$Z^{\aaa_{s-1}}{}_{\aaa_{s}}$ in \eq{sabel} are the same as the given gauge
generator $R^{i}{}_{\aaa_{0}}$ and gauge(--for--gauge)$^s$--generators
$Z^{\aaa_{s-1}}{}_{\aaa_{s}}$ in the Noether identities \e{noether1}--\e{noether2},
respectively.

\noi
With the above identification, the rank of the $\Squad$ part \e{sabel} of the 
Hessian is
\beq
\left. \rank(\papal{\Gamma^{A}}\Squad\papar{\Gamma^{B}})\right|_{T=0}
\~\equi{\e{rankcondition1}+\e{rankcondition2}}\~
\MMM_{-1}+2\sum_{s=0}^{\infty}\MMM_{s}
\=\sum_{s=-1}^{\infty}(\MMM_{s}+\MMM_{s+1})
\~\equi{\e{mandm}}\~\sum_{s=-1}^{\infty}\mmm_{s}\=N \label{bvmagic}
\eeq
at the stationary $\varphi$-surface
\beq
\EOM_{i}\~\equiv\~(S^{}_{0}\papar{\varphi^{i}}) \= 0\~, 
\eeq
as a result of the rank conditions \es{rankcondition1}{rankcondition2}. So the
full action $S\!=\!\Squad\!+\!S^{\rm non-quad}$ is then guaranteed to be proper.
The properness condition for the solution $S$ at the stationary
$\varphi$-surface implies, by continuity, properness of the $S$ solution in
some sufficiently small $\varphi$-neighborhood of the stationary
$\varphi$-surface.

\subsection{Classical BRST Operator}
\label{secbrst}

\noi
If a proper solution $S$ to eqs.\ \e{ghostcons}--\e{properness1} exists, the
corresponding classical BRST operator is defined as
\beq
\brst\~:=\~(S,\~\cdot\~)\~. \label{clasbrstop}
\eeq
It is nilpotent
\beq
\brst{}^{2}\=(S,(S,\~\cdot\~))\=\Hf ((S,S),\~\cdot\~)\=0\~,
\eeq
as a consequence of the Jacobi identity for the antibracket and the classical 
master equation \e{cmej}.

\begin{theorem}[Acyclicity of the BRST operator]
Let $S$ be a proper solution defined in a tubular $\varphi$-neighborhood of the 
stationary $\varphi$-surface. Then the cohomology of the BRST operator
$\brst\!=\!(S,\cdot)$ is acyclic, \ie
\beq
\forall {\it~functions~}f:\~\~\~\~\brst(f)=0 \~\~\wedge\~\~ \gh(f)<0
\~\~\~\~\Rightarrow\~\~\~\~
\exists g:\~ f= \brst(g) \label{brstacyclic2}
\eeq
in some tubular $\varphi$-neighborhood of the stationary $\varphi$-surface. 
\label{theorembrst2}
\end{theorem}

\noi
{\sc Proof of Theorem~\ref{theorembrst2}}:~~
Use Theorem~\ref{theoremkt1}, Theorem~\ref{theoremkt15},
Lemma~\ref{theoremlemma}, and the fact that antifield number is (weakly)
greater than minus ghost number, \ie
$\afn(f)\geq-\gh(f)>0$, since $\gh(f)\!+\!\afn(f)=\puregh(f)\geq 0$.
\proofbox

\section{Standard Methods}
\label{secfakeexist}

\subsection{Introduction}
\label{secfakeintro}

\noi
In this Section~\ref{secfakeexist} we try to attack the problem of existence 
of a field--antifield formulation for reducible theories by applying the
standard method that is known to work in the irreducible case
\cite{vt82,bv85,fishen90,hentei92}, \ie generating a proper action $S$ from a 
Koszul--Tate operator $\brst_{-1}$ and keeping track of antifield number
``$\afn$''. Unfortunately, it is cumbersome to directly verify the existence
and acyclicity of the Koszul--Tate operator $\brst_{-1}$ in the reducible case
using this standard approach. The standard method will nevertheless serve as a
simplified template, on which we will develop a complete existence proof in the
next Section~\ref{secrealexist}, using a reduced (shifted) Koszul--Tate 
operator $\brst_{(-1)}$ and a (shifted) antifield number ``$\safn$'', 
respectively.

\subsection{Antifield Number}
\label{secantigh}

\noi
The antifield number ``$\afn$'' is defined as zero for fields $\Phi^{\alpha}$,
\ie $\afn(\Phi^{\alpha})\!=\!0$, and it is defined as minus the ghost number
for antifields $\Phi^{*}_{\alpha}$, \ie
$\afn(\Phi^{*}_{\alpha})\!=\!-\!\gh(\Phi^{*}_{\alpha})$. See
Table~\ref{multtable1}. 
Any action $S$ of ghost number zero can be expanded \wrt antifield number.
\beq
S\=\sum_{r=0}^{\infty}S_{r}\~,\~\~\~\~\~\~\~\~\afn(S_{r})\=r\~.
\eeq 
Let us also expands the antibracket $(\cdot,\cdot)$ according to antifield 
number.
\beq
(f,g)\=\sum_{s=1}^{L+2}(f,g)^{}_{-s}\=\sum_{s=-1}^{L}(f,g)^{}_{-s-2}
\~,\~\~\~\~\~\~\~\~
\afn(f,g)^{}_{r}\=\afn(f)+\afn(g)+r\~,
\eeq
\beq
(f,g)^{}_{-s-2}
\~:=\~f\left(\papar{\Phi^{\aaa_{s}}}\papal{\Phi^{*}_{\aaa_{s}}}
-\papar{\Phi^{*}_{\aaa_{s}}}\papal{\Phi^{\aaa_{s}}}\right)g 
\~,\~\~\~\~\~\~\~\~ s\in\{-1,\ldots,L\}\~.
\eeq
Elementary considerations reveal the following useful Lemma~\ref{lemmass1}.

\begin{lemma}
Let $f,g$ be two functions of definite antifield number. Then
\beq
(f,g)^{}_{-s}\~\neq\~0
\~\~\~\~\~\~\~\~\Rightarrow\~\~\~\~\~\~\~\~
\afn(f)\~\geq\~s\~\~\vee\~\~\afn(g)\~\geq\~s\~. \label{lemmass1a}  
\eeq
Assume that $f$ also has definite ghost number. Then
\beq
(f,g)^{}_{-s}\~\neq\~0\~\~\wedge\~\~\gh(f)\~\geq\~-1
\~\~\~\~\~\~\~\~\Rightarrow\~\~\~\~\~\~\~\~
\puregh(f)\~\equiv\~\afn(f)+\gh(f)\~\geq\~s\!-\!1\~.\label{lemmass1ab} 
\eeq
In particular,
\beq
(f,g)^{}_{-s}\~\neq\~0\~\~\wedge\~\~\gh(f)\=0
\~\~\~\~\~\~\~\~\Rightarrow\~\~\~\~\~\~\~\~
\afn(f)\~\geq\~s\!-\!1\~.\label{lemmass1b} 
\eeq
\label{lemmass1}
\end{lemma}

\noi
{\sc Proof of Lemma~\ref{lemmass1}}:~~
There are only two possibilities. The first case is
\beq
\exists \aaa_{s-2}\~:\~\~\~\~
(\papal{\Phi^{*}_{\aaa_{s-2}}}f)\~\neq\~0\~\~\~\~\wedge\~\~\~\~
(\papal{\Phi^{\aaa_{s-2}}}g)\~\neq\~0\~, \label{firstcase}
\eeq
and the second case is with the r\^oles of $f$ and $g$ interchanged, 
\beq
\exists \aaa_{s-2}\~:\~\~\~\~
(\papal{\Phi^{\aaa_{s-2}}}f)\~\neq\~0\~\~\~\~\wedge\~\~\~\~
(\papal{\Phi^{*}_{\aaa_{s-2}}}g)\~\neq\~0\~. \label{secondcase}
\eeq
In the first case, $\afn(f)\!\geq\!\afn(\Phi^{*}_{\aaa_{s-2}})\!=\!s$.
If also $\gh(f)\!\geq\!-1$, then 
$\puregh(f)\!\equiv\!\afn(f)\!+\!\gh(f)\!\geq\!s-1$.
In the second case, $\afn(g)\!\geq\!\afn(\Phi^{*}_{\aaa_{s-2}})\!=\!s$,
and $\puregh(f)\!\geq\!\puregh(\Phi^{\aaa_{s-2}})\!=\!s\!-\!1$.
\proofbox

\subsection{Koszul--Tate Operator $\brst_{-1}$}
\label{seckt1}

\noi
If the proper solution $S$ exists, the Koszul--Tate operator $\brst_{-1}$ is 
then defined as the leading antifield number sector for the classical BRST
operator,
\beq
\brst\~:=\~(S,\~\cdot\~)\equi{\e{lemmass1b}}\sum_{p=-1}^{\infty}\brst_{p}\~,
\eeq
where
\beq
\brst_{p}\=\sum_{r=0}^{\infty}(S^{}_{r},\~\cdot\~)^{}_{p-r}
\~,\~\~\~\~\~\~\~\~\afn(\brst_{p})\=p
\~,\~\~\~\~\~\~\~\~p\~\in\~\{-1,0,1,2,\ldots\}\~.
\eeq
The Koszul--Tate operator $\brst_{-1}$ is of the form
\beq
\brst_{-1}\=
\sum_{r=0}^{L+1}(S^{}_{r},\~\cdot\~)^{}_{-r-1}
\=V^{}_{\alpha}\papal{\Phi^{*}_{\alpha}}
\=\EOM_{i}\papal{\varphi^{*}_{i}}
+\sum_{s=0}^{L}\Phi^{*}_{\aaa_{s-1}}Z^{\aaa_{s-1}}{}_{\aaa_{s}}
\papal{c^{*}_{\aaa_{s}}}
+M^{}_{\alpha}\papal{\Phi^{*}_{\alpha}}\~,\label{kt1}
\eeq
where the functions $M^{}_{\alpha}\!=\!\cO((\Phi^{*})^{2})$ are at least
quadratic in the antifields. It follows from the requirement
\beq
\puregh(V^{}_{\alpha})\=\puregh(\brst_{-1})
\=\gh(\brst_{-1})\!+\!\afn(\brst_{-1})\=1\!-\!1\=0
\eeq
that the functions $V^{}_{\alpha}\!=\!V^{}_{\alpha}(\varphi,\Phi^{*})$ cannot
depend on the ghosts ``$c$''. In more detail, the $V_{\alpha}$ functions read
\beq
V^{}_{\aaa_{-1}}\equiv V^{}_{i}\=\EOM_{i}\equiv(S^{}_{0}\papar{\varphi^{i}})
\~,\~\~\~\~\~\~\~\~
V^{}_{\aaa_{0}}\=\varphi^{*}_{i}R^{i}{}_{\aaa_{0}}\~,\label{vee1}
\eeq
\beq
V^{}_{\aaa_{s}}\=\Phi^{*}_{\aaa_{s-1}}Z^{\aaa_{s-1}}{}_{\aaa_{s}}
+M^{}_{\aaa_{s}}\~,\~\~\~\~\~\~\~\~s\in\{0,\ldots,L\}\~,\label{vee2}
\eeq
\beq
M^{}_{\alpha}\= \cO((\Phi^{*})^{2})\~,\~\~\~\~\~\~\~\~
\afn(V^{}_{\alpha})\=\afn(M^{}_{\alpha})
\=\afn(\Phi^{*}_{\alpha})\!-\!1\~.\label{vee3}
\eeq
(See also eqs.\ \e{ktm2}--\e{ktp3}.)
Phrased differently, the classical BRST operator ``$\brst$'' is a deformation
of the Koszul--Tate operator, when one uses the antifield number ``$\afn$'' as
a resolution degree. The classical master \eq{cmej} implies that the classical
BRST operator ``$\brst$'' is nilpotent $\brst{}^{2}\!=\!0$, and hence that the
Koszul--Tate operator $\brst_{-1}$ is nilpotent,
\beq
\left(\brst_{-1}\right)^{2}\=0\~\~\~\~\Leftrightarrow\~\~\~\~
V_{\alpha}(\papal{\Phi^{*}_{\alpha}}V_{\beta})\=0
\~\~\~\~\Leftrightarrow\~\~\~\~
\brst_{-1}\left(S^{}_{0}
+\sum_{s=0}^{L}V_{\aaa_{s}}c^{\aaa_{s}}\right)\=0\~.\label{ktnilp}
\eeq
The last rewriting shows that the proper action 
$S\!=\!S^{}_{0}\!+\!\sum_{s=0}^{L}V^{}_{\aaa_{s}}c^{\aaa_{s}}\!+\!\cO(c^{2})$ is
$\brst_{-1}$-invariant modulo terms that are at least quadratic in the ghosts 
``$c$''. Explicitly, the nilpotency of $\brst_{-1}$ implies the Noether
identities \e{noether1}--\e{noether2}, and their consistency relations, \eg
\eqs{bconrelb2}{bconrelb3}. 

\begin{theorem}[Local acyclicity of the Koszul--Tate operator]
Assume that the K--T operator
$\brst_{-1}\!=\!V_{\alpha}\papal{\Phi^{*}_{\alpha}}$ is nilpotent in some
$\varphi$-neighborhood of a $\varphi$-point on the stationary 
$\varphi$-surface, where the structure functions $V^{}_{\alpha}$ are of the 
form \e{vee1}--\e{vee3}. Then the cohomology of the Koszul--Tate operator 
$\brst_{-1}$ is acyclic in some $\varphi$-neighborhood of the $\varphi$-point, 
\ie 
\beq
\forall{\it~functions~}f:\~\~\~\~\brst_{-1}f=0 \~\~\wedge\~\~ 
\afn(f)>0\~\~\~\~\Rightarrow\~\~\~\~ 
\exists g:\~ f= \brst_{-1}g\~. \label{ktacyclic1}
\eeq 
\label{theoremkt1}
\end{theorem}

\noi
{\sc Proof of Theorem~\ref{theoremkt1} using reduced antifield number}:~~
Use Theorem~\ref{theoremkt5}, Lemma~\ref{theoremlemmaop}, and the fact that 
strictly positive antifield number implies strictly positive reduced antifield
number $\afn(f)>0\Rightarrow\rafn(f)>0 $, see Section~\ref{secrealexist}.
\proofbox

\noi
{\sc Proof of Theorem~\ref{theoremkt1} using shifted antifield number}:~~
Use Corollary~\ref{corollarykt2}, Lemma~\ref{theoremlemma}, and the fact that 
shifted antifield number is (weakly) greater than antifield number
$\safn(f)\geq\afn(f)>0$, see Section~\ref{secrealexist}.
\proofbox

\begin{theorem}[Globalization (Theorem 2 in \Ref{fishen89})]
Assume that for every $\varphi$-point on the stationary $\varphi$-surface 
there exists some nilpotent and acyclic Koszul--Tate operator in a local 
$\varphi$-neigh\-bor\-hood around the $\varphi$-point. Then there exists a
smooth, nilpotent and acyclic Koszul--Tate operator in a tubular 
$\varphi$-neighborhood of the whole stationary $\varphi$-surface.
\label{theoremkt15}
\end{theorem}

\noi
{\sc Sketched Proof of Theorem~\ref{theoremkt15}}:~~
Cover the stationary $\varphi$-surface with sufficiently small
$\varphi$-neighborhoods $U_{i}$ such that the local Koszul--Tate operators 
(which may depend on the $\varphi$-neigh\-bor\-hoods $U_{i}$) are nilpotent
and acyclic. Use a smooth partition of unity $\sum_{i}f_{i}\!=\!1$, 
$f_{i}\!=\!f_{i}(\varphi)$, ${\rm supp}f_{i}\subseteq U_{i}$, to define a global
Koszul--Tate operator $\brst_{-1}$. Here it is used that the Koszul--Tate 
operators do not act on the original $\varphi$-variables. 
\proofbox

\subsection{Existence of $S$}
\label{secexists1}

\noi
Now we would like to deduce a proper solution from the Koszul--Tate operator.

\begin{theorem}[Existence theorem for $S$ action \cite{bv85}]
Let there be given a nilpotent, acyclic Kos\-zul--Tate ope\-ra\-tor
$\brst_{-1}\!=\!V^{}_{\alpha}\papal{\Phi^{*}_{\alpha}}$, that is defined in a
$\varphi$-neighborhood $U$, with antifield number minus one, and where
$V^{}_{\alpha}\!=\!V^{}_{\alpha}(\varphi,\Phi^{*})$. This guarantees the 
existence of a proper solution 
$S\!=\!S^{}_{0}\!+\!\sum_{s=0}^{L}V^{}_{\aaa_{s}}c^{\aaa_{s}}\!+\!\cO(c^{2})$ 
to the classical master equation \e{cmej} in $U$.
\label{theoremss1}
\end{theorem}

\noi
{\sc Proof of Theorem~\ref{theoremss1}}:~~Define the classical master
expression
\beq
\CME\~:=\~\Hf(S,S) \label{cmxj}
\eeq 
and the Jacobiator 
\beq
\JI\~:=\~\Hf(S,(S,S))\~\equi{\e{clasbrstop}}\~\brst(\CME)\~. \label{jij}
\eeq
It follows from Lemma~\ref{lemmass1} that the $r$th classical master 
expression $\CME_{r}$ can be written as
\bea
\CME_{r}\~:=\~\Hf \sum_{p,q \geq 0} (S^{}_{p},S^{}_{q})^{}_{r-p-q}
\equi{\e{lemmass1b}}
\Hf \sum_{0\leq p,q\leq r+1} (S^{}_{p},S^{}_{q})^{}_{r-p-q} 
\equi{\e{lemmass1a}}
\brst_{-1}S^{}_{r+1}+\cB^{}_{r}\~,
\label{cmxdecomp1}
\eea
where 
\beq
\cB^{}_{r}\~:=\~\Hf \sum_{0\leq p,q\leq r} (S^{}_{p},S^{}_{q})^{}_{r-p-q}
\~\equi{r-p-q<0}\~\Hf \sum_{1\leq p,q\leq r} (S^{}_{p},S^{}_{q})^{}_{r-p-q}
\=\cO(c^{2})\~,\~\~\~\~r\~\geq\~0\~.
\label{bdef1}
\eeq
The second equality of \eq{bdef1} follows because the antibracket itself has 
negative antifield number. The third equality of \eq{bdef1} follows from the
$\brst_{-1}$ nilpotency \e{ktnilp}. The proof of the main statement is an
induction in the antifield number $r$. Assume that there exist
\beq
S^{}_{0}\~,\~\~\~\~S_{1}\=V^{}_{\aaa_{0}}c^{\aaa_{0}}\~,\~\~\~\~
S^{}_{2}\=V^{}_{\aaa_{1}}c^{\aaa_{1}}+\cO(c^{2})
\~,\~\~\~\~\ldots\~,\~\~\~\~
S^{}_{r}\=V^{}_{\aaa_{r-1}}c^{\aaa_{r-1}}+\cO(c^{2})\~,
\eeq 
such that
\beq
0\=\CME_{0}\=\CME_{1}\=\ldots\=\CME_{r-1}\~.\label{cmxassump1}
\eeq
It follows that the $\cB^{}_{r}$ function \e{bdef1} exists as well, because 
$\cB^{}_{r}$ only depends on the previous $S^{}_{\leq r}$. We want to prove that
there exists $S^{}_{r+1}$ such that $\CME_{r}\!=\!0$. The Jacobi identity
$\JI\!=\!0$ gives
\beq
0\=\JI_{r-1}\~\equi{\e{jij}}\~\sum_{p=-1}^{\infty}\brst_{p}\CME_{r-p-1}
\~\equi{\e{cmxassump1}}\~\brst_{-1}\CME_{r}
\~\equi{\e{cmxdecomp1}}\~\brst_{-1}\cB^{}_{r}\~,
\eeq
so $\cB^{}_{r}$ is $\brst_{-1}$-closed. If $r\!=\!0$, one defines
\beq
S^{}_{1}\~:=\~V^{}_{\aaa_{0}}c^{\aaa_{0}}
\=\varphi^{*}_{i}R^{i}{}_{\aaa_{0}}c^{\aaa_{0}}\~.
\eeq
Then
\beq
\CME_{0}\equi{\e{cmxdecomp1}}\brst_{-1}S^{}_{1}+\cB^{}_{0}\=0+0\=0\~,
\eeq
because of the Noether identity \e{noether1}. If $r\!>\!0$, then the
acyclicity condition \e{ktacyclic1} shows that there exists a function 
$S^{}_{r+1}\!=\!V^{}_{\aaa_{r}}c^{\aaa_{r}}\!+\!\cO(c^{2})$ such that 
$-\brst_{-1}S^{}_{r+1}\!=\!\cB^{}_{r}\!=\!\cO(c^{2})$. Here we used that
Koszul--Tate operator $\brst_{-1}$ does not depend on the ghosts
$c^{\aaa_{s}}$, $s\!\in\!\{0,\ldots,L\}$.
\proofbox

\subsection{Anticanonical Transformations}
\label{secanticantrans}

\noi
Consider finite anticanonical transformations $e^{(\Psi,\~\cdot\~)}$, where
$\Psi$ is a Grassmann--odd generator with ghost number minus one, 
$\gh(\Psi)\!=\!-1$. Such transformations form a group under composition
\beq
e^{(\Psi_{1},\~\cdot\~)} e^{(\Psi_{2},\~\cdot\~)}
\=e^{({\rm BCH}(\Psi_{1},\Psi_{2}),\~\cdot\~)} \~,
\eeq
where
\bea
{\rm BCH}(\Psi_{1},\Psi_{2})
&=&\Psi_{1}+\int_{0}^{1}\! dt \sum_{n=0}^{\infty}\frac{(-1)^{n}}{n+1}
\left[e^{-t(\Psi_{2},\~\cdot\~)}e^{-(\Psi_{1},\~\cdot\~)}-1\right]^{n}\Psi_{2}
\cr
&=&\Psi_{1}+\Psi_{2}+\Hf(\Psi_{1},\Psi_{2})+
\frac{1}{12}(\Psi_{1},(\Psi_{1},\Psi_{2}))
+\frac{1}{12}((\Psi_{1},\Psi_{2}),\Psi_{2}))+\cO(\Psi_{i}^{4})
\eea
is the Baker--Campbell--Hausdorff series expansion (with Lie brackets replaced 
by antibrackets). 

\noi
Note that an anticanonical transformation $e^{(\Psi,\~\cdot\~)}S$ of a solution
$S$ to the classical master \eq{cmej} is again a solution to the classical 
master \eq{cmej}. We shall now show that any two proper solutions are related 
via an anticanonical transformation modulo global obstructions.

\begin{theorem}[Natural arbitrariness/ambiguity of proper solution] Let $S$
and $\bar{S}$ be two pro\-per solutions to the classical master \eq{cmej}
with ghost number zero, with correct original limit \e{origlimit}, and
defined in some $\varphi$-neighborhood of a $\varphi$-point on the stationary 
$\varphi$-surface.
Then there exists a Grassmann--odd generator $\Psi$ with ghost
number minus one, $\gh(\Psi)\!=\!-1$, and of order $\Psi\!=\!\cO(c)$, such that
\beq
\bar{S}\=e^{(\Psi,\~\cdot\~)}S 
\eeq
in some $\varphi$-neighborhood of the $\varphi$-point.
If $S$ and $\bar{S}$ also have the same gauge(--for--gauge)$^s$--generators,
$Z^{\aaa_{s-1}}{}_{\aaa_{s}}\!=\!\bar{Z}^{\aaa_{s-1}}{}_{\aaa_{s}}$,
$s\!\in\!\{0,\ldots,L\}$, then the Grassmann--odd generator $\Psi$ can be 
chosen of the form $\Psi\!=\!\cO((\Phi^{*})^{2},c^{2})$.
\label{theorembrst3}
\end{theorem}

\noi
{\sc Remark}:~~The essential ingredient of the proof of  
Theorem~\ref{theorembrst3}, is the acyclicity of the Koszul--Tate operator
$\brst_{-1}$ associated with the proper solution $S$, where 
$\brst\!=\!(S,\cdot)=\brst_{-1}\!+\!\ldots$.

\noi
{\sc Proof of Theorem~\ref{theorembrst3}}:~~
The main proof is an induction in the number $m\!\geq\!1$ of ghosts $c$.
The induction assumption is that there exists an anticanonical transformation 
$e^{-(\Psi,\~\cdot\~)}\bar{S}$ of $\bar{S}$, such that one has
(after renaming $e^{-(\Psi,\~\cdot\~)}\bar{S}\to\bar{S}$)
\beq
\bar{S}-S\=\cO(c^{m})\~. \label{ambiinductionassump}
\eeq
The induction assumption is true for $m\!=\!1$, because one assumes that the 
correct original limit \e{origlimit} is fulfilled,
$\bar{S}^{}_{0}\!=\!S^{}_{0}$. Now consider $m\!=\!2$. One has that
\beq 
S\=S^{}_{0}+\sum_{s=0}^{L}V^{}_{\aaa_{s}}c^{\aaa_{s}}+\cO(c^{2})
\~,\~\~\~\~\~\~\~\~
\bar{S}\=S^{}_{0}+\sum_{s=0}^{L}\bar{V}^{}_{\aaa_{s}}c^{\aaa_{s}}+\cO(c^{2})\~.
\eeq
One now uses (nested) induction in antifield number $0\!\leq\!r\!\leq\!L$.
(This is a finite induction, if the number $L$ of stages is finite.) 
The induction assumption is that there exists an anticanonical transformation 
$e^{-(\Psi,\~\cdot\~)}\bar{S}$ of $\bar{S}$, with $\Psi\!=\!\cO(c)$, such that 
one has (after renaming $e^{-(\Psi,\~\cdot\~)}\bar{S}\to\bar{S}$)
\beq
\bar{V}^{}_{\aaa_{-1}}=V^{}_{\aaa_{-1}}\~,\~\~\
\bar{V}^{}_{\aaa_{0}}=V^{}_{\aaa_{0}}\~,\~\~\
\bar{V}^{}_{\aaa_{1}}=V^{}_{\aaa_{1}}\~,\~\~\ldots\~,\~\~\
\bar{V}^{}_{\aaa_{r-1}}=V^{}_{\aaa_{r-1}}\~. \label{finiteinductionassump1}
\eeq
{}From nilpotency \e{ktnilp} of the two Koszul--Tate operators, one knows that
\beq
\brst_{-1}V^{}_{\aaa_{r}}\~\equi{\e{ktnilp}}\~0\~,\~\~\~\~\~\~\~\~
\bbrst_{-1}\bar{V}^{}_{\aaa_{r}}\~\equi{\e{ktnilp}}\~0\~.
\eeq
Since the function $V^{}_{\alpha}\!=\!V^{}_{\alpha}(\varphi,\Phi^{*})$ does not
depend on the ghost variables $c^{\aaa_{s}}$, $s\!\in\!\{0,\ldots,L\}$, the 
antifield number is 
\beq
\afn(V^{}_{\aaa_{r}})\=-\gh(V^{}_{\aaa_{r}})
\=\gh(c^{\aaa_{r}})\=r+1\~.
\eeq
It follows that the structure function 
$V^{}_{\aaa_{r}}\!=\!V^{}_{\aaa_{r}}(\varphi^{j};\varphi^{*}_{j},
c^{*}_{\bbbb_{0}},\ldots,c^{*}_{\bbbb_{r-1}})$
cannot depend on antifields $c^{*}_{\aaa_{s}}$, for 
$r\!\leq\!s\!\leq\!L$, because their antifield number
$\afn(c^{*}_{\aaa_{s}})\!=\!s\!+\!2$ is too big. Similarly, for
$\bar{V}^{}_{\aaa_{r}}\!=\!\bar{V}^{}_{\aaa_{r}}(\varphi^{j};\varphi^{*}_{j},
c^{*}_{\bbbb_{0}},\ldots,c^{*}_{\bbbb_{r-1}})$. 
{}From the induction assumption \e{finiteinductionassump1}, one concludes that
\beq
0\~\equi{\e{ktnilp}}\~\bbrst_{-1}\bar{V}^{}_{\aaa_{r}}
\=\sum_{s=-1}^{r-1}\bar{V}^{}_{\aaa_{s}}
(\papal{\Phi^{*}_{\aaa_{s}}}\bar{V}^{}_{\aaa_{r}})
\~\equi{\e{finiteinductionassump1}}\~
\sum_{s=-1}^{r-1}V^{}_{\aaa_{s}}
(\papal{\Phi^{*}_{\aaa_{s}}}\bar{V}^{}_{\aaa_{r}})
\=\brst_{-1}\bar{V}^{}_{\aaa_{r}}\~.
\eeq
Hence the difference
$\brst_{-1}(V^{}_{\aaa_{r}}\!-\!\bar{V}^{}_{\aaa_{r}})\!=\!0$ is zero, so by
acyclicity \e{ktacyclic1} of the Koszul--Tate operator $\brst_{-1}$, there
exists
\beq
U^{}_{\aaa_{r}}\=U^{}_{\aaa_{r}}(\varphi^{j};\varphi^{*}_{j},
c^{*}_{\bbbb_{0}},\ldots,c^{*}_{\bbbb_{r}})
\=c^{*}_{\bbbb_{r}}\~\tilde{U}^{\bbbb_{r}}{}_{\aaa_{r}}(\varphi^{j}) 
+\cO\left((\Phi^{*})^{2}\right)
\eeq
with $\afn(U^{}_{\aaa_{r}})\!=\!-\gh(U^{}_{\aaa_{r}})\!=\!r\!+\!2$, such
that 
\beq
V^{}_{\aaa_{r}}-\bar{V}^{}_{\aaa_{r}}\=\brst_{-1}U^{}_{\aaa_{r}}\~.\label{vu}
\eeq
One is allowed to change 
\beq
U^{}_{\aaa_{r}}\~\longrightarrow\~U^{}_{\aaa_{r}}+\brst_{-1}W^{}_{\aaa_{r}}\~, 
\label{uw}
\eeq
where
\beq
W^{}_{\aaa_{r}}\=W^{}_{\aaa_{r}}(\varphi^{j};\varphi^{*}_{j},
c^{*}_{\bbbb_{0}},\ldots,c^{*}_{\bbbb_{r+1}})
\=c^{*}_{\bbbb_{r+1}}\~\tilde{W}^{\bbbb_{r+1}}{}_{\aaa_{r+1}}(\varphi^{j})
+\cO\left((\Phi^{*})^{2}\right)
\eeq
with $\afn(W^{}_{\aaa_{r}})\!=\!-\gh(W^{}_{\aaa_{r}})\!=\!r\!+\!3$.
The leading contribution proportional to $c^{*}_{\aaa_{r-1}}$ in \eq{vu} 
and $c^{*}_{\aaa_{r}}$ in \eq{uw} read
\beq
Z^{\aaa_{r-1}}{}_{\aaa_{r}}-\bar{Z}^{\aaa_{r-1}}{}_{\aaa_{r}}
\=Z^{\aaa_{r-1}}{}_{\bbbb_{r}}\tilde{U}^{\bbbb_{r}}{}_{\aaa_{r}}
\~,\~\~\~\~\~\~\~\~
\tilde{U}^{\bbbb_{r}}{}_{\aaa_{r}}\~\longrightarrow\~
\tilde{U}^{\bbbb_{r}}{}_{\aaa_{r}}
+Z^{\bbbb_{r}}{}_{\bbbb_{r+1}}\tilde{W}^{\bbbb_{r+1}}{}_{\aaa_{r+1}}\~.
\eeq
Alternatively, if one defines 
$\Delta^{\bbbb_{r}}{}_{\aaa_{r}}\!
:=\!(1\!-\!\tilde{U})^{\bbbb_{r}}{}_{\aaa_{r}}$, one has
\beq
\bar{Z}^{\aaa_{r-1}}{}_{\aaa_{r}}
\=Z^{\aaa_{r-1}}{}_{\bbbb_{r}}\Delta^{\bbbb_{r}}{}_{\aaa_{r}}
\~,\~\~\~\~\~\~\~\~
\Delta^{\bbbb_{r}}{}_{\aaa_{r}}
\~\longrightarrow\~\Delta^{\bbbb_{r}}{}_{\aaa_{r}}
-Z^{\bbbb_{r}}{}_{\bbbb_{r+1}}\tilde{W}^{\bbbb_{r+1}}{}_{\aaa_{r+1}}\~.
\eeq
It follows from the rank conditions for $\bar{Z}^{\aaa_{s-1}}{}_{\aaa_{s}}$
and $Z^{\aaa_{s-1}}{}_{\aaa_{s}}$, that one may assume that 
$\Delta^{\bbbb_{r}}{}_{\aaa_{r}}$ is a regular invertible matrix, possibly after
an allowed change \e{uw}. (This is the only place in the proof where one uses
that the $\bar{S}$ solution is proper.) Therefore one may assume that the
matrix $\tilde{U}^{\bbbb_{r}}{}_{\aaa_{r}}\!=\!(1\!-\!\Delta)^{\bbbb_{r}}{}_{\aaa_{r}}$
has no eigenvalues equal to $1$.
Next apply the anticanonical transformation $e^{-(\Psi,\~\cdot\~)}$ to 
$\bar{S}$, with $\Psi\!=\!\Psi^{}_{\aaa_{r}}c^{\aaa_{r}}$, where
\bea
\Psi^{}_{\aaa_{r}}&:=&U^{}_{\bbbb_{r}}f(\tilde{U})^{\bbbb_{r}}{}_{\aaa_{r}}\~,
\label{psidef} \\
\Psi^{}_{\aaa_{r}}&=&\Psi^{}_{\aaa_{r}}(\varphi^{j};\varphi^{*}_{j},
c^{*}_{\bbbb_{0}},\ldots,c^{*}_{\bbbb_{r}})
\=c^{*}_{\bbbb_{r}}\~\tilde{\Psi}^{\bbbb_{r}}{}_{\aaa_{r}}(\varphi^{j})
+\cO\left((\Phi^{*})^{2}\right)\~, \label{psidef2}
\eea
and where the holomorphic function
\beq
f(z)\~:=\~-\frac{{\rm Ln}(1-z)}{z}\=\sum_{n=0}^{\infty}\frac{z^{n}}{n+1}
\eeq
has a logarithmic singularity at $z\!=\!1$. One may place the branch--cut of 
$f$ away from the eigenvalue spectrum of the matrix 
$\tilde{U}^{\bbbb_{r}}{}_{\aaa_{r}}$, so that $f(\tilde{U})$ is well-defined. 
The leading tilde piece of the $\Psi$ function \e{psidef2} is 
\beq
\tilde{\Psi}^{\bbbb_{r}}{}_{\aaa_{r}}
\=-\ln(1-\tilde{U})^{\bbbb_{r}}{}_{\aaa_{r}}
\~\~\~\~\Leftrightarrow\~\~\~\~
\tilde{U}^{\bbbb_{r}}{}_{\aaa_{r}}
\=(1-e^{-\tilde{\Psi}})^{\bbbb_{r}}{}_{\aaa_{r}}\~.\label{leadingtildepiece}
\eeq
One calculates
\beq
(\~\cdot\~,\Psi)
\=(\~\cdot\~,\Psi)^{}_{-1}
+\sum_{s=0}^{r}(\~\cdot\~\papar{c^{\bbbb_{s}}})
(\papal{c^{*}_{\bbbb_{s}}}\Psi^{}_{\aaa_{r}})c^{\aaa_{r}}
-(\~\cdot\~\papar{c^{*}_{\aaa_{r}}})\Psi^{}_{\aaa_{r}}\~,
\eeq
so that
\beq
(\bar{S},\Psi)+\cO(c^{2})\=
(S^{}_{0}+\sum_{s=0}^{r}\bar{V}^{}_{\aaa_{s}}c^{\aaa_{s}},\Psi)
\=\bbrst_{-1}\Psi^{}_{\aaa_{r}}\~c^{\aaa_{r}}+\cO(c^{2})\~,
\eeq
and hence
\bea \lefteqn{ 
\left[e^{-(\Psi,\~\cdot\~)}-1\right]\bar{S}+\cO(c^{2}) } \cr
&=&\left[e^{-(\Psi,\~\cdot\~)}-1\right]\left[
S^{}_{0}+\sum_{s=0}^{r}\bar{V}^{}_{\aaa_{s}}c^{\aaa_{s}}\right]+\cO(c^{2}) 
\=\bbrst_{-1}\Psi^{}_{\bbbb_{r}}\~
E(\tilde{\Psi})^{\bbbb_{r}}{}_{\aaa_{r}}\~c^{\aaa_{r}} \cr
&\equi{\e{finiteinductionassump1}}&\brst_{-1}\Psi^{}_{\bbbb_{r}}\~
E(\tilde{\Psi})^{\bbbb_{r}}{}_{\aaa_{r}}\~c^{\aaa_{r}}
+(\bar{V}^{}_{\cccc_{r}}\!-\!V^{}_{\cccc_{r}})
(\papal{c^{*}_{\cccc_{r}}}\Psi^{}_{\bbbb_{r}})
E(\tilde{\Psi})^{\bbbb_{r}}{}_{\aaa_{r}}\~c^{\aaa_{r}} \cr
&\equi{\e{vu}+\e{psidef2}}&\brst_{-1}\Psi^{}_{\bbbb_{r}}\~
E(\tilde{\Psi})^{\bbbb_{r}}{}_{\aaa_{r}}\~c^{\aaa_{r}}
-\brst_{-1}U^{}_{\cccc_{r}}\~\tilde{\Psi}^{\cccc_{r}}{}_{\bbbb_{r}}\~
E(\tilde{\Psi})^{\bbbb_{r}}{}_{\aaa_{r}}\~c^{\aaa_{r}} \cr
&\equi{\e{psidef}+\e{leadingtildepiece}}&\brst_{-1}U^{}_{\cccc_{r}}\~
f(1\!-\!e^{-\tilde{\Psi}})^{\cccc_{r}}{}_{\bbbb_{r}}\~
E(\tilde{\Psi})^{\bbbb_{r}}{}_{\aaa_{r}}\~c^{\aaa_{r}}
-\brst_{-1}U^{}_{\cccc_{r}}\~
(e^{\tilde{\Psi}}\!-\!1)^{\cccc_{r}}{}_{\aaa_{r}}\~c^{\aaa_{r}} \cr
&=&\brst_{-1}U^{}_{\aaa_{r}}\~c^{\aaa_{r}}\~,
\eea
where
$E(z)\!:=\!\frac{e^{z}-1}{z}\!=\!\sum_{n=0}^{\infty}\frac{z^{n}}{(n+1)!}$.
One has (after renaming $e^{-(\Psi,\~\cdot\~)}\bar{S}\to\bar{S}$)
\beq
\bar{V}^{}_{\aaa_{-1}}=V^{}_{\aaa_{-1}}\~,\~\~\
\bar{V}^{}_{\aaa_{0}}=V^{}_{\aaa_{0}}\~,\~\~\
\bar{V}^{}_{\aaa_{1}}=V^{}_{\aaa_{1}}\~,\~\~\ldots\~,\~\~\
\bar{V}^{}_{\aaa_{r}}=V^{}_{\aaa_{r}}\~, \label{finiteinductionassump2}
\eeq
which is the nested induction assumption \e{finiteinductionassump1} with 
$r\!\to\!r\!+\!1$. This finishes the proof that the induction assumption 
\e{ambiinductionassump} is true for $m\!=\!2$, and that the two Koszul--Tate 
operators $\bbrst_{-1}\!=\!\brst_{-1}$ are equal.

\noi
Now assume that $m\!\geq\!3$ and that the induction assumption
\e{ambiinductionassump} is true up to previous number  ``$m\!-\!1$'' of ghosts.
We would like to prove \eq{ambiinductionassump} for ``$m$''. One now uses
nested induction in antifield number $0\!\leq\!r\!\leq\!m(L\!+\!2)$.
Assume that there exists an anticanonical transformation 
$e^{-(\Psi,\~\cdot\~)}\bar{S}$ of $\bar{S}$, such that one has
(after renaming $e^{-(\Psi,\~\cdot\~)}\bar{S}\to\bar{S}$)
\beq
\bar{S}^{}_{0}-S^{}_{0}=\cO(c^{m})\~,\~\~
\bar{S}^{}_{1}-S^{}_{1}=\cO(c^{m})\~,\~\~\ldots\~,\~\~
\bar{S}^{}_{r}-S^{}_{r}=\cO(c^{m})\~,
\eeq
while $\bar{S}^{}_{r+1}\!-\!S^{}_{r+1}\!=\!\cO(c^{m-1})$.
The two classical master equations yield
\beq
0\=\CME_{r}\~\equi{\e{cmxdecomp1}}\~
\brst_{-1}S^{}_{r+1}+\cB^{}_{r}\~,\~\~\~\~\~\~\~\~
0\=\bCME_{r}
\~\equi{\e{cmxdecomp1}}\~\bbrst_{-1}\bar{S}^{}_{r+1}+\bar{\cB}^{}_{r}\~,
\eeq
where
\beq 
\cB^{}_{r}\~\equi{\e{bdef1}}\~
\Hf \sum_{1\leq p,q\leq r} (S^{}_{p},S^{}_{q})^{}_{r-p-q}\~,\~\~\~\~\~\~\~\~
\bar{\cB}^{}_{r}\~\equi{\e{bdef1}}\~
\Hf \sum_{1\leq p,q\leq r} (\bar{S}^{}_{p},\bar{S}^{}_{q})^{}_{r-p-q}\~.
\label{bbbdef1}
\eeq
Hence the difference is
\beq
\brst_{-1}(S^{}_{r+1}-\bar{S}^{}_{r+1})\=\bar{\cB}^{}_{r}-\cB^{}_{r}
\=\Hf \sum_{1\leq p,q\leq r}\left[
(\bar{S}^{}_{p}-S^{}_{p},\bar{S}^{}_{q})^{}_{r-p-q}
+(\bar{S}^{}_{p}-S^{}_{p},S^{}_{q})^{}_{r-p-q}\right]
\=\cO(c^{m})\~.
\eeq
Since the Koszul--Tate operator $\brst_{-1}$ is acyclic \e{ktacyclic1} and 
preserves the number of $c$'s, there exists a 
$\Psi^{}_{r+2}\!=\!\cO(c^{m-1})$ with $\afn(\Psi^{}_{r+2})\!=\!r\!+\!2$ such
that
$S^{}_{r+1}\!-\!\bar{S}^{}_{r+1}\!-\!\brst_{-1}\Psi^{}_{r+2}\!=\!\cO(c^{m})$.
Next apply the anticanonical transformation
$e^{-(\Psi^{}_{r+2},\~\cdot\~)}$ to $\bar{S}$. One calculates
\bea
\left[e^{-(\Psi^{}_{r+2},\~\cdot\~)}-1\right]\bar{S}
&=&\left[e^{-(\Psi^{}_{r+2},\~\cdot\~)}-1\right]
\left[S^{}_{0}+\sum_{s=0}^{L}V^{}_{\aaa_{s}}c^{\aaa_{s}}
+\cO(c^{2})\right] \cr
&\equi{m\geq3}&
\left(S^{}_{0}+\sum_{s=0}^{L}V^{}_{\aaa_{s}}c^{\aaa_{s}},\~\Psi^{}_{r+2}\right)
+\cO(c^{m})
\=\brst_{-1}\Psi^{}_{r+2}+\cO(c^{m})\~.
\eea
One has (after renaming $e^{-(\Psi^{}_{r+2},\~\cdot\~)}\bar{S}\to\bar{S}$)
that $\bar{S}\!-\!S\!=\!\cO(c^{m-1})$, and
\beq
\bar{S}^{}_{0}-S^{}_{0}=\cO(c^{m})\~,\~\~
\bar{S}^{}_{1}-S^{}_{1}=\cO(c^{m})\~,\~\~\ldots\~,\~\~
\bar{S}^{}_{r+1}-S^{}_{r+1}=\cO(c^{m})\~.
\eeq
\proofbox

\section{Existence of Proper Action}
\label{secrealexist}

\subsection{Transversal and Longitudinal Fields}
\label{secltfields}

\noi
Because of the Noether identity \e{noether1}, there exist locally $\MMM_{-1}$
independent on--shell gauge--invariants $\xi^{I}\!=\!\xi^{I}(\varphi)$, which
we will call the {\em transversal} fields. They satisfy in terms of formulas 
\beq
(\xi^{I}\papar{\varphi^{i}}) R^{i}{}_{\aaa_{0}}\=\EOM_{i}K^{iI}_{\aaa_{0}}
\~\approx\~0
\~,\~\~\~\~\~\~\~\~I\in\{1,\ldots,\MMM_{-1}\}\~, \label{deftransv}
\eeq 
near the stationary $\varphi$-surface, where ``$\approx$'' means equality 
modulo equations of motion
\beq
\EOM_{i}\~\equiv\~(S_{0}\papar{\varphi^{i}})\~,\label{eoma}
\eeq
and where $K^{iI}_{\aaa_{0}}\!=\!K^{iI}_{\aaa_{0}}(\varphi)$ are some
structure functions. Consider now an arbitrary $\varphi$-neighborhood $U$, 
where the transversal fields $\xi^{I}$ are defined, and where $U$ is 
sufficiently close so that it intersects the stationary $\varphi$-surface. 
It follows from \eq{deftransv} that the values of the gauge--invariants 
$\xi^{I}(\varphi)\!=\!\xi^{I}_{\cl}$ do not depend on the point $\varphi$ if
one only varies the point $\varphi$ {\em within} the $\MMM_{0}$-dimensional 
stationary subsurface $\EOM_{i}(\varphi)\!=\!0$. By a redefinition
$\xi^{I}\to\xi^{I}\!-\!\xi^{I}_{\cl}$ it is possible to assume (and we will
do so from now on), that
$\forall \varphi\!\in\!U:\EOM_{i}(\varphi)\!=\!0 
\Rightarrow \xi^{I}(\varphi)\!=\!0$.
Since transversal fields $\xi^{I}$ are independent, it follows that the 
equations of motion are equivalent to the vanishing of the transversal 
fields
\beq
\forall \varphi\in U\~:\~\EOM_{i}(\varphi)\=0\~\~\~\~\Leftrightarrow\~\~\~\~
\xi^{I}(\varphi)\=0\~.\label{eomb}
\eeq
The transversal fields $\xi^{I}$ can locally be complemented with so--called 
{\em longitudinal} fields $\theta^{\AAA_{0}}\!=\!\theta^{\AAA_{0}}(\varphi)$, 
such that the change of coordinates 
\beq
\varphi^{i}\~\~\~\~\longrightarrow\~\~\~\~
\bar{\varphi}^{i}\~\equiv\~\twobyone{\xi^{I}}{\theta^{\AAA_{0}}}
\label{varphibarvarphi}
\eeq
is a non-singular coordinate transformation. Here the indices runs over
\beq
i\in\{1,\ldots,n\}\~,\~\~\~\~\~\~\~\~
I\in\{1,\ldots,\MMM_{-1}\}\~,\~\~\~\~\~\~\~\~
\AAA_{0}\in\{1,\ldots,\MMM_{0}\}\~,\~\~\~\~\~\~\~\~\MMM_{-1}+\MMM_{0}\=n\~.
\eeq
By definition 
\beq
(S_{0}\papar{\xi^{I}}) 
\=(S_{0}\papar{\varphi^{i}})(\varphi^{i}\papar{\xi^{I}}) 
\~\equi{\e{eoma}}\~\cO(\EOM)\~.\label{eom0} 
\eeq
The rectangular matrix 
\beq
{\cal R}^{\AAA_{0}}{}_{\aaa_{0}}
\~:=\~(\theta^{\AAA_{0}}\papar{\varphi^{i}}) R^{i}{}_{\aaa_{0}}
\~,\~\~\~\~\~\~\~\~\aaa_{0}\in\{1,\ldots,\mmm_{0}\}\~,\~\~\~\~\~\~\~\~
\mmm_{0}\geq\MMM_{0}\~,
\eeq
must have maximal rank near the stationary $\varphi$-surface
\beq
0\~\leq\~\rank({\cal R}^{\AAA_{0}}{}_{\aaa_{0}})\=\MMM_{0} \~.
\eeq
(If it didn't have maximal rank, it would signal the possibility to define
at least one more transversal coordinate that satisfies \eq{deftransv}.)
Therefore there exists a right inverse matrix ${\cal N}^{\aaa_{0}}{}_{\AAA_{0}}$, 
such that
\beq
{\cal R}^{\AAA_{0}}{}_{\aaa_{0}}{\cal N}^{\aaa_{0}}{}_{\BBBB_{0}}\= 
\delta^{\AAA_{0}}_{\BBBB_{0}}\~.
\eeq 
The Noether identity \e{noether1} yields
\bea
0&=&(S_{0}\papar{\varphi^{i}})R^{i}{}_{\aaa_{0}}
\=(S_{0}\papar{\theta^{\AAA_{0}}})
(\theta^{\AAA_{0}}\papar{\varphi^{i}})R^{i}{}_{\aaa_{0}}
+(S_{0}\papar{\xi^{J}})(\xi^{J}\papar{\varphi^{i}})R^{i}{}_{\aaa_{0}} \cr
&\equi{\e{deftransv}}&
(S_{0}\papar{\theta^{\AAA_{0}}}){\cal R}^{\AAA_{0}}{}_{\aaa_{0}}
+\EOM_{i}K^{iJ}_{\aaa_{0}} (\papal{\xi^{J}}S_{0})
(-1)^{\eps_{J}\eps_{\aaa_{0}}}  \label{krewrittings1} \\
&=&(S_{0}\papar{\theta^{\AAA_{0}}})\tilde{\cal R}^{\AAA_{0}}{}_{\aaa_{0}}
+(S_{0}\papar{\xi^{I}})(\xi^{I}\papar{\varphi^{i}})
K^{iJ}_{\aaa_{0}}(\papal{\xi^{J}}S_{0})(-1)^{\eps_{J}\eps_{\aaa_{0}}}
\~, \label{krewrittings2}
\eea
where
\beq
\tilde{\cal R}^{\AAA_{0}}{}_{\aaa_{0}}
\~:=\~{\cal R}^{\AAA_{0}}{}_{\aaa_{0}}
+(\theta^{\AAA_{0}}\papar{\varphi^{i}})
K^{iJ}_{\aaa_{0}}(\papal{\xi^{J}}S_{0})(-1)^{\eps_{J}\eps_{\aaa_{0}}}
\~\approx\~{\cal R}^{\AAA_{0}}{}_{\aaa_{0}}\~.
\eeq
The expression \e{krewrittings2} is for later convenience. {}From the
expression \e{krewrittings1}, one sees that  
\beq
(S_{0}\papar{\theta^{\AAA_{0}}}) \~\equi{\e{eom0}+\e{krewrittings1}}\~ 
\cO(\EOM{}^{2})\~. \label{eom1}
\eeq
By differentiating the Noether identity \e{noether1} \wrt $\varphi^{i}$,
one derives that
\beq
0\~\approx\~(\papal{\varphi^{i}}S_{0}\papar{\varphi^{j}})R^{j}{}_{\aaa_{0}}
\~\approx\~(\papal{\varphi^{i}}S_{0}\papar{\theta^{\AAA_{0}}})
{\cal R}^{\AAA_{0}}{}_{\aaa_{0}}\~,
\eeq
and hence,
\beq
(\papal{\varphi^{i}}S_{0}\papar{\theta^{\AAA_{0}}})\~\approx\~0\~.
\eeq
Therefore the rank condition \e{rankcondition1} implies that
\beq
0\~\leq\~\rank(\papal{\xi^{I}}S_{0}
\papar{\xi^{J}})\= \MMM_{-1}
\eeq
near the stationary $\varphi$-surface. The transversal and longitudinal fields
are not uniquely defined.

\noi
If the $K^{iI}_{\aaa_{0}}$ structure functions additionally satisfy the
integrability condition
\beq
K^{IJ}_{\aaa_{0}}\~:=\~(\xi^{I}\papar{\varphi^{i}}) K^{iJ}_{\aaa_{0}}
\=-(-1)^{\eps_{I}\eps_{J}}(I \leftrightarrow J)\~, \label{bconsistrel}
\eeq
then the second term of the expression \e{krewrittings2} vanishes identically,
so that 
\beq
(S_{0}\papar{\theta^{\AAA_{0}}})\=0~,
\eeq
instead of just \eq{eom1}. Therefore, in case of the integrability condition
\e{bconsistrel}, one has the following.

\begin{principle}[The Gauge Principle] Locally near the stationary
$\varphi$-surface, the original action $S_{0}\!=\!S_{0}(\xi)$ depends on only
$\MMM_{-1}$ independent quantities $\xi^{I}\!=\!\xi^{I}(\varphi)$,
$I\!\in\!\{1,\ldots,\MMM_{-1}\}$.
\label{gaugeprinciple}
\end{principle}

\noi
In fact, the Gauge Principle~\ref{gaugeprinciple} is precisely equivalent to
the pair of \eqs{deftransv}{bconsistrel}. It was shown in \Ref{bv84} by
integrating the generalized Lie equations that finite gauge transformations do
exist, and in particular, that the gauge principle \e{gaugeprinciple} holds,
and hence that there exist $\xi^{I}$ such that both
\eqs{deftransv}{bconsistrel} are satisfied. This implies, among other things,
that the set of stationary points for $S_{0}$ constitute a smooth submanifold.
However, since $\xi^{I}$ generically will be space--time non--local, and since
we will not actually need the integrability condition \e{bconsistrel} in the
following, we are reluctant to unnecessarily enforce \eq{bconsistrel} on
$\xi^{I}$ in what follows.

\subsection{Reduced and Shifted Antifield Number}
\label{secsafn}

\noi
We would like to redefine the Koszul--Tate operator and the resolution degree,
so that the Koszul--Tate operator is more directly related to the $\Squad$ part
\e{sabel} of the action $S$ in the reducible case. The new resolution degrees 
will be the so--called reduced and shifted antifield number. Transversal 
fields $\xi^{I}$ and antifields $\Phi^{*}_{\alpha}$ are charged under reduced
and shifted antifield number
\beq
\rafn(\xi^{I})\=1\=\safn(\xi^{I})\~,\~\~\~\~\~\~\~\~
\rafn(\Phi^{*}_{\alpha})\=1\~,\~\~\~\~\~\~\~\~
\safn(\Phi^{*}_{\alpha})\=1+\afn(\Phi^{*}_{\alpha})
\=1-\gh(\Phi^{*}_{\alpha})\~,
\eeq
\beq
\safn(\Phi^{*}_{\aaa_{s}})\=s+3\~,\~\~\~\~\~\~\~\~s\~\in\~\{-1,\ldots,L\}\~.
\eeq
All the other variables, \ie the longitudinal fields $\theta^{\AAA_{0}}$ and 
ghosts $c^{\aaa_{s}}$, $s\!\in\!\{0,\ldots,L\}$, carry no shifted or reduced 
antifield number, see Table~\ref{multtable1}. Reduced and shifted antifield 
number are not independent, because in general
\beq
\safn\=\rafn+\afn\~.
\eeq
Notice also that
\beq
\forall {\rm ~functions~}f=f(\Gamma):\~\~\~\~
\safn(f)\=0\~\~\Leftrightarrow\~\~\rafn(f)\=0\~.
\eeq
On one hand, when considering only the Koszul--Tate operator, 
$\brst_{-1}\!=\!V^{}_{\alpha}\papal{\Phi^{*}_{\alpha}}$ ,
$V^{}_{\alpha}\!=\!V^{}_{\alpha}(\varphi,\Phi^{*})$, where ghosts ``$c$'' are
passive spectators, then the reduced antifield number ``$\rafn$'' is the 
simplest resolution degree to work with, \cf Theorem~\ref{theoremss5}. Reduced
antifield number is also easy to transcribe into the Hamiltonian framework. On
the other hand, when considering the whole BRST operator ``$\brst$'', where
ghosts ``$c$'' are active, then one needs the shifted antifield number
``$\safn$'', \cf Theorem~\ref{theoremss2}. (For instance, some parts of the
longitudinal derivative \cite{hentei92} would have leading resolution degree,
if one only uses reduced antifield number as resolution degree, which would be
devastating. On the other hand, shifted antifield number appropriately pushes
the longitudinal derivative down the resolution hierarchy.)

\noi
Reduced and shifted antifield number will depend on the local choice of
transversal fields, so they are {\em not} globally defined. 

\noi
Let us now decompose the action and the antibracket \wrt shifted antifield 
number. {}First rewrite the Jacobian matrix of the transformation 
\e{varphibarvarphi},
\beq
(\bar{\varphi}^{i}\papar{\varphi^{j}})\~=:\~\Lambda^{i}{}_{j}
\label{lambdam1}
\eeq 
as functions $\Lambda^{i}{}_{j}\!=\!\Lambda^{i}{}_{j}(\bar{\varphi})$ of 
transversal and longitudinal fields 
$\bar{\varphi}^{i}\!\equiv\!\{\xi^{I};\theta^{\AAA_{0}}\}$. One can expand
these functions in shifted antifield number.
\beq
\Lambda^{i}{}_{j}\=\sum_{r=0}^{\infty} \Lambda^{i}_{(r)j}
\~,\~\~\~\~\~\~\~\~\safn(\Lambda^{i}_{(r)j})\=r\~. 
\eeq
One next expands the classical master action $S$ according to the shifted 
antifield number.
\beq
S\=\sum_{r=0}^{\infty}S^{}_{(r)}\~,\~\~\~\~\~\~\~\~\safn(S^{}_{(r)})\=r\~,
\eeq 
\beq
S^{}_{(0)}\=S^{}_{0(0)}\~\equi{\e{eom1}}\~{\rm const}.\~,\~\~\~\~\~\~\~\~
S^{}_{(1)}\=S^{}_{0(1)}\~\equi{\e{eom0}}\~0\~,\~\~\~\~\~\~\~\~
S^{}_{(2)}\=S^{}_{0(2)}+\varphi^{*}_{i}R^{i}_{(0)\aaa_{0}}\~c^{\aaa_{0}}\~.
\label{lowestesses}
\eeq
Here $S_{0(2)}$ is of the form
\beq
S^{}_{0(2)}\=\Hf \xi^{I}H^{}_{IJ}\xi^{I}\~,\~\~\~\~\~\~\~\~
H^{}_{JI}\=-(-1)^{(\eps_{I}+1)(\eps_{J}+1)}H^{}_{IJ}\~,\~\~\~\~\~\~\~\~
H^{}_{IJ}\=H^{}_{IJ}(\theta)\~.
\eeq
The antibracket $(\cdot,\cdot)$ expands as
\beq
(f,g)\=(f,g)^{\xi}+(f,g)^{\theta}+(f,g)^{c}
\=\sum_{r=-(L+3)}^{\infty}(f,g)^{}_{(r)}\~,
\eeq
\beq
(f,g)^{}_{(r)}\=(f,g)_{(r)}^{\xi}+(f,g)_{(r)}^{\theta}+(f,g)^{c}_{(r)}
\~,\~\~\~\~\~\~\~\~
\safn(f,g)^{}_{(r)}\=\safn(f)+\safn(g)+r\~,
\eeq
\beq
(f,g)^{\xi}\=\sum_{r=-3}^{\infty}(f,g)^{\xi}_{(r)}
\~,\~\~\~\~\~\~\~\~
(f,g)^{\theta}\=\sum_{r=-2}^{\infty}(f,g)^{\theta}_{(r)}
\~,\~\~\~\~\~\~\~\~
(f,g)^{c}\=\sum_{s=0}^{L}(f,g)^{c}_{(-s-3)}\~,
\eeq
\beq
(f,g)^{\xi}_{(r-3)}\~:=\~(f\papar{\xi^{I}})
\Lambda^{I}_{(r)i}(\papal{\varphi^{*}_{i}}g) 
-(-1)^{(\eps_{f}+1)(\eps_{g}+1)}(f \leftrightarrow g)
\~,\~\~\~\~\~\~\~\~r\in\{0,1,2,\ldots\}\~,
\eeq
\beq
(f,g)^{\theta}_{(r-2)}\~:=\~(f\papar{\theta^{\AAA_{0}}})
\Lambda^{\AAA_{0}}_{(r)i}(\papal{\varphi^{*}_{i}}g) 
-(-1)^{(\eps_{f}+1)(\eps_{g}+1)}(f \leftrightarrow g)
\~,\~\~\~\~\~\~\~\~r\in\{0,1,2,\ldots\}\~,
\eeq
\beq
(f,g)^{c}_{(-s-3)}\~:=\~f\left(\papar{c^{\aaa_{s}}}\papal{c^{*}_{\aaa_{s}}}
-\papar{c^{*}_{\aaa_{s}}}\papal{c^{\aaa_{s}}}\right)g 
\~,\~\~\~\~\~\~\~\~ s\in\{0,\ldots,L\}\~.
\eeq
Elementary considerations reveal the following useful Lemma~\ref{lemmass2}
for the ghost sector $(\cdot,\cdot)^{c}$.

\begin{lemma}
Let $f,g$ be two functions of definite shifted antifield number. Then
\beq
(f,g)^{c}_{(-s)}\~\neq\~0
\~\~\~\~\~\~\~\~\Rightarrow\~\~\~\~\~\~\~\~
\safn(f)\~\geq\~s\~\~\vee\~\~\safn(g)\~\geq\~s\~. \label{lemmass2a} 
\eeq
Assume that $f$ also has definite ghost number. Then
\beq
(f,g)^{c}_{(-s)}\~\neq\~0\~\~\wedge\~\~\gh(f)\~\geq\~-1
\~\~\~\~\~\~\~\~\Rightarrow\~\~\~\~\~\~\~\~
\safn(f)+\gh(f)\~\geq\~s\!-\!1\~.\label{lemmass2ab} 
\eeq
In particular,
\beq
(f,g)^{c}_{(-s)}\~\neq\~0\~\~\wedge\~\~\gh(f)\=0
\~\~\~\~\~\~\~\~\Rightarrow\~\~\~\~\~\~\~\~
\safn(f)\~\geq\~s\!-\!1\~.\label{lemmass2b} 
\eeq
\label{lemmass2}
\end{lemma}

\noi
The equations of motion expand as
\beq
\EOM_{i}\=\sum_{r=0}^{\infty}\EOM_{(r)i}
\~,\~\~\~\~\~\~\~\~\safn(\EOM_{(r)i})\=r\~. 
\eeq
Eq.\ \e{eomb} implies that
\beq
\EOM_{(0)i}\~\equi{\e{eomb}}\~0\~.\label{eom2}
\eeq
Together with the Noether identity \e{noether1}, this implies that
\beq
\EOM_{(1)i}R^{i}_{(0)\aaa_{0}}\=0\~.\label{eom4}
\eeq
We also have
\beq
\Lambda^{I}_{(0)i}R^{i}_{(0)\aaa_{0}}\~\equi{\e{deftransv}}\~0\~.
\label{deftransv2}
\eeq
Eq.\ \e{deftransv2} implies that
\beq
\Hf(S_{(2)},S_{(2)})^{}_{(-3)}\=\Hf(S_{(2)},S_{(2)})^{\xi}_{(-3)}
\=(S_{0(2)}\papar{\xi^{I}})\Lambda^{I}_{(0)i} 
R^{i}_{(0)\aaa_{0}}\~c^{\aaa_{0}}\~\equi{\e{deftransv2}}\~0\~.\label{eom3}
\eeq

\subsection{Transversal and Longitudinal Ghosts and Antifields}
\label{secltghosts}

\noi
The only purpose of this Section~\ref{secltghosts} is to device rotated 
auxiliary variables, in which the acyclicity of a shifted Koszul--Tate 
operator \e{kt2} becomes apparent, see Section~\ref{seckt2}. We stress that
we do not use the rotated auxiliary variables thereafter. 
In particular, the rotated auxiliary variables do not enter the existence
proofs in Sections~\ref{secexists5}--\ref{secexists2}.

\noi
We will first show by induction in the stage $s\!\in\!\{0,\ldots,L\}$, 
that there exists a sequence of invertible rotation matrices  
$\Lambda^{\bbbb_{s}}{}_{\aaa_{s}}\!=\!\Lambda^{\bbbb_{s}}{}_{\aaa_{s}}(\varphi)$,
such that the first $\MMM_{s-1}$ rows in the rotated 
gauge(--for--gauge)$^s$--generator
\beq
\Lambda^{\bbbb_{s-1}}{}_{\aaa_{s-1}}Z^{\aaa_{s-1}}{}_{\aaa_{s}}
\~\approx\~\twobyone{{\bf 0}^{}_{\MMM_{s-1}\times\mmm_{s}}}
{{\bf *}^{}_{\MMM_{s}\times\mmm_{s}}}^{}_{\mmm_{s-1}\times\mmm_{s}}
\~,\~\~\~\~\~\~\~\~s\!\in\!\{0,\ldots,L\}\~, \label{inductionassump}
\eeq
vanishes weakly in some $\varphi$-neighborhood of the stationary
$\varphi$-surface. The rotation matrix $\Lambda^{\bbbb_{s-1}}{}_{\aaa_{s-1}}$ in
\eq{inductionassump} is fixed at the previous stage $s\!-\!1$ of the induction
proof. Eq.\ \e{inductionassump} for $s\!=\!0$ is just \eq{deftransv}. {}From
the rank condition \e{rankcondition2}, it follows that it is possible to find
an invertible rotation matrix 
$\Lambda^{\bbbb_{s}}{}_{\aaa_{s}}\!=\!\Lambda^{\bbbb_{s}}{}_{\aaa_{s}}(\varphi)$,
such that the rotated gauge(--for--gauge)$^s$--generator is
\beq
\bar{Z}^{\bbbb_{s-1}}{}_{\bbbb_{s}}
\~:=\~\Lambda^{\bbbb_{s-1}}{}_{\aaa_{s-1}}Z^{\aaa_{s-1}}{}_{\aaa_{s}}
(\Lambda^{-1})^{\aaa_{s}}{}_{\bbbb_{s}}
\~\approx\~\twobytwo{{\bf 0}^{}_{\MMM_{s-1}\times\MMM_{s}}}
{{\bf 0}^{}_{\MMM_{s-1}\times\MMM_{s+1}}}
{{\bf 1}^{}_{\MMM_{s}\times\MMM_{s}}}{{\bf 0}^{}_{\MMM_{s-1}\times\MMM_{s+1}}}
^{}_{\mmm_{s-1}\times\mmm_{s}}\~,\~\~\~\~\~\~\~\~s\!\in\!\{0,\ldots,L\}\~.
\label{z0010}
\eeq
Due to \eq{z0010} and the Noether relation \e{noether2}, the induction 
assumption \e{inductionassump} is then satisfied for the next stage, and so 
forth.

\noi
One next defines rotated ghosts and antifields
\bea
\twobyone{\bar{c}_{\perp}^{\BBBB_{s}}}
{\rule{0ex}{3.5ex}\bar{c}_{\parallel}^{\BBBB_{s+1}}}
&\equiv&
\bar{c}^{\bbbb_{s}}\~:=\~\Lambda^{\bbbb_{s}}_{(0)\aaa_{s}}c^{\aaa_{s}}
\~,\~\~\~\~\~\~\~\~s\!\in\!\{0,\ldots,L\}\~, \\
\onebytwo{\bar{\Phi}^{*\perp}_{\BBBB_{-1}}}
{\bar{\Phi}^{*\parallel}_{\BBBB_{0}}}
&\equiv&\onebytwo{\xi^{*}_{J}}{\theta^{*}_{\BBBB_{0}}}\~\equiv\~
\bar{\varphi}^{*}_{j}\~:=\~\varphi^{*}_{i}(\Lambda^{-1})^{i}_{(0)j}\~, \\
\onebytwo{\bar{\Phi}^{*\perp}_{\BBBB_{s}}}
{\bar{\Phi}^{*\parallel}_{\BBBB_{s+1}}}
&\equiv&\bar{\Phi}^{*}_{\bbbb_{s}}
\~:=\~\Phi^{*}_{\aaa_{s}}(\Lambda^{-1})^{\aaa_{s}}_{(0)\bbbb_{s}}
\~,\~\~\~\~\~\~\~\~s\!\in\!\{-1,\ldots,L\}\~.
\eea
The first $\MMM_{s}$ rotated ghosts are called transversal ghosts
$\bar{c}^{}_{\perp}$ and the last $\MMM_{s+1}$ rotated ghosts are called
longitudinal ghosts $\bar{c}^{}_{\parallel}$, where
$\MMM_{s}\!+\!\MMM_{s+1}\!=\!\mmm_{s}$. (Please do not confuse the notation for
rotated ghosts $\bar{c}$ with antighosts, which we do not consider here.) The 
antibrackets of rotated auxiliary variables are not completely on standard
Darboux form, so the rotated antifields are not antifields in the strict sense
of the word. In more detail, the antibrackets of rotated variables read
\beq
(\bar{\varphi}^{i},\bar{\varphi}^{*}_{j})\~\approx\~\delta^{i}_{j}
\~,\~\~\~\~\~\~\~\~
(\bar{c}^{\aaa_{s}},\bar{c}^{*}_{\bbbb_{r}})\=
\delta^{\aaa_{s}}_{\bbbb_{r}}\delta^{s}_{r}~,
\eeq
and other fundamental antibrackets vanish, except for the 
$\theta^{*}_{\AAA_{0}}$ sector, which can have non-trivial antibrackets with
rotated auxiliary variables $\bar{c}$ and $\bar{\Phi}^{*}$, because of
$\theta$-dependence of the rotation matrix 
$\Lambda^{\aaa_{s}}_{(0)\bbbb_{s}}\!=\!\Lambda^{\aaa_{s}}_{(0)\bbbb_{s}}(\theta)$.
Rotated auxiliary variables $\bar{c}$ and $\bar{\Phi}^{*}$ carry the
same shifted and reduced antifield number as the unrotated auxiliary variables 
$c$ and $\Phi^{*}$, respectively.

\subsection{Shifted Koszul--Tate Operator $\brst_{(-1)}$ and Reduced  Koszul--Tate 
Operator $\brst_{-1[0]}$ }
\label{seckt2}

\noi
The shifted Koszul--Tate operator $\brst_{(-1)}$ is defined as the leading 
shifted antifield number sector for the classical BRST operator
\beq
\brst\~:=\~(S,\~\cdot\~)\~\equi{\e{lemmass2b}}\~
\sum_{p=-1}^{\infty}\brst_{(p)}\~,
\eeq
where
\beq
\brst_{(p)}\=\sum_{r=0}^{\infty}(S^{}_{(r)},\~\cdot\~)^{}_{(p-r)}
\~\equi{\e{lowestesses}}\~\sum_{r=2}^{\infty}(S^{}_{(r)},\~\cdot\~)^{}_{(p-r)}
\~,\~\~\~\~\~\~\~\~\safn(\brst_{(p)})\=p
\~,\~\~\~\~\~\~\~\~p\~\in\~\{-1,0,1,2,\ldots\}\~.
\eeq
Similarly, the reduced Koszul--Tate operator $\brst_{-1[0]}$ is defined
as the leading reduced antifield number sector for the Koszul--Tate operator 
\beq
\brst_{-1}\=\sum_{p=0}^{\infty}\brst_{-1[p]}\~.
\eeq
The shifted and reduced Koszul--Tate operator $\brst_{(-1)}$ and $\brst_{-1[0]}$
are equal, and of the form
\bea
\brst_{(-1)}&=&
\sum_{r=0}^{\infty}(S^{}_{(r)},\~\cdot\~)^{}_{(-r-1)}
\~\equi{\e{lowestesses}}\~\sum_{r=2}^{\infty}(S^{}_{(r)},\~\cdot\~)^{}_{(-r-1)}
\=\sum_{s=-1}^{L}V^{}_{(s+2)\aaa_{s}}\papal{\Phi^{*}_{\aaa_{s}}}
\cr &=&\EOM_{(1)i}\papal{\varphi^{*}_{i}}
+\sum_{s=0}^{L}\Phi^{*}_{\aaa_{s-1}}Z^{\aaa_{s-1}}_{(0)\aaa_{s}}
\papal{c^{*}_{\aaa_{s}}} 
\=\EOM_{[1]i}\papal{\varphi^{*}_{i}}
+\sum_{s=0}^{L}\Phi^{*}_{\aaa_{s-1}}Z^{\aaa_{s-1}}_{[0]\aaa_{s}}
\papal{c^{*}_{\aaa_{s}}} \cr
&\equi{\e{kt1}}&\brst_{-1[0]} \~.\label{kt2}
\eea
Here we have introduced shifted structure functions $V^{}_{(s+2)\aaa_{s}}$ with
\beq
\gh(V^{}_{(s+2)\aaa_{s}})
\=-(s\!+\!1)\~,\~\~\~\~\~\~\~\~
\afn(V^{}_{(s+2)\aaa_{s}})\~<\~\safn(V^{}_{(s+2)\aaa_{s}})\=s\!+\!2\~\geq\~1
\~,\~\~\~\~\~\~\~\~s\in\{-1,\ldots,L\}\~.
\eeq
It follows from the requirement
\beq
0\~\leq\~\puregh(V^{}_{(s+2)\aaa_{s}})
\=\gh(V^{}_{(s+2)\aaa_{s}})+\afn(V^{}_{(s+2)\aaa_{s}})
\~<\~\gh(V^{}_{(s+2)\aaa_{s}})+\safn(V^{}_{(s+2)\aaa_{s}})\=1\~,
\eeq
that the $V^{}_{(s+2)\aaa_{s}}$ functions cannot depend on the ghosts ``$c$'', 
and must contain precisely one antifield or transversal field in each term. In
more detail, the $V_{\alpha}$ functions read
\beq
V^{}_{(1)\aaa_{-1}}\equiv V^{}_{(1)i}\=\EOM_{(1)i}
\~,\~\~\~\~\~\~\~\~
V^{}_{(2)\aaa_{0}}\=\varphi^{*}_{i}R^{i}_{(0)\aaa_{0}}\~,\label{svee1}
\eeq
\beq
V^{}_{(s+2)\aaa_{s}}\=\Phi^{*}_{\aaa_{s-1}}Z^{\aaa_{s-1}}_{(0)\aaa_{s}}
\~,\~\~\~\~\~\~\~\~s\in\{0,\ldots,L\}\~.\label{svee2}
\eeq
Let us use the notation $\delta$ for the shifted (=reduced) Koszul--Tate 
operator $\delta\!:=\!\brst_{(-1)}\!=\!\brst_{-1[0]}$. It 
is nilpotent because of the Noether identities \e{noether1}--\e{noether2}.

\begin{theorem}[Local acyclicity of the reduced Koszul--Tate operator]
The ope\-ra\-tor co\-ho\-mo\-lo\-gy of the reduced Koszul--Tate operator
$\delta\!:=\!\brst_{-1[0]}$ is acyclic in a local $\varphi$-neighborhood 
$U$ of each $\varphi$-point on the stationary $\varphi$-surface, \ie
\beq
\forall {~\it operators~}X:\~\~\~\~[\delta,X]=0 \~\~\wedge\~\~ \rafn(X)\neq 0
\~\~\~\~\Rightarrow\~\~\~\~\exists Y:\~ X= [\delta,Y]\~. \label{ktacyclic5}
\eeq
\label{theoremkt5}
\end{theorem}

\begin{corollary}[Local acyclicity of the shifted Koszul--Tate operator]
The cohomology of the shifted Koszul--Tate operator 
$\delta\!:=\!\brst_{(-1)}$ is acyclic in a local $\varphi$-neighborhood 
$U$ of each $\varphi$-point on the stationary $\varphi$-surface, \ie
\beq
\forall {~\it functions~}f:\~\~\~\~(\delta f)=0 \~\~\wedge\~\~ \safn(f)> 0
\~\~\~\~\Rightarrow\~\~\~\~\exists g:\~ f= (\delta g)\~. \label{ktacyclic2}
\eeq
\label{corollarykt2}
\end{corollary}

\noi
{\sc Remark}:~~
Note that $(\delta f)$ or $\delta f$ are shorthand notations for $[\delta,f]$. 
The function $g$ in \eq{ktacyclic2} is unique up to an $\delta$-exact term, 
because of $\safn(g)\!=\!\safn(f)\!+\!1\!>\!1$ and the acyclicity condition
\e{ktacyclic2}.

\noi
{\sc Proof of Corollary~\ref{corollarykt2}}:~~Use that one can identify a 
function $f$ with the operator $L^{}_{f}$ that multiplies with $f$ from the 
left, $L^{}_{f}g\!:=\!fg$. Now use Theorem~\ref{theoremkt5} and the fact that
strictly positive shifted antifield number implies strictly positive reduced
antifield number $\safn(f)>0\Rightarrow\rafn(f)>0$.
\proofbox

\noi
{\sc Proof of Theorem~\ref{theoremkt5}}:~~
In rotated variables, the shifted (=reduced) Koszul--Tate operator $\delta$ 
reads
\beq
\delta
\=\bar{\xi}_{I}\papal{\xi^{*}_{I}}
+\sum_{s=0}^{L}\bar{c}^{*\parallel}_{\AAA_{s}}
\papal{\bar{c}^{*\perp}_{\AAA_{s}}}\~,
\label{kt2rotated}
\eeq
where we have defined rotated transversal fields
\beq
\bar{\xi}^{}_{I}\~:=\~(S_{0(2)} \papar{\xi^{I}})
\=\xi^{J}H_{JI}\~,\~\~\~\~\~\~\~\~
\rafn(\bar{\xi}^{}_{I})\=1\~.
\eeq
The dual operator reads
\beq
\tilde{\delta}\=\xi^{*}_{I}\papal{\bar{\xi}_{I}}
+\sum_{s=0}^{L}\bar{c}^{*\perp}_{\AAA_{s}}
\papal{\bar{c}^{*\parallel}_{\AAA_{s}}}
\~,\~\~\~\~\~\~\~\~\rafn(\tilde{\delta})\=0\~.
\label{dualoperator}
\eeq
The commutator is a Euler/conformal vector field 
\beq
K\~:=\~[\delta,\tilde{\delta}]\=
\bar{\xi}_{I}\papal{\bar{\xi}_{I}}+
\bar{\Phi}^{*}_{\alpha}\papal{\bar{\Phi}^{*}_{\alpha}}
\~,\~\~\~\~\~\~\~\~\rafn(K)\=0\~.
\eeq
The commutator $K$ commutes with both $\delta$ and $\tilde{\delta}$. 
{}For instance, 
$[\delta,K]=[\delta,[\delta,\tilde{\delta}]]
=\Hf [[\delta,\delta],\tilde{\delta}]=0$,
due to the Jacobi identity. It is useful to define the operator
\beq
k\~:=\~[K,\~\cdot\~]\~,\~\~\~\~\~\~\~\~[K,X]\=\rafn(X)X\~,\~\~\~\~\~\~\~\~
k(\ln f)\=\frac{1}{f}k(f)\=\rafn(f)\~.
\eeq
Moreover, the kernel of the $k$ operator is precisely the sector with zero 
reduced antifield number,
\beq
k(X)\=0\~\~\~\~ \Leftrightarrow\~\~\~\~\rafn(X)\=0\~.
\eeq
The contracting homotopy operator $\delta^{-1}$ is defined as
\beq
\delta^{-1}X\~:=\~
\left\{\begin{array}{lcl}
k^{-1}[\tilde{\delta},X]
=[\tilde{\delta},k^{-1}X]
&\for&\rafn(X)\neq 0\~, \cr 0&\for&\rafn(X)= 0\~.
\end{array}\right. \label{definvbrst}
\eeq
In contrast to the dual operator $\tilde{\delta}$, the contracting homotopy
operator $\delta^{-1}$ is {\em not} a linear derivation, \ie $\delta^{-1}$ does
{\em not} satisfy a (Grassmann--graded) Leibniz rule. One has
\beq
\delta^{-1}[\delta, X]+[\delta,\delta^{-1}X]\=
\left\{\begin{array}{lcl}
X&\for&\rafn(X)\neq 0\~, \cr 0&\for&\rafn(X)= 0\~.
\end{array}\right. \label{ssm1}
\eeq
In particular, one may choose $Y\!:=\!\delta^{-1}X$ in \eq{ktacyclic5}.
\proofbox

\noi
{\sc Remark}:~~The operator $k$, when applied to an operator $X$, does not 
change the auxiliary variables in $X$. The operators $\delta$ and 
$\tilde{\delta}$ act trivially on ghosts $c$, but may annihilate or create
a single original antifield $\varphi^{*}$ (or its derivative).

\subsection{Existence of Koszul--Tate Operator $\brst_{-1}$}
\label{secexists5}

\noi
Let there be given an original action $S^{}_{0}$,
gauge(--for--gauge)$^{s}$--generators $Z^{\aaa_{s-1}}{}_{\aaa_{s}}$,
$s\!\in\!\{0,\ldots,L\}$, an antisymmetric first--stage stucture function 
$B^{ij}_{\aaa_{1}}$, and higher--stage stucture functions
$B^{i\aaa_{s-2}}_{\aaa_{s}}$, $s\!\in\!\{2,\ldots,L\}$, that satisfy the
Noether identities \e{noether1}--\e{noether2}. In particular, there are given
action parts 
\bea
\Snoeth&:=&\Sfixed-\sum_{s=2}^{L}
(-1)^{\eps(c^{*}_{\aaa_{s-2}})}c^{*}_{\aaa_{s-2}}
\varphi^{*}_{i}B^{i\aaa_{s-2}}_{\aaa_{s}}\~c^{\aaa_{s}}\~,\label{snoeth5} \\
\Sfixed&:=&\Squad+\frac{(-1)^{\eps_{j}}}{2}\varphi^{*}_{j}\varphi^{*}_{i}
B^{ij}_{\aaa_{1}}\~c^{\aaa_{1}}\~,\label{sfixed5} \\
\Squad&:=&S^{}_{0}+\sum_{s=0}^{L}\Phi^{*}_{\aaa_{s-1}}
Z^{\aaa_{s-1}}{}_{\aaa_{s}}\~c^{\aaa_{s}}\~.\label{squad5}
\eea
Define
\bea
\Vnoeth_{\aaa_{s}}&:=&(\Snoeth\papar{c^{\aaa_{s}}})
\=\Vfixed_{\aaa_{s}}-\left\{\begin{array}{rcc} 
(-1)^{\eps(c^{*}_{\aaa_{s-2}})}c^{*}_{\aaa_{s-2}}
\varphi^{*}_{i}B^{i\aaa_{s-2}}_{\aaa_{s}}
&\for& 2\leq s \leq L\~, \cr 0&\for& 0\leq s \leq 1\~, \end{array} \right. 
\label{vnoeth5}\\
\Vfixed_{\aaa_{s}}&:=&(\Sfixed\papar{c^{\aaa_{s}}})
\=\Vquad_{\aaa_{s}}+\left\{\begin{array}{rcc} 
\frac{(-1)^{\eps_{j}}}{2}\varphi^{*}_{j}\varphi^{*}_{i}B^{ij}_{\aaa_{1}}
&\for& s=1\~, \cr 0&\for& s\neq \pm 1\~, \end{array} \right.\label{vfixed5} \\
\Vquad_{\aaa_{s}}&:=&(\Squad\papar{c^{\aaa_{s}}})
\= \Phi^{*}_{\aaa_{s-1}}Z^{\aaa_{s-1}}{}_{\aaa_{s}}\~\for\~0\leq s\leq L\~,
\label{vquad5} \\
\Vnoeth_{i}&:=&\Vfixed_{i}\~:=\~\Vquad_{i}\~:=\~\EOM_{i}\~:=\~
(S^{}_{0}\papar{\varphi^{i}})\~. \label{vlast5}
\eea

\begin{theorem}[Local existence of Koszul--Tate operator $\brst_{-1}$]
Let there be given an a\-cyc\-lic, nil\-po\-tent, reduced Koszul--Tate operator
\beq
\delta\~:=\~\brst_{-1[0]}\= \EOM_{[1]i}\papal{\varphi^{*}_{i}}
+\sum_{s=0}^{L}\Phi^{*}_{\aaa_{s-1}}Z^{\aaa_{s-1}}_{[0]\aaa_{s}}
\papal{c^{*}_{\aaa_{s}}}\~, \label{kt3op}
\eeq
that is defined in some $\varphi$-neighborhood of a $\varphi$-point on the 
stationary  $\varphi$-surface, and with reduced antifield number zero. This
guarantees the local existence (in some $\varphi$-neighborhood of the
$\varphi$-point) of a nilpotent, acyclic Koszul--Tate operator 
\beq
\bbrst_{-1}\= \bar{V}^{}_{\alpha}\papal{\Phi^{*}_{\alpha}}\~,
\eeq
with antifield number minus one, and that satisfies the boundary condition
\beq 
\bar{V}^{}_{\alpha}
\=\Vfixed_{\alpha}+\cO\left(\Phi^{*}c^{*},(\Phi^{*})^{3}\right)\~.\label{bc3op}
\eeq
All such operators are of the form
\beq
\bar{V}^{}_{\aaa_{s}}\=\Vfixed_{\aaa_{s}}
-(-1)^{\eps(c^{*}_{\aaa_{s-2}})}c^{*}_{\aaa_{s-2}}
\varphi^{*}_{i}\left(B^{i\aaa_{s-2}}_{\aaa_{s}}+R^{i}{}_{\aaa_{0}}
X^{\aaa_{0}\aaa_{s-2}}_{\aaa_{s}}+\EOM_{j}Y^{ji\aaa_{s-2}}_{\aaa_{s}}\right)
+\cO\left((c^{*})^{2},(\Phi^{*})^{3}\right)\~,
\eeq
where $Y^{ji\aaa_{s-2}}_{\aaa_{s}}\!=\!-(-1)^{\eps_{i}\eps_{j}}Y^{ij\aaa_{s-2}}_{\aaa_{s}}$.
\label{theoremss5}
\end{theorem}

\noi
{\sc Proof of Theorem~\ref{theoremss5}}:~~Let us, for notational reasons, put a 
bar on top of the sought--for Koszul--Tate operator $\bbrst_{-1}$, and no bar
on quantities associated with the given boundary conditions
\e{snoeth5}--\e{vlast5}. We use $\Delta$ to denote differences, \eg
$\Delta\Vnoeth_{\alpha}\!:=\!\bVnoeth_{\alpha}-\Vnoeth_{\alpha}$, 
$\Delta\Vfixed_{\alpha}\!:=\!\bVfixed_{\alpha}-\Vfixed_{\alpha}$, and so
forth. We shall below inductively define the bar solution $\bbrst_{-1}$ to all
orders in the reduced antifield number, but initially, we only fix the
zeroth--order part $\bbrst_{-1[0]}\~:=\~\delta$ to be equal to the reduced
Koszul--Tate operator $\delta$. The $r$th nilpotency expression $\bNI_{[r]}$ 
for a bar solution $\bbrst_{-1}$ can then be written as
\beq
\bNI_{[r]}\~:=\~\Hf \sum_{p=0}^{r}[\bbrst_{-1[p]},\bbrst_{-1[r-p]}]
\=\left\{\begin{array}{lcl}
\Hf[\delta,\delta]\=0&\for&r\!=\!0\~, \cr 
[\delta,\bbrst_{-1[r]}]+\bar{\cB}^{}_{[r]} &\for&r\!\geq\!1\~,
\end{array}\right.
\label{nidecomp5}
\eeq
where 
\beq
\bar{\cB}^{}_{[r]}
\~:=\~\Hf \sum_{p=1}^{r-1}[\bbrst_{-1[p]},\bbrst_{-1[r-p]}]
\~,\~\~\~\~r\~\geq\~1\~.
\label{bdef5}
\eeq
The $\bar{\cB}^{}_{[r]}$ operator \e{bdef5} is a linear derivation, since it is
a commutator of linear derivations. It cannot contain derivatives \wrt 
original antifields $\varphi^{*}_{i}$, since $\afn(\bar{\cB}^{}_{[r]})\!=\!-2$. 
Hence the $\bar{\cB}^{}_{[r]}$ operator \e{bdef5} is of the form
\beq
\bar{\cB}^{}_{[r]}\=
\sum_{s=0}^{L}\bar{\cB}^{}_{[r+1]\aaa_{s}}\papal{c^{*}_{\aaa_{s}}}\~.
\eeq
The proof of the main statement is an induction in the reduced antifield 
number $r\!\geq\!1$. Assume that there exists a bar solution
\beq
\bbrst_{-1[p]}\=\bar{V}^{}_{[p+1]\alpha}\papal{\Phi^{*}_{\alpha}}
\~,\~\~\~\~\~\~\~\~\rafn(\bbrst_{-1[p]})\=p
\~,\~\~\~\~\~\~\~\~p\in\{0,1,\ldots,r\!-\!1\}\~,
\eeq
\beq
\bar{V}^{}_{[p]\alpha}\=\bVnoeth_{[p]\alpha}
+\cO\left((c^{*})^{2},(\Phi^{*})^{3}\right)
\~,\~\~\~\~\~\~\~\~\rafn(\bar{V}^{}_{[p]\alpha})\=p
\~,\~\~\~\~\~\~\~\~p\in\{1,2,\ldots,r\}\~, \label{inductassump1op}
\eeq
such that the boundary condition 
\beq
\Delta\Vfixed_{[p]\alpha}\=0
\~,\~\~\~\~\~\~\~\~p\in\{1,2,\ldots,r\}\~, \label{inductassump2op}
\eeq
is fulfilled, such that
\beq
\Delta B^{i\aaa_{s-2}}_{[p]\aaa_{s}}
\=\sum_{q=0}^{p}R^{i}_{[p-q]\aaa_{0}}X^{\aaa_{0}\aaa_{s-2}}_{[q]\aaa_{s}}
+\sum_{q=0}^{p-1}\EOM_{[p-q]j}Y^{ji\aaa_{s-2}}_{[q]\aaa_{s}}
\~,\~\~\~\~\~\~\~\~p\in\{0,1,\ldots,r\!-\!2\}\~, \label{inductassump3op}
\eeq
\beq 
Y^{ji\aaa_{s-2}}_{[p]\aaa_{s}}
\=-(-1)^{\eps_{i}\eps_{j}}Y^{ij\aaa_{s-2}}_{[p]\aaa_{s}}
\~,\~\~\~\~\~\~\~\~p\in\{0,1,\ldots,r\!-\!3\}
\~,\~\~\~\~\~\~\~\~s\in\{2,3,\ldots,L\}\~,
\eeq
and such that the nilpotency holds up to the order $r\!-\!1$ in reduced
antifield number
\beq
0\=\bNI_{[0]}\=\bNI_{[1]}\=\ldots\=\bNI_{[r-1]}\~.
\label{niassump5}
\eeq
It follows from the induction assumption that the $\bar{\cB}^{}_{[r]}$ operator
\e{bdef5} exists. The Jacobi identity $\bJI$ gives
\beq
0\=\bJI_{[r]}\=\Hf[\bbrst_{-1},[\bbrst_{-1},\bbrst_{-1}]]^{}_{[r]}\=
\sum_{p=0}^{\infty}[\bbrst_{-1[p]},\bNI_{[r-p]}]
\~\equi{\e{niassump5}}\~[\delta,\bNI_{[r]}]
\~\equi{\e{nidecomp5}}\~[\delta,\bar{\cB}^{}_{[r]}]\~. 
\label{jacobicalc5}
\eeq
Hence the $\bar{\cB}^{}_{[r]}$ operator \e{bdef5} is $\delta$-closed.
Let us tentatively define
\beq
\bbrst_{-1[r]}\~:=\~-\delta^{-1}\bar{\cB}^{}_{[r]}\~,\label{deltadef5}
\eeq
\cf definition \e{definvbrst}. It follows from \eq{ssm1} that the $r$th
nilpotency relation \e{nidecomp5} is fulfilled
\beq
\bNI_{[r]}\~\equi{\e{nidecomp5}}\~\bar{\cB}^{}_{[r]}+[\delta,\bbrst_{-1[r]}]
\~\equi{\e{deltadef5}}\~
\bar{\cB}^{}_{[r]}-[\delta,\delta^{-1}\bar{\cB}^{}_{[r]}]
\~\equi{\e{ssm1}}\~
\delta^{-1}[\delta,\bar{\cB}^{}_{[r]}]\~\equi{\e{jacobicalc5}}\~0\~,
\label{niceguy5}
\eeq
since $r\!\neq\!0$. It is easy to see that the $\bbrst_{-1[r]}$ operator
\e{deltadef5} is a linear derivation. It cannot contain derivatives \wrtt
rotated transversal coordinates $\bar{\xi}_{I}$, since
$\afn(\bbrst_{-1[r]})\!=\!-1$. 
Hence the $\bbrst_{-1[r]}$ operator \e{deltadef5} is of the form
\beq
\bbrst_{-1[r]}\=\bar{V}^{}_{[r+1]\alpha}\papal{\Phi^{*}_{\alpha}}\~.
\eeq
This choice $\bbrst_{-1[r]}$ may not meet the prescribed boundary condition 
\e{bc3op}. Let us probe the difference in terms of cohomology.
\bea
\lefteqn{[\delta,\Delta V^{}_{[r+1]i}]\papal{\varphi^{*}_{i}}
\=0\~, \label{cv0} }\\
\lefteqn{ \Delta V^{}_{[r+1]i}[\papal{\varphi^{*}_{i}},\delta]+
[\delta,\Delta V^{}_{[r+1]\aaa_{0}}]\papal{c^{*}_{\aaa_{0}}}  }\cr
&=&\left(\Delta\EOM_{[r+1]i}\~R^{i}_{[0]\aaa_{0}}+
\EOM_{[1]i}\Delta R^{i}_{[r]\aaa_{0}}\right)\papal{c^{*}_{\aaa_{0}}} 
\~\equi{\e{noether1}}\~
-\Delta\sum_{p=2}^{r}\EOM_{[p]i}R^{i}_{[r+1-p]\aaa_{0}}\papal{c^{*}_{\aaa_{0}}}
\~\equi{\e{inductassump2op}}\~0 \~, \label{cv1} \\
\lefteqn{ \Delta V^{}_{[r+1]\aaa_{0}}[\papal{c^{*}_{\aaa_{0}}},\delta]+
[\delta,\Delta V^{}_{[r+1]\aaa_{1}}]\papal{c^{*}_{\aaa_{1}}} }\cr
&=&\varphi^{*}_{i}\left(\Delta R^{i}_{[r]\aaa_{0}}\~Z^{\aaa_{0}}_{[0]\aaa_{1}}
+R^{i}_{[0]\aaa_{0}}\Delta Z^{\aaa_{0}}_{[r]\aaa_{1}}
-\EOM_{[1]j}\Delta B^{ji}_{[r-1]\aaa_{1}}\right)\papal{c^{*}_{\aaa_{1}}} \cr
&\equi{\e{noether15}}&\varphi^{*}_{i}\Delta\left(
\sum_{p=2}^{r}\EOM_{[p]j} B^{ji}_{[r-p]\aaa_{1}}
-\sum_{p=1}^{r-1}R^{i}_{[p]\aaa_{0}} Z^{\aaa_{0}}_{[r-p]\aaa_{1}}
\right)\papal{c^{*}_{\aaa_{1}}} 
\~\equi{\e{inductassump2op}}\~0\~, \label{cv2} \\
\lefteqn{\Delta\Vnoeth_{[r+1]\aaa_{s-1}}
[\papal{c^{*}_{\aaa_{s-1}}},\delta]
+[\delta,\Delta\Vnoeth_{[r+1]\aaa_{s}}]\papal{c^{*}_{\aaa_{s}}} 
+\cO\left((\Phi^{*})^{2}\papal{c^{*}}\right) }\cr
&=&c^{*}_{\aaa_{s-2}}\Delta Z^{\aaa_{s-2}}_{[r]\aaa_{s-1}}
[\papal{c^{*}_{\aaa_{s-1}}},\delta]
+[\delta,\Delta\Vnoeth_{[r+1]\aaa_{s}}]\papal{c^{*}_{\aaa_{s}}}  \cr
&=&c^{*}_{\aaa_{s-2}}\left(
\Delta Z^{\aaa_{s-2}}_{[r]\aaa_{s-1}}\~Z^{\aaa_{s-1}}_{[0]\aaa_{s}}
+Z^{\aaa_{s-2}}_{[0]\aaa_{s-1}}\Delta Z^{\aaa_{s-1}}_{[r]\aaa_{s}}
-\EOM_{[1]i}\Delta B^{i\aaa_{s-2}}_{[r-1]\aaa_{s}}
\right)\papal{c^{*}_{\aaa_{s}}} \cr
&\equi{\e{noether2}}&c^{*}_{\aaa_{s-2}}\Delta\left(
\sum_{p=2}^{r}\EOM_{[p]i} B^{i\aaa_{s-2}}_{[r-p]\aaa_{s}}
-\sum_{p=1}^{r-1}Z^{\aaa_{s-2}}_{[p]\aaa_{s-1}} 
Z^{\aaa_{s-1}}_{[r-p]\aaa_{s}}\right)\papal{c^{*}_{\aaa_{s}}} 
\~\equi{\e{inductassump2op}}\~c^{*}_{\aaa_{s-2}}\sum_{p=2}^{r}\EOM_{[p]i}
\Delta B^{i\aaa_{s-2}}_{[r-p]\aaa_{s}}\papal{c^{*}_{\aaa_{s}}} \cr
&\equi{\e{inductassump3op}}&c^{*}_{\aaa_{s-2}}\left(\sum_{p=2}^{r}\EOM_{[p]i}
\sum_{q=0}^{r-p}R^{i}_{[r-p-q]\aaa_{0}}X^{\aaa_{0}\aaa_{s-2}}_{[q]\aaa_{s}}
+\sum_{p=2}^{r-1}\EOM_{[p]i}\sum_{q=0}^{r-1-p}
\EOM_{[r-p-q]j}Y^{ji\aaa_{s-2}}_{[q]\aaa_{s}} 
\right)\papal{c^{*}_{\aaa_{s}}} \cr
&=&c^{*}_{\aaa_{s-2}}\left(\sum_{q=0}^{r-2}\sum_{p=2}^{r-q}
\EOM_{[p]i}R^{i}_{[r-q-p]\aaa_{0}}X^{\aaa_{0}\aaa_{s-2}}_{[q]\aaa_{s}}
+\sum_{q=0}^{r-3}\sum_{p=2}^{r-1-q}
\EOM_{[p]i}\EOM_{[r-q-p]j}Y^{ji\aaa_{s-2}}_{[q]\aaa_{s}} 
\right)\papal{c^{*}_{\aaa_{s}}} \cr
&\equi{\e{noether1}}&-c^{*}_{\aaa_{s-2}}\EOM_{[1]i}
\tilde{B}^{i\aaa_{s-2}}_{[r-1]\aaa_{s}}\papal{c^{*}_{\aaa_{s}}}
\= -c^{*}_{\aaa_{s-2}}[\delta,\varphi^{*}_{i}
\tilde{B}^{i\aaa_{s-2}}_{[r-1]\aaa_{s}}]\papal{c^{*}_{\aaa_{s}}}\~, \label{cvs} 
\eea
where
\beq
\tilde{B}^{i\aaa_{s-2}}_{[r-1]\aaa_{s}}\~:=\~
\sum_{q=0}^{r-2}R^{i}_{[r-1-q]\aaa_{0}}X^{\aaa_{0}\aaa_{s-2}}_{[q]\aaa_{s}}
+\sum_{q=0}^{r-3}\EOM_{[r-1-q]j}Y^{ji\aaa_{s-2}}_{[q]\aaa_{s}} \~.
\eeq
If one adds together eqs.\ \e{cv0}--\e{cvs}, one gets
\beq
\left[\delta,\~\Delta\brst{}^{\rm Noether}_{[r]}+\tilde{B}^{}_{[r]}\right]
\=\cO\left((\Phi^{*})^{2}\papal{c^{*}}\right)
\~,\label{altogethercv}
\eeq
\beq
\Delta\brst{}^{\rm Noether}_{[r]}
\~:=\~\Delta\Vnoeth_{[r+1]\alpha}\papal{\Phi^{*}_{\alpha}}
\~,\~\~\~\~\~\~\~\~
\tilde{B}^{}_{[r]}
\~:=\~\sum_{s=2}^{L}(-1)^{\eps(c^{*}_{\aaa_{s-2}})}
c^{*}_{\aaa_{s-2}}\varphi^{*}_{i}\~
\tilde{B}^{i\aaa_{s-2}}_{[r-1]\aaa_{s}}\papal{c^{*}_{\aaa_{s}}}
\~.\label{altogethercv2}
\eeq
Now redefine the $\bbrst_{-1[r]}$ solution by a $\delta$-exact amount 
\beq
\bbrst_{-1[r]}\~\~\longrightarrow\~\~\bbrst_{-1[r]}
-\left[\delta,\~\delta^{-1}\left(\Delta\brst{}^{\rm Noether}_{[r]}
+\tilde{B}^{}_{[r]}\right)\right]\~, \label{newsolution2op}
\eeq
\cf definition \e{definvbrst}. The new $\bbrst_{-1[r]}$ operator
\e{newsolution2op} is still a linear derivation that satisfies \eq{niceguy5}, 
but now 
\bea
\Delta\brst{}^{\rm Noether}_{[r]}&\longrightarrow&
\Delta\brst{}^{\rm Noether}_{[r]}
-\left[\delta,\~\delta^{-1}\left(\Delta\brst{}^{\rm Noether}_{[r]}
+\tilde{B}^{}_{[r]}\right)\right] \cr
&&\~\equi{\e{ssm1}}\~\delta^{-1}\left[\delta,\~
\Delta\brst{}^{\rm Noether}_{[r]}+\tilde{B}_{[r]}\right]-\tilde{B}_{[r]}
\~\equi{\e{altogethercv}}\~\cO\left((\Phi^{*})^{2}\papal{\Phi^{*}}\right)\~,
\eea
so now the following boundary condition is satisfied as well
\beq
\Delta\Vquad_{[r+1]\alpha}
\~:=\~\bVquad_{[r+1]\alpha}-\Vquad_{[r+1]\alpha}\=0\~,\label{bcbc1}
\eeq
or equivalently,
\beq
\Delta\EOM_{[r+1]i}\=0\~,\~\~\~\~\~\~\~\~
\Delta R^{i}_{[r]\aaa_{0}}\=0\~,\~\~\~\~\~\~\~\~
\Delta Z^{\aaa_{s-1}}_{[r]\aaa_{s}}\=0\~,\~\~\~\~\~\~\~\~
s\in\{1,\ldots,L\}\~,\label{bcbc2}
\eeq
\cf \eq{ssm1}. It still remains to show that the boundary condition
$\Delta B^{ij}_{[r-1]\aaa_{1}}\!=\!0$ can be achieved (after an appropriate 
$\delta$-exact shift of the $\bbrst_{-1[r]}$ solution \e{newsolution2op}).
To this end, repeat the calculation \e{cv2} with the knowledge \e{bcbc2}, and 
conclude that
\beq
[\delta,\Delta V^{}_{[r+1]\aaa_{1}}]\=0\~,\~\~\~\~\~\~\~\~
\Delta V^{}_{[r+1]\aaa_{1}}
\=\frac{(-1)^{\eps_{j}}}{2}\varphi^{*}_{j}\varphi^{*}_{i}
\Delta B^{ij}_{[r-1]\aaa_{1}}
\~.\label{bcbc3consequences}
\eeq
Now redefine the $\bbrst_{-1[r]}$ solution \e{newsolution2op} by a 
$\delta$-exact amount 
\bea
\bbrst_{-1[r]}&\longrightarrow&\bbrst_{-1[r]}
-\left[\delta,\~\delta^{-1}\left(\Delta V^{}_{[r+1]\aaa_{1}}\right)
\papal{c^{*}_{\aaa_{1}}}\right] \cr
&=&\bbrst_{-1[r]}
-\left[\delta,\~\delta^{-1}\Delta V^{}_{[r+1]\aaa_{1}}\right]
\papal{c^{*}_{\aaa_{1}}}
+\left(\delta^{-1}\Delta V^{}_{[r+1]\aaa_{1}}\right)
[\papal{c^{*}_{\aaa_{1}}},\delta]\~. \label{newsolution3op}
\eea
The second and third term on the \rhs of \eq{newsolution3op} will change 
the structure functions $\bar{B}^{ij}_{[r-1]\aaa_{1}}$ and
$\bar{B}^{i\aaa_{0}}_{[r-1]\aaa_{2}}$, respectively. In detail,
\beq
\Delta V^{}_{[r+1]\aaa_{1}}\~\~\longrightarrow\~\~
\Delta V^{}_{[r+1]\aaa_{1}}
-\left[\delta,\~\delta^{-1}\Delta V^{}_{[r+1]\aaa_{1}}\right]
\~\equi{\e{ssm1}}\~
\delta^{-1}\left[\delta,\Delta V^{}_{[r+1]\aaa_{1}}\right]
\~\equi{\e{bcbc3consequences}}\~0\~,
\eeq
so now at least $\Delta B^{ij}_{[r-1]\aaa_{1}}\!=\!0$, and therefore the 
boundary condition \e{bc3op} is fulfilled,
\beq
\Delta\Vfixed_{[r+1]\alpha}
\~:=\~\bVfixed_{[r+1]\alpha}-\Vfixed_{[r+1]\alpha}\=0\~.\label{bcbc4}
\eeq
{}Finally, repeat calculation \e{cvs} with the knowledge \e{bcbc2}, and
conclude that
\beq
[\delta,\varphi^{*}_{i}\Delta B^{i\aaa_{s-2}}_{[r-1]\aaa_{s}}]
\=[\delta,\varphi^{*}_{i}\tilde{B}^{i\aaa_{s-2}}_{[r-1]\aaa_{s}}]
\~,\~\~\~\~\~\~\~\~s\in\{2,\ldots,L\}\~.\label{bcbc5consequences}
\eeq
In other words, there exist structure functions
$X^{\aaa_{0}\aaa_{s-2}}_{[r-1]\aaa_{s}}$ and
$Y^{ji\aaa_{s-2}}_{[r-2]\aaa_{s}}
\!=\!-(-1)^{\eps_{i}\eps_{j}}Y^{ij\aaa_{s-2}}_{[r-2]\aaa_{s}}$ such that
\beq
\Delta B^{i\aaa_{s-2}}_{[r-1]\aaa_{s}}
\=\tilde{B}^{i\aaa_{s-2}}_{[r-1]\aaa_{s}}
+R^{i}_{[0]\aaa_{0}}X^{\aaa_{0}\aaa_{s-2}}_{[r-1]\aaa_{s}}
+\EOM_{[1]j}Y^{ji\aaa_{s-2}}_{[r-2]\aaa_{s}}
\=\sum_{q=0}^{r-1}R^{i}_{[r-1-q]\aaa_{0}}X^{\aaa_{0}\aaa_{s-2}}_{[q]\aaa_{s}}
+\sum_{q=0}^{r-2}\EOM_{[r-1-q]j}Y^{ji\aaa_{s-2}}_{[q]\aaa_{s}}\~,\label{bxy5}
\eeq
which is induction assumption \e{inductassump3op} for the next step 
$p\!=\!r\!-\!1$.
\proofbox

\subsection{Existence of Proper Action $S$}
\label{secexists2}

\begin{table}[t]
\caption{Table over the antibracket $(f,g)$ of various functions 
$f$ and $g$ with ghost number zero, $\gh(f)\!=\!0\!=\!\gh(g)$. }

\label{multtable2}
\begin{center}
\begin{tabular}{|l||c|c|c|c|}  \hline
$\!\downarrow f\~\~\~\~g\to\!$&$S_{0}$&$\cO(\Phi^{*}c)$&$\cO((\Phi^{*})^{2})$
&$\cO(c^{2})$ \rule{0ex}{3ex} \\ \hline\hline
$S_{0}$&$0$&$\cO(c)$&$\cO(\Phi^{*}\xi c)$&$\cO(c^{2})$\rule{0ex}{3ex} 
\\ \hline
$\cO(\Phi^{*}c)$&&$\cO(\Phi^{*}c)$&$\cO((\Phi^{*})^{2})$
&$\cO(c^{2})$ \rule{0ex}{3ex} \\ \hline
$\cO((\Phi^{*})^{2})$&&&$\cO((\Phi^{*})^{2})$&$\cO((\Phi^{*})^{2}c^{2})$ 
\rule{0ex}{3ex} \\ \hline
$\cO(c^{2})$&&&&$\cO(c^{2})$ \rule{0ex}{3ex} \\ \hline
\end{tabular}
\end{center}
\end{table}

\noi
Let there be given an original action $S^{}_{0}$,
gauge(--for--gauge)$^{s}$--generators $Z^{\aaa_{s-1}}{}_{\aaa_{s}}$,
$s\!\in\!\{0,\ldots,L\}$, an antisymmetric first--stage stucture function 
$B^{ij}_{\aaa_{1}}$, and higher--stage stucture functions
$B^{i\aaa_{s-2}}_{\aaa_{s}}$, $s\!\in\!\{2,\ldots,L\}$, that satisfy the
Noether identities \e{noether1}--\e{noether2}.

\begin{theorem}[Local existence of proper action $S$] Let there be given
a nilpotent, acyclic shifted Koszul--Tate operator
\beq
\delta\~:=\~\brst_{(-1)}\= \EOM_{(1)i}\papal{\varphi^{*}_{i}}
+\sum_{s=0}^{L}\Phi^{*}_{\aaa_{s-1}}Z^{\aaa_{s-1}}_{(0)\aaa_{s}}
\papal{c^{*}_{\aaa_{s}}}\~, \label{kt3}
\eeq
that is defined in some $\varphi$-neighborhood of a $\varphi$-point on the
stationary $\varphi$-surface, and with shifted antifield number minus one. 
This guarantees the local existence (in some $\varphi$-neighborhood of the
$\varphi$-point) of a proper solution $\bar{S}$ to the classical master
equation \e{cmej}, that satisfies the boundary condition
\beq 
\bar{S}\=\Sfixed+\cO\left(\Phi^{*}c^{*},(\Phi^{*})^{3},c^{2}\right)\~. 
\label{bc3}
\eeq
All such solutions are of the form
\beq
\bar{S}\=\Sfixed -\sum_{s=2}^{L}(-1)^{\eps(c^{*}_{\aaa_{s-2}})}
c^{*}_{\aaa_{s-2}}\varphi^{*}_{i}\left(B^{i\aaa_{s-2}}_{\aaa_{s}}
+R^{i}{}_{\aaa_{0}}X^{\aaa_{0}\aaa_{s-2}}_{\aaa_{s}}
+\EOM_{j}Y^{ji\aaa_{s-2}}_{\aaa_{s}}\right)c^{\aaa_{s}}
+\cO\left((c^{*})^{2},(\Phi^{*})^{3},c^{2}\right)\~,
\eeq
where $Y^{ji\aaa_{s-2}}_{\aaa_{s}}\!=\!-(-1)^{\eps_{i}\eps_{j}}Y^{ij\aaa_{s-2}}_{\aaa_{s}}$.
\label{theoremss2}
\end{theorem}

\noi
{\sc First proof of Theorem~\ref{theoremss2} using reduced antifield number}:~~ 
Combine Theorem~\ref{theoremss1} and Theorem~\ref{theoremss5}.
\proofbox

\noi
{\sc Second proof of Theorem~\ref{theoremss2} using shifted antifield 
number}:~~Let us, for notational reasons, put a bar on top of the sought--for
proper action $\bar{S}$, and no bar on quantities associated with the given
boundary conditions \e{snoeth5}--\e{squad5}. We use $\Delta$ to denote 
differences, \eg $\Delta\Snoeth\!:=\!\bSnoeth\!-\!\Snoeth$, 
$\Delta\Sfixed\!:=\!\bSfixed\!-\!\Sfixed$, and so forth. We shall below
inductively define the bar solution $\bar{S}$ to all orders in the shifted 
antifield number, but initially, we only fix the zeroth--order, first--order 
and second--order part as
\beq
\bar{S}^{}_{(0)}\~:=\~S^{}_{(0)}\~,\~\~\~\~\~\~\~\~
\bar{S}^{}_{(1)}\~:=\~S^{}_{(1)}\~,\~\~\~\~\~\~\~\~
\bar{S}^{}_{(2)}\~:=\~S^{}_{(2)}\~, \label{lowestbesses}
\eeq
\cf \eq{lowestesses}.
The $r$th classical master expression $\bCME_{(r)}$ for a bar solution
$\bar{S}$ can be written as
\bea
\bCME_{(r)}&:=&\Hf \sum_{p,q \geq 0}
(\bar{S}^{}_{(p)},\bar{S}^{}_{(q)})^{}_{(r-p-q)}
\~\equi{\e{lowestesses}}\~
\Hf \sum_{p,q \geq 2} (\bar{S}^{}_{(p)},\bar{S}^{}_{(q)})^{}_{(r-p-q)} \cr
&\equi{\e{lemmass2b}}&
\Hf \sum_{2\leq p,q\leq r+1} (\bar{S}^{}_{(p)},\bar{S}^{}_{(q)})^{}_{(r-p-q)}
\~\equi{\e{lemmass2a}+\e{eom3}}\~\left\{\begin{array}{l}
\bbrst_{(-1)}\bar{S}^{}_{(r+1)}+\bar{\cB}^{}_{(r)}\~, \cr
0\~\for\~r\in\{0,1\}\~,\end{array} \right.
\label{cmxdecomp2}
\eea
where 
\beq
\bar{\cB}^{}_{(r)}\~:=\~
\Hf \sum_{2\leq p,q\leq r}(\bar{S}^{}_{(p)},\bar{S}^{}_{(q)})^{}_{(r-p-q)} 
\~,\~\~\~\~r\~\geq\~0\~.
\label{bdef2}
\eeq
The proof of the main statement is an induction in the shifted antifield 
number $r\!\geq\!2$. Assume that there exist a bar solution
\beq
\bar{S}^{}_{(p)}\=\bSnoeth_{(p)}
+\cO\left((c^{*})^{2},(\Phi^{*})^{3},c^{2}\right)
\~,\~\~\~\~\~\~\~\~p\in\{0,1,\ldots,r\}\~, \label{inductassump1}
\eeq
such that the boundary condition 
\beq
\Delta\Sfixed_{(p)}\=0
\~,\~\~\~\~\~\~\~\~p\in\{0,1,\ldots,r\}\~, \label{inductassump2}
\eeq
is fulfilled, such that
\beq
\Delta B^{i\aaa_{s-2}}_{(p)\aaa_{s}}
\=\sum_{q=0}^{p}R^{i}_{(p-q)\aaa_{0}}X^{\aaa_{0}\aaa_{s-2}}_{(q)\aaa_{s}}
+\sum_{q=0}^{p-1}\EOM_{(p-q)j}Y^{ji\aaa_{s-2}}_{(q)\aaa_{s}}
\~,\~\~\~\~\~\~\~\~p\in\{0,1,\ldots,r\!-\!s\!-\!3\}\~, \label{inductassump3}
\eeq
\beq 
Y^{ji\aaa_{s-2}}_{(p)\aaa_{s}}
\=-(-1)^{\eps_{i}\eps_{j}}Y^{ij\aaa_{s-2}}_{(p)\aaa_{s}}
\~,\~\~\~\~\~\~\~\~p\in\{0,1,\ldots,r\!-\!s\!-\!4\}
\~,\~\~\~\~\~\~\~\~s\in\{2,3,\ldots,L\}\~,
\eeq
and such that the classical master equation holds up to a order $r\!-\!1$ in
shifted antifield number
\beq
0\=\bCME_{(0)}\=\bCME_{(1)}\=\ldots\=\bCME_{(r-1)}\~.
\label{cmxassump2}
\eeq
The action
$\bar{S}^{}_{(p)}\!=\!\bar{S}^{}_{(p)}(\Phi^{\alpha};\varphi^{*}_{i},
c^{*}_{\aaa_{0}},\ldots,c^{*}_{\aaa_{p-3}})$
cannot depend on antifields $c^{*}_{\aaa_{q}}$, for 
$p-3\!<\!q\!\leq\!L$, because their shifted antifield number
$\safn(c^{*}_{\aaa_{q}})\!=\!q\!+\!3$ is too big. 
It follows from the induction assumption that the $\bar{\cB}^{}_{(r)}$ function
\e{bdef2} exists, and that it is a function 
$\bar{\cB}^{}_{(r)}\!=\!\bar{\cB}^{}_{(r)}(\Phi^{\alpha};\varphi^{*}_{i},
c^{*}_{\aaa_{0}},\ldots,c^{*}_{\aaa_{r-3}})$. 
We want to prove that there exists 
$\bar{S}^{}_{(r+1)}\!=\!\bar{S}^{}_{(r+1)}(\Phi^{\alpha};\varphi^{*}_{i},
c^{*}_{\aaa_{0}},\ldots,c^{*}_{\aaa_{r-2}})$, such that $\bCME_{(r)}\!=\!0$. 
The Jacobi identity $\bJI$ gives
\beq
0\=\bJI_{(r-1)}\~\equi{\e{jij}}\~
\sum_{p=-1}^{\infty}\bbrst_{(p)}\bCME_{(r-p-1)}
\~\equi{\e{cmxassump2}}\~\bbrst_{(-1)}\bCME_{(r)}
\~\equi{\e{cmxdecomp2}}\~\bbrst_{(-1)}\bar{\cB}^{}_{(r)}
\~\equi{\e{hjaelpendehaand}}\~
\delta\bar{\cB}^{}_{(r)}\~. \label{jacobicalc}
\eeq
In the last equality of \eq{jacobicalc} is used that the two shifted 
Koszul--Tate operators $\delta\!\equiv\!\brst_{(-1)}$ and $\bbrst_{(-1)}$ agree
on functions
$f\!=\!f(\Phi^{\alpha};\varphi^{*}_{i},c^{*}_{\aaa_{0}},\ldots,c^{*}_{\aaa_{r-2}})$,
due to the induction assumption \e{inductassump2}.
\beq
\Delta\brst_{(-1)}
\=\sum_{p=0}^{\infty}(\Delta \Squad_{(p)},\~\cdot\~)^{}_{(-p-1)}
\~\equi{\e{inductassump2}}\~
\sum_{p=r+1}^{\infty}(\Delta \Squad_{(p)},\~\cdot\~)^{}_{(-p-1)}
\=\sum_{p=r+1}^{\infty}(\Delta \Squad_{(p)}\papar{c^{\aaa_{p-2}}})
\papal{c^{*}_{\aaa_{p-2}}}\~.\label{hjaelpendehaand}
\eeq
Hence the $\bar{\cB}^{}_{(r)}$ function \e{bdef2} is $\delta$-closed. 
The acyclicity condition \e{ktacyclic2} then shows that there exists a function 
$\bar{S}_{(r+1)}$ such that 
\beq
-\delta\bar{S}^{}_{(r+1)}\=\bar{\cB}^{}_{(r)}\~,\label{niceguy}
\eeq
because $r\!>\!0$. The $r$th classical master equation is then satisfied
\beq
\bCME_{(r)}
\~\equi{\e{cmxdecomp2}}\~
\bbrst_{(-1)}\bar{S}^{}_{(r+1)}+\bar{\cB}^{}_{(r)}
\~\equi{\e{hjaelpendehaand}}\~
\delta\bar{S}^{}_{(r+1)}+\bar{\cB}^{}_{(r)}
\~\equi{\e{niceguy}}\~0\~.
\label{cmxr}
\eeq
This choice $\bar{S}^{}_{(r+1)}$ may not meet the prescribed
boundary condition \e{bc3}. Let us probe the difference in terms of cohomology.
\bea
\delta\Delta S^{}_{0(r+1)}&=&0\~, \label{cl0} \\
\delta\Delta S^{}_{1(r+1)}
&=&\EOM_{(1)i}\Delta R^{i}_{(r-1)\aaa_{0}}c^{\aaa_{0}}
\~\equi{\e{noether1}}\~
-\Delta\sum_{p=2}^{r}\EOM_{(p)i} R^{i}_{(r-p)\aaa_{0}}c^{\aaa_{0}}
\~\equi{\e{inductassump2}}\~0 \~, \label{cl1} \\
\delta\Delta S^{}_{2(r+1)}
&=&\varphi^{*}_{i}\left(R^{i}_{(0)\aaa_{0}}
\Delta Z^{\aaa_{0}}_{(r-2)\aaa_{1}}
-\EOM_{(1)j}\Delta B^{ji}_{(r-3)\aaa_{1}}\right)c^{\aaa_{1}} \cr
&\equi{\e{noether15}}&\varphi^{*}_{i}\Delta\left(
\sum_{p=2}^{r-2}\EOM_{(p)j} B^{ji}_{(r-p-2)\aaa_{1}}
-\sum_{p=1}^{r-2}R^{i}_{(p)\aaa_{0}} Z^{\aaa_{0}}_{(r-p-2)\aaa_{1}}
\right)c^{\aaa_{1}}
\~\equi{\e{inductassump2}}\~0\~, \label{cl2} \\
\lefteqn{\delta\Delta\Snoeth_{s(r+1)}+\cO((\Phi^{*})^{2})} \cr
&=&\delta\Delta\Sfixed_{s(r+1)}
-c^{*}_{\aaa_{s-3}}\delta\varphi^{*}_{i}\~
\Delta B^{i\aaa_{s-3}}_{(r-s-1)\aaa_{s-1}}\~c^{\aaa_{s-1}} \cr
&=&c^{*}_{\aaa_{s-3}}\left(Z^{\aaa_{s-3}}_{(0)\aaa_{s-2}}
\Delta Z^{\aaa_{s-2}}_{(r-s)\aaa_{s-1}}
-\EOM_{(1)i}\Delta B^{i\aaa_{s-3}}_{(r-s-1)\aaa_{s-1}}\right)c^{\aaa_{s-1}} \cr
&\equi{\e{noether2}}&c^{*}_{\aaa_{s-3}}\Delta\left(\sum_{p=2}^{r-s}
\EOM_{(p)i} B^{i\aaa_{s-3}}_{(r-s-p)\aaa_{s-1}}
-\sum_{p=1}^{r-s}Z^{\aaa_{s-3}}_{(p)\aaa_{s-2}}
Z^{\aaa_{s-2}}_{(r-s-p)\aaa_{s-1}}\right)c^{\aaa_{s-1}} \cr
&\equi{\e{inductassump2}}&c^{*}_{\aaa_{s-3}}\sum_{p=2}^{r-s}
\EOM_{(p)i}\Delta B^{i\aaa_{s-3}}_{(r-s-p)\aaa_{s-1}}\~c^{\aaa_{s-1}} \cr
&\equi{\e{inductassump3}}&c^{*}_{\aaa_{s-3}}\left(\sum_{p=2}^{r-s}\EOM_{(p)i}
\sum_{q=0}^{r-s-p}R^{i}_{(r-s-p-q)\aaa_{0}}
X^{\aaa_{0}\aaa_{s-3}}_{(q)\aaa_{s-1}}\right. \cr
&&+\left.\sum_{p=2}^{r-s-1}\EOM_{(p)i}\sum_{q=0}^{r-s-1-p}
\EOM_{(r-s-p-q)j}Y^{ji\aaa_{s-3}}_{(q)\aaa_{s-1}}\right)c^{\aaa_{s-1}} \cr
&=&c^{*}_{\aaa_{s-3}}\left(\sum_{q=0}^{r-s-2}\sum_{p=2}^{r-s-q}\EOM_{(p)i}
R^{i}_{(r-s-q-p)\aaa_{0}}X^{\aaa_{0}\aaa_{s-3}}_{(q)\aaa_{s-1}}\right. \cr
&&+\left.\sum_{q=0}^{r-s-3}\sum_{p=2}^{r-s-q-1}\EOM_{(p)i}
\EOM_{(r-s-q-p)j}Y^{ji\aaa_{s-3}}_{(q)\aaa_{s-1}}\right)c^{\aaa_{s-1}} \cr
&\equi{\e{noether1}}&-c^{*}_{\aaa_{s-3}}\EOM_{(1)i}\~
\tilde{B}^{i\aaa_{s-3}}_{(r-s-1)\aaa_{s-1}}\~c^{\aaa_{s-1}}
\=-c^{*}_{\aaa_{s-3}}\delta\varphi^{*}_{i}\~
\tilde{B}^{i\aaa_{s-3}}_{(r-s-1)\aaa_{s-1}}\~c^{\aaa_{s-1}}\~, 
\label{cls}
\eea
where
\beq
\tilde{B}^{i\aaa_{s-3}}_{(r-s-1)\aaa_{s-1}}\~:=\~
\sum_{q=0}^{r-s-2}R^{i}_{(r-s-q-1)\aaa_{0}}
X^{\aaa_{0}\aaa_{s-3}}_{(q)\aaa_{s-1}}
+\sum_{q=0}^{r-s-3}\EOM_{(r-s-q-1)j}Y^{ji\aaa_{s-3}}_{(q)\aaa_{s-1}}\~.
\eeq
If one adds together eqs.\ \e{cl0}--\e{cls}, one gets
\beq
\delta\left(\Delta\Snoeth_{(r+1)}+\tilde{B}_{(r+1)}\right)
\=\cO\left((\Phi^{*})^{2}\right)
\~,\~\~\~\~\~\~\~\~
\tilde{B}_{(r+1)}\~:=\~
\sum_{s=2}^{L}(-1)^{\eps(c^{*}_{\aaa_{s-2}})}
c^{*}_{\aaa_{s-2}}\varphi^{*}_{i}\~
\tilde{B}^{i\aaa_{s}}_{(r-s-2)}\~c^{*}_{\aaa_{s}}\~.\label{altogethercl}
\eeq
Now redefine the $\bar{S}^{}_{(r+1)}$ solution by an $\delta$-exact amount 
\beq
\bar{S}^{}_{(r+1)}\~\longrightarrow\~
\bar{S}^{}_{(r+1)}-\delta\delta^{-1}
\left(\Delta\Snoeth_{(r+1)}+\tilde{B}_{(r+1)}\right)\~, \label{newsolution}
\eeq
\cf definition \e{definvbrst}. The new $\bar{S}^{}_{(r+1)}$ solution
\e{newsolution} still satisfies \eq{niceguy}, but now
\bea
\Delta\Snoeth_{(r+1)}&\longrightarrow&\Delta\Snoeth_{(r+1)}
-\delta\delta^{-1}\left(\Delta\Snoeth_{(r+1)}+\tilde{B}_{(r+1)}\right) \cr
&&\~\equi{\e{ssm1}}\~\delta^{-1}\delta\left(
\Delta\Snoeth_{(r+1)}+\tilde{B}_{(r+1)}\right)-\tilde{B}_{(r+1)}
\~\equi{\e{altogethercl}}\~\cO\left((\Phi^{*})^{2}\right)\~,
\eea
so now the boundary condition \e{bc3} is satisfied as well
\beq
\Delta\Sfixed_{(r+1)}\~:=\~\bSfixed_{(r+1)}-\Sfixed_{(r+1)}\=0\~.
\eeq
One may now repeat the calculation \e{cls} with 
$\Delta Z^{\aaa_{s}}_{(r-s-2)\aaa_{s+1}}\!=\!0$. It follows that
\beq
\delta\varphi^{*}_{i}\~\Delta B^{i\aaa_{s-2}}_{(r-s-2)\aaa_{s}}\=
\delta\varphi^{*}_{i}\~\tilde{B}^{i\aaa_{s-2}}_{(r-s-2)\aaa_{s}}\~.
\label{bcbc2consequences}
\eeq
In other words, there exist structure functions
$X^{\aaa_{0}\aaa_{s-2}}_{(r-s-2)\aaa_{s}}$ and
$Y^{ji\aaa_{s-2}}_{(r-s-3)\aaa_{s}}
\!=\!-(-1)^{\eps_{i}\eps_{j}}Y^{ij\aaa_{s-2}}_{(r-s-3)\aaa_{s}}$ such that
\bea
\Delta B^{i\aaa_{s-2}}_{(r-s-2)\aaa_{s}}
&=&\tilde{B}^{i\aaa_{s-2}}_{(r-s-2)\aaa_{s}}
+R^{i}_{(0)\aaa_{0}}X^{\aaa_{0}\aaa_{s-2}}_{(r-s-2)\aaa_{s}}
+\EOM_{(1)j}Y^{ji\aaa_{s-2}}_{(r-s-3)\aaa_{s}} \cr
&=&\sum_{q=0}^{r-s-2}R^{i}_{(r-s-2-q)\aaa_{0}}
X^{\aaa_{0}\aaa_{s-2}}_{(q)\aaa_{s}}
+\sum_{q=0}^{r-s-3}\EOM_{(r-s-2-q)j}Y^{ji\aaa_{s-2}}_{(q)\aaa_{s}}\~,
\label{bxy2}
\eea
which is induction assumption \e{inductassump3} for the next step 
$p\!=\!r\!-\!s\!-\!2$.
\proofbox

\vspace{0.8cm}

\noi
{\sc Acknowledgement:}~We thank Poul Henrik Damgaard and Marc Henneaux for 
discussion. We would like to thank Poul Henrik Damgaard, the Niels Bohr
Institute and the Niels Bohr International Academy for warm hospitality.
I.A.B. would like to thank Marc Henneaux, Glenn Barnich and Universite 
Libre de Bruxelles for warm hospitality. K.B.\ would also like to thank
M.~Vasiliev and the Lebedev Physics Institute for warm hospitality. The work
of I.A.B.\ is supported by grants RFBR 08--01--00737, RFBR 08--02--01118 and
LSS--1615.2008.2. The work of K.B.\ is supported by the Ministry of Education
of the Czech Republic under the project MSM 0021622409. 

\appendix

\section{From Acyclicity to Nilpotency of Koszul--Tate Operator}
\label{secacycnilp}

\noi
Let the $r$th {\em complex} be the set of functions
$f\!=\!f(\varphi^{i};\varphi^{*}_{i},c^{*}_{\aaa_{0}},\ldots,c^{*}_{\aaa_{r}})$ that
does not depend on $c^{*}_{\aaa_{r+1}}$, $c^{*}_{\aaa_{r+2}}$, $c^{*}_{\aaa_{r+3}}$,
$\ldots$.

\begin{lemma}[Nilpotent extension to the next stage]
Let there be given a positive integer $r\!\geq\!1$. Let there be given a 
nilpotent Koszul--Tate operator $\brst_{-1}$ on the $r$th complex in a
tubular $\varphi$-neighborhood of the stationary $\varphi$-surface, such that 
it is acyclic on the $(r\!-\!2)$th complex. Let there also be given a
$(r\!+\!1)$th stage Noether identity
\beq
Z^{\aaa_{r-1}}{}_{\aaa_{r}}Z^{\aaa_{r}}{}_{\aaa_{r+1}}
\=\EOM_{i}B^{i\aaa_{r-1}}_{\aaa_{r+1}}
\~,\~\~\~\~\~\~\~\~\EOM_{i}\~:=\~(S^{}_{0}\papar{\varphi^{i}})
\~.\label{noetherr1}
\eeq
Then there exists a nilpotent extension of the Koszul--Tate operator
$\brst_{-1}$ to the $(r\!+\!1)$th complex in some tubular
$\varphi$-neighborhood of the stationary $\varphi$-surface, so that
\beq
\brst_{-1}c^{*}_{\aaa_{r+1}}\=c^{*}_{\aaa_{r}}Z^{\aaa_{r}}{}_{\aaa_{r+1}}
+\cO\left((\Phi^{*})^{2}\right)\~.
\eeq
All such extensions must be of the form
\beq
\brst_{-1}c^{*}_{\aaa_{r+1}}
=c^{*}_{\aaa_{r}}Z^{\aaa_{r}}{}_{\aaa_{r+1}}
-(-1)^{\eps(c^{*}_{\aaa_{r-1}})}c^{*}_{\aaa_{r-1}}
\varphi^{*}_{i}\left(B^{i\aaa_{r-1}}_{\aaa_{r+1}}+R^{i}{}_{\aaa_{0}}
X^{\aaa_{0}\aaa_{r-1}}_{\aaa_{r+1}}+\EOM_{j}Y^{ji\aaa_{r-1}}_{\aaa_{r+1}}
\right)+\cO\left((c^{*})^{2},(\Phi^{*})^{3}\right)\~, 
\label{lemmaacycnilpbc}
\eeq
where $Y^{ji\aaa_{r-1}}_{\aaa_{r+1}}
\!=\!-(-1)^{\eps_{i}\eps_{j}}Y^{ij\aaa_{r-1}}_{\aaa_{r+1}}$. 
\label{lemmaacycnilp}
\end{lemma}

\noi
{\sc Remark}~~
The Lemma~\ref{lemmaacycnilp} says nothing about if the nilpotent 
Koszul--Tate extension is also acyclic on the $(r\!-\!1)$th complex.
Instead, we find that the shifted Koszul--Tate
operator is better suited to address the issue of acyclicity. Nevertheless, 
the Lemma~\ref{lemmaacycnilp} makes perfectly clear that nilpotency is not 
the bottleneck in the Koszul--Tate construction (acyclicity is!), and that the
higher--stage gauge(--for--gauge)$^{r+1}$--generators $Z^{\aaa_{r}}{}_{\aaa_{r+1}}$
can be preserved in the Koszul--Tate operator $\brst_{-1}$. The latter point 
has received very little attention in the literature, see \eg Theorem 3 and 
Theorem 4 in \Ref{fishen89}.

\noi
{\sc Proof of Lemma~\ref{lemmaacycnilp}}:~~ 
Define 
\bea
U^{}_{\aaa_{r+1}}&:=&c^{*}_{\aaa_{r}}Z^{\aaa_{r}}{}_{\aaa_{r+1}}
-(-1)^{\eps(c^{*}_{\aaa_{r-1}})}c^{*}_{\aaa_{r-1}}
\varphi^{*}_{i}B^{i\aaa_{r-1}}_{\aaa_{r+1}} \~,\label{wr1} \\
\cB^{}_{\aaa_{r+1}}&:=&\brst_{-1}U^{}_{\aaa_{r+1}}
\~\equi{\e{wr1}}\~\brst_{-1}\left(c^{*}_{\aaa_{r}}Z^{\aaa_{r}}{}_{\aaa_{r+1}}
-(-1)^{\eps(c^{*}_{\aaa_{r-1}})}c^{*}_{\aaa_{r-1}}
\varphi^{*}_{i}B^{i\aaa_{r-1}}_{\aaa_{r+1}}\right) \cr
&=&\left(c^{*}_{\aaa_{r-1}}Z^{\aaa_{r-1}}{}_{\aaa_{r}}+M^{}_{\aaa_{r}}\right)
Z^{\aaa_{r}}{}_{\aaa_{r+1}}
-c^{*}_{\aaa_{r-1}}\EOM_{i}B^{i\aaa_{r-1}}_{\aaa_{r+1}}
-(-1)^{\eps(c^{*}_{\aaa_{r-1}})}V^{}_{\aaa_{r-1}}\~
\varphi^{*}_{i}B^{i\aaa_{r-1}}_{\aaa_{r+1}} \cr
&\equi{\e{noetherr1}}&M^{}_{\aaa_{r}}\~Z^{\aaa_{r}}{}_{\aaa_{r+1}}
-(-1)^{\eps(c^{*}_{\aaa_{r-1}})}V_{\aaa_{r-1}}\~
\varphi^{*}_{i}B^{i\aaa_{r-1}}_{\aaa_{r+1}}
\=\cO\left((\Phi^{*})^{2}\right)\~.\label{br1}
\eea
The $\cB^{}_{\aaa_{r+1}}$ function \e{br1} belongs to the $(r\!-\!2)$th complex, 
since both the functions
\beq
M^{}_{\aaa_{r}}\=M^{}_{\aaa_{r}}
(\varphi^{i};\varphi^{*}_{i},c^{*}_{\aaa_{0}},\ldots,c^{*}_{\aaa_{r-2}})
\~\~{\rm and}\~\~ V^{}_{\aaa_{r-1}}\=V^{}_{\aaa_{r-1}}
(\varphi^{i};\varphi^{*}_{i},c^{*}_{\aaa_{0}},\ldots,c^{*}_{\aaa_{r-2}})
\eeq
belong to that. Moreover $\afn(\cB^{}_{\aaa_{r+1}})\!=\!r\!+\!3$.
Because of acyclicity, there exists a function
$\tilde{M}^{}_{\aaa_{r+1}}\!=\!\tilde{M}^{}_{\aaa_{r+1}}
(\varphi^{i};\varphi^{*}_{i},c^{*}_{\aaa_{0}},\ldots,c^{*}_{\aaa_{r-1}})$ in the
$(r\!-\!1)$th complex, such that 
$\brst_{-1}\tilde{M}^{}_{\aaa_{r+1}}\!=\!\cB^{}_{\aaa_{r+1}}$.
It follows moreover that 
$\afn(\tilde{M}^{}_{\aaa_{r+1}})\!=\!r\!+\!2\!=\!\afn(c^{*}_{\aaa_{r}})$,
so that $\tilde{M}^{}_{\aaa_{r+1}}$ cannot contain terms that are first or
zeroth order in antifields. Hence
$\tilde{M}^{}_{\aaa_{r+1}}\!=\!\cO((\Phi^{*})^{2})$ is of the form
\beq
\tilde{M}^{}_{\aaa_{r+1}}\=(-1)^{\eps(c^{*}_{\aaa_{r-1}})}c^{*}_{\aaa_{r-1}}
\varphi^{*}_{i}\tilde{B}^{i\aaa_{r-1}}_{\aaa_{r+1}}
+\cO\left((c^{*})^{2},(\Phi^{*})^{3}\right)\~.
\eeq
Therefore 
\beq
\cB^{}_{\aaa_{r+1}}
\=\brst_{-1}\tilde{M}^{}_{\aaa_{r+1}}\=c^{*}_{\aaa_{r-1}}
\EOM_{i}\tilde{B}^{i\aaa_{r-1}}_{\aaa_{r+1}}+\cO\left((\Phi^{*})^{2}\right)\~.
\eeq
Since $\cB^{}_{\aaa_{r+1}}\!=\!\cO((\Phi^{*})^{2})$, it follows that
\beq
\EOM_{i}\tilde{B}^{i\aaa_{r-1}}_{\aaa_{r+1}}\=0\~,
\eeq
and therefore there exist functions $X^{\aaa_{0}\aaa_{r-1}}_{\aaa_{r+1}}$ and
$Y^{ji\aaa_{r-1}}_{\aaa_{r+1}}\!=\!-(-1)^{\eps_{i}\eps_{j}}Y^{ij\aaa_{r-1}}_{\aaa_{r+1}}$
such that
\beq
\tilde{B}^{i\aaa_{r-1}}_{\aaa_{r+1}}
\=R^{i}{}_{\aaa_{0}}X^{\aaa_{0}\aaa_{r-1}}_{\aaa_{r+1}}
+\EOM_{j}Y^{ji\aaa_{r-1}}_{\aaa_{r+1}}\~.\label{bexact2} 
\eeq
Now define
\beq
\brst_{-1}c^{*}_{\aaa_{r+1}}\~:=\~
U^{}_{\aaa_{r+1}}-\tilde{M}^{}_{\aaa_{r+1}}\~.
\eeq
It is nilpotent, because 
\beq
\brst{}^{2}_{-1}c^{*}_{\aaa_{r+1}}
\=\brst_{-1}\left(U^{}_{\aaa_{r+1}}-\tilde{M}^{}_{\aaa_{r+1}}\right)
\=\cB^{}_{\aaa_{r+1}}-\cB^{}_{\aaa_{r+1}}\=0\~.
\eeq
\proofbox

\section{Elimination of $B$-Terms by Off--Shell Change of
Generators}
\label{secrot}

\noi
If the number of stages is finite $L\!<\!\infty$, it is possible to apply 
off-shell changes to the gauge(--for--gauge)$^{s}$--generators
\beq
\bar{Z}^{\aaa_{s-1}}{}_{\bbbb_{s}}
\~:=\~Z^{\aaa_{s-1}}{}_{\aaa_{s}}P^{\aaa_{s}}{}_{\bbbb_{s}}\~\approx\~
Z^{\aaa_{s-1}}{}_{\bbbb_{s}}  
\~,\~\~\~\~\~\~\~\~s\~\in\~\{0,\ldots,L\}\~, \label{bzbardef}
\eeq
so that the higher--stage Noether identities \e{noether2} becomes strong 
\beq
\bar{Z}^{\aaa_{s-2}}{}_{\aaa_{s-1}}\bar{Z}^{\aaa_{s-1}}{}_{\aaa_{s}}\=0
\~,\~\~\~\~\~\~\~\~s\in\{1,\ldots,L\}\~,\label{noether2strong}
\eeq
or equivalently, that the $B^{i\aaa_{s-2}}_{\aaa_{s}}\!=\!0$ vanish from the
higher--stage Noether identities \e{noether2}. We should stress that changes to
the gauge(--for--gauge)$^{s}$--generators \e{bzbardef} goes against the paper's
main policy of preserving the original gauge algebra as it is.

\noi
{\sc Proof of Strong Noether Identities \e{noether2strong}}:~~Define that the 
new $s$-stage generator $\bar{Z}^{\aaa_{s-1}}{}_{\bbbb_{s}}\!:=\!0$, the 
$s$-stage gauge condition matrix $\chi^{\aaa_{s}}{}_{\aaa_{s-1}}\!:=\!0$, and
the $s$-stage Faddeev--Popov propagator $D^{\aaa_{s}}{}_{\bbbb_{s}}\!:=\!0$ 
vanish for $s\!>\!L$. Define 
$P^{\aaa_{L}}{}_{\bbbb_{L}}\!:=\!\delta^{\aaa_{L}}_{\bbbb_{L}}$, so that the top
stage is unchanged
$\bar{Z}^{\aaa_{L-1}}{}_{\bbbb_{L}}\!:=\!Z^{\aaa_{L-1}}{}_{\bbbb_{L}}$.

\noi
One starts at the top stage $s\!=\!L$ and goes down successively to the zeroth 
stage $s\!=\!0$ by induction. Assume that the strong Noether identities 
\e{noether2strong} holds for earlier ($\equiv$ higher) stages. 
Choose the $s$-stage gauge condition matrix
$\chi^{\aaa_{s}}{}_{\aaa_{s-1}}\!=\!\chi^{\aaa_{s}}{}_{\aaa_{s-1}}(\varphi)$,
such that 
\beq
\rank(\chi^{\aaa_{s}}{}_{\bbbb_{s-1}}\bar{Z}^{\bbbb_{s-1}}{}_{\cccc_{s}})
\=\MMM_{s}\~.
\eeq
Define the $s$-stage Faddeev--Popov propagator 
$D^{\aaa_{s}}{}_{\bbbb_{s}}\!=\!D^{\aaa_{s}}{}_{\bbbb_{s}}(\varphi)$ so that
\beq
D^{\aaa_{s}}{}_{\bbbb_{s}}
\chi^{\bbbb_{s}}{}_{\bbbb_{s-1}}\bar{Z}^{\bbbb_{s-1}}{}_{\cccc_{s}}
\=\delta^{\aaa_{s}}_{\cccc_{s}}
-\bar{Z}^{\aaa_{s}}{}_{\aaa_{s+1}}D^{\aaa_{s+1}}{}_{\bbbb_{s+1}}
\chi^{\bbbb_{s+1}}{}_{\cccc_{s}}\~.\label{bfppropdef}
\eeq
The feasibility to meet condition \e{bfppropdef} for $s\!<\!L$ can be seen from
the induction assumption \e{noether2strong} (with substitution
$s\!\to\!s\!+\!1$). The Faddeev--Popov propagator $D^{\aaa_{s}}{}_{\bbbb_{s}}$
is typically space--time non-local. Define
\beq
P^{\aaa_{s-1}}{}_{\bbbb_{s-1}}\~:=\~\delta^{\aaa_{s-1}}_{\bbbb_{s-1}}
-\bar{Z}^{\aaa_{s-1}}{}_{\aaa_{s}}D^{\aaa_{s}}{}_{\bbbb_{s}}
\chi^{\bbbb_{s}}{}_{\bbbb_{s-1}}\~.\label{blambdadef}
\eeq
Next calculate the new $s$-stage generator 
$\bar{Z}^{\aaa_{s-2}}{}_{\bbbb_{s-1}}$ from the definition \e{bzbardef}. 
The weak Noether identities \e{noether2} guarantee that the generator only
changes off-shell, 
$\bar{Z}^{\aaa_{s-2}}{}_{\bbbb_{s-1}}\!\approx\!Z^{\aaa_{s-2}}{}_{\bbbb_{s-1}}$.
Calculate
\bea
\bar{Z}^{\aaa_{s-2}}{}_{\bbbb_{s-1}}\bar{Z}^{\bbbb_{s-1}}{}_{\cccc_{s}}
&\equi{\e{bzbardef}}&Z^{\aaa_{s-2}}{}_{\aaa_{s-1}}
P^{\aaa_{s-1}}{}_{\bbbb_{s-1}}\bar{Z}^{\bbbb_{s-1}}{}_{\cccc_{s}} \cr
&\equi{\e{blambdadef}}&Z^{\aaa_{s-2}}{}_{\aaa_{s-1}}
\left(\delta^{\aaa_{s-1}}_{\bbbb_{s-1}}
-\bar{Z}^{\aaa_{s-1}}{}_{\aaa_{s}}D^{\aaa_{s}}{}_{\bbbb_{s}}
\chi^{\bbbb_{s}}{}_{\bbbb_{s-1}} \right)
\bar{Z}^{\bbbb_{s-1}}{}_{\cccc_{s}} \cr
&\equi{\e{bfppropdef}}&Z^{\aaa_{s-2}}{}_{\bbbb_{s-1}}
\bar{Z}^{\bbbb_{s-1}}{}_{\cccc_{s}}
-Z^{\aaa_{s-2}}{}_{\aaa_{s-1}}\bar{Z}^{\aaa_{s-1}}{}_{\aaa_{s}}
\left(\delta^{\aaa_{s}}_{\cccc_{s}}
-\bar{Z}^{\aaa_{s}}{}_{\aaa_{s+1}}D^{\aaa_{s+1}}{}_{\bbbb_{s+1}}
\chi^{\bbbb_{s+1}}{}_{\cccc_{s}}\right) \cr
&\equi{\e{noether2strong}}&0\~, \label{bcalculation}
\eea
where the induction assumption \e{noether2strong} (with substitution 
$s\!\to\!s\!+\!1$) was used in the last step.
\proofbox

\section{Antisymmetric $B^{ij}_{\aaa_{1}}$ Exists.}
\label{secbanti}

\begin{proposition}
If the structure function $B^{ij}_{\aaa_{1}}$ locally satisfies the first--stage
Noether identity \e{noether15}, then there exists an antisymmetric structure
function $ \tilde{B}^{ij}_{\aaa_{1}}
\!=\!-(-1)^{\eps_{i}\eps_{j}}\tilde{B}^{ji}_{\aaa_{1}}$
that does the same.
\label{bantiprop}
\end{proposition}

\noi
Proof of Proposition~\ref{bantiprop}:~~Recall that there exists transversal and
longitudinal fields
$\bar{\varphi}^{\bar{\imath}}\!\equiv\!\{\xi^{I};\theta^{\AAA_{0}}\}$
so that the original action $S_{0}$ only depends on $\xi^{I}$, \cf the Gauge 
Principle~\ref{gaugeprinciple}. Define 
\beq
R^{\bar{\imath}}{}_{\aaa_{0}}
\~:=\~\Lambda^{\bar{\imath}}{}_{i}R^{i}{}_{\aaa_{0}}\~,\~\~\~\~\~\~\~\~
B^{\bar{\imath}\bar{\jmath}}_{\aaa_{1}}\~:=\~
\Lambda^{\bar{\imath}}{}_{i}B^{ij}_{\aaa_{1}}\Lambda^{\bar{\jmath}}{}_{j}
(-1)^{(\eps_{j}+\eps_{\aaa_{1}})(\eps_{j}+\eps_{\bar{\jmath}})}\~,
\label{barrb}
\eeq
where we have used the Jacobian $\Lambda^{\bar{\imath}}{}_{i}$ matrix
\e{lambdam1}. Then
\beq
(S^{}_{0}\papar{\xi^{J}})B^{J\bar{\imath}}_{\aaa_{1}}
\=(S^{}_{0}\papar{\bar{\varphi}^{\bar{\jmath}}})
B^{\bar{\jmath}\bar{\imath}}_{\aaa_{1}}
\~\equi{\e{noether15}}\~
R^{\bar{\imath}}{}_{\aaa_{0}}Z^{\aaa_{0}}{}_{\aaa_{1}}\~,
\label{barnoether15}
\eeq
\beq
R^{I}{}_{\aaa_{0}}
\~\equi{\e{barrb}}\~(\xi^{I}\papar{\varphi^{i}})R^{i}{}_{\aaa_{0}}
\~\equi{\e{deftransv}}\~(S^{}_{0}\papar{\varphi^{j}})K^{jI}_{\aaa_{0}}
\~\equi{\e{bconsistrel}}\~(S^{}_{0}\papar{\xi^{J}})K^{JI}_{\aaa_{0}}
\~,\~\~\~\~\~\~\~\~
K^{JI}_{\aaa_{0}}\=-(-1)^{\eps_{I}\eps_{J}}K^{IJ}_{\aaa_{0}}\~.
\eeq
Now define antisymmetric tilde structure functions 
$\tilde{B}^{\bar{\imath}\bar{\jmath}}_{\aaa_{1}}
\!=\!-(-1)^{\eps_{\bar{\imath}}\eps_{\bar{\jmath}}}
\tilde{B}^{\bar{\jmath}\bar{\imath}}_{\aaa_{1}}$ as follows.
\beq
\tilde{B}^{IJ}_{\aaa_{1}}
\~:=\~K^{IJ}_{\aaa_{0}}Z^{\aaa_{0}}{}_{\aaa_{1}}\~,\~\~\~\~\~\~\~\~
\tilde{B}^{I\AAA_{0}}_{\aaa_{1}}\~:=\~
B^{I\AAA_{0}}_{\aaa_{1}}
\~=:\~-(-1)^{\eps_{I}\eps_{\AAA_{0}}}\tilde{B}^{\AAA_{0}I}_{\aaa_{1}}
\~,\~\~\~\~\~\~\~\~
\tilde{B}^{\AAA_{0}B^{}_{0}}_{\aaa_{1}}\~:=\~0\~.
\eeq
It is easy to see that the antisymmetric structure functions 
$\tilde{B}^{\bar{\imath}\bar{\jmath}}_{\aaa_{1}}$ also satisfies the first--stage
Noether identity \e{barnoether15}.
\proofbox

\noi
{\sc Remark}:~~If the structure functions $B^{ij}_{\aaa_{1}}$ is a tensor under
change of coordinates (as is normally assumed), the antisymmetric 
$\tilde{B}^{ij}_{\aaa_{1}}$ in the proof of Proposition~\ref{bantiprop}
is not necessarily also a tensor.

\section{Deformation of Acyclicity}

\noi
Acyclicity is stabile under deformations in the following sense.

\begin{lemma}[Function version]
Let there be given a nilpotent Grassmann--odd operator $\delta$, 
$\delta^{2}\!=\!0$, with resolution expansion 
$\delta\!=\!\sum_{k=-1}^{\infty}\delta_{(k)}$, $\deg(\delta_{(k)})\!=\!k$, 
\wrt an integer resolution degree ``$\deg$''. Assume the leading nilpotent 
operator $\delta_{(-1)}$ is acyclic, \ie  
 \beq
\forall {\it~functions~}f:\~\~\~\~\delta_{(-1)}f=0 \~\~\wedge\~\~ \deg(f)> 0
\~\~\~\~\Rightarrow\~\~\~\~
\exists g:\~ f= \delta_{(-1)}g\~. \label{dmoneacyclic}
\eeq
Then the operator $\delta$ itself is acyclic as well, \ie
 \beq
\forall {\it~functions~}f:\~\~\~\~\delta f=0 \~\~\wedge\~\~ \deg(f)> 0
\~\~\~\~\Rightarrow\~\~\~\~ 
\exists g:\~ f= \delta g\~. \label{dacyclic}
\eeq
\label{theoremlemma}
\end{lemma}

\noi
{\sc Proof of Lemma~\ref{theoremlemma}}:~~The $n$th nilpotency relation reads
\beq
0\=(\delta^{2})_{(n)} \=\sum_{k=-1}^{n+1}\delta_{(k)}\delta_{(n-k)}\~.
\label{d2nilrel}
\eeq
Let there be given $\delta$-closed function $f\!=\!\sum_{k=1}^{\infty}f_{(k)}$,
$(\delta f)\!=\!0$, with $\deg(f)\!>\!0$. We would like to find a function 
$g\!=\!\sum_{k=2}^{\infty}g_{(k)}$, such that $f\!=\!(\delta g)$, \ie
\beq
f_{(m)}\=\sum_{k=-1}^{m-2}\delta_{(k)}g_{(m-k)}
\=\sum_{k=2}^{m+1}\delta_{(m-k)}g_{(k)}
\=\sum_{k=0}^{m-1}\delta_{(m-k-2)}g_{(k+2)}\~.   \label{deltaexact}
\eeq
The proof of the main statement is an induction in the resolution degree.
Assume that there exist functions $g_{(2)}$, $g_{(3)}$, $g_{(4)}$, $\ldots$,
$g_{(n+1)}$, such that $f_{(1)}$, $f_{(2)}$, $f_{(3)}$, $\ldots$, $f_{(n)}$,
satisfy $\delta$-exactness relation \e{deltaexact}, where $n\!\geq\!0$. We
would like to find a function $g_{(n+2)}$, such that $f_{(n+1)}$ satisfies
$\delta$-exactness relation \e{deltaexact}. The following two functions
$A_{(n+1)}$ and $B_{(n)}$ are well-defined by the induction assumption.
\bea
A_{(n+1)}&:=&\sum_{k=0}^{n-1}\delta_{(k)}g_{(n+1-k)}\~, \label{ahelpful} \\ 
B_{(n)}&:=&
\sum_{k=0}^{n-1}\delta_{(k)}f_{(n-k)}
\~\equi{\e{deltaexact}}\~\sum_{k=0}^{n-1}\delta_{(k)}
\sum_{\ell=0}^{n-k-1}\delta_{(n-k-\ell-2)}g_{(\ell+2)}
\=\sum_{\ell=0}^{n-1}\left[\sum_{k=0}^{n-\ell-1}
\delta_{(k)}\delta_{(n-\ell-k-2)}\right]g_{(\ell+2)} \cr
&=&\sum_{\ell=0}^{n-1}\left[(\delta^{2})_{(n-\ell-2)}- 
\delta_{(-1)}\delta_{(n-\ell-1)}\right]g_{(\ell+2)}
\~\equi{\e{d2nilrel}+\e{ahelpful}}\~-\delta_{(-1)}A_{(n+1)}\~.
\label{bhelpful}
\eea
The $n$th closeness relation is
\beq
0\=(\delta f)_{(n)} 
\=\sum_{k=-1}^{n-1}\delta_{(k)}f_{(n-k)}
\=\delta_{(-1)}f_{(n+1)}+B_{(n)}
\~\equi{\e{bhelpful}}\~\delta_{(-1)}\left[f_{(n+1)}-A_{(n+1)}\right]
\~,\label{deltaclosed}
\eeq
so $f_{(n+1)}\!-\!A_{(n+1)}$ is $\delta_{(-1)}$-closed. Because $n\!\geq\!0$, and
because of the acyclicity \e{dmoneacyclic}, there exists a function $g_{(n+2)}$
such that 
$f_{(n+1)}\!=\!\delta_{(-1)}g_{(n+2)}\!+\!A_{(n+1)}
\!=\!\sum_{k=-1}^{n-1}\delta_{(k)}g_{(n+1-k)}$,
which is just the sought--for $\delta$-exactness relation \e{deltaexact}.
\proofbox

\begin{lemma}[Operator version]
Let there be given a nilpotent Grassmann--odd operator $\delta$, 
$\left[\delta,\delta\right]\!=\!0$, with resolution expansion 
$\delta\!=\!\sum_{k=0}^{\infty}\delta_{[k]}$, $\deg(\delta_{[k]})\!=\!k$, 
\wrt an integer resolution degree ``$\deg$''. Assume the leading nilpotent 
operator $\delta_{[0]}$ is acyclic, \ie  
 \beq
\forall {\it~operators~}X:\~\~\~\~\left[\delta_{[0]},X\right]=0 
\~\~\wedge\~\~ \deg(X) > 0\~\~\~\~\Rightarrow\~\~\~\~
\exists Y:\~ X= \left[\delta_{[0]},Y\right]\~. \label{dmoneacyclicop}
\eeq
Then the operator $\delta$ itself is acyclic as well, \ie
 \beq
\forall {\it~operators~}X:\~\~\~\~\left[\delta, X\right]=0 
\~\~\wedge\~\~ \deg(X) > 0 \~\~\~\~\Rightarrow\~\~\~\~ 
\exists Y:\~ X= \left[\delta, Y\right]\~. \label{dacyclicop}
\eeq
\label{theoremlemmaop}
\end{lemma}

\noi
{\sc Proof of Lemma~\ref{theoremlemmaop}}:~~The $n$th nilpotency relation reads
\beq
0\=\left[\delta,\delta\right]_{[n]} 
\=\sum_{k=0}^{n}\left[\delta_{[k]},\delta_{[n-k]}\right]\~.
\label{d2nilrelop}
\eeq
Let there be given $\delta$-closed operator $X\!=\!\sum_{k=1}^{\infty}X_{[k]}$,
$\left[\delta, X\right]\!=\!0$, with $\deg(X)\!>\!0$. We would like to find 
an operator $Y\!=\!\sum_{k=1}^{\infty}Y_{[k]}$, such that 
$X\!=\!\left[\delta, Y\right]$, \ie
\beq
X_{[m]}\=\sum_{k=0}^{m-1}\left[\delta_{[k]},Y_{[m-k]}\right]
\=\sum_{k=1}^{m}\left[\delta_{[m-k]},Y_{[k]}\right]\~.   \label{deltaexactop}
\eeq
The proof of the main statement is an induction in the resolution degree.
Assume that there exist operators $Y_{[1]}$, $Y_{[2]}$, $Y_{[3]}$, $\ldots$,
$Y_{[n-1]}$, such that $X_{[1]}$, $X_{[2]}$, $X_{[3]}$, $\ldots$, $X_{[n-1]}$,
satisfy $\delta$-exactness relation \e{deltaexactop}, where $n\!\geq\!1$. We
would like to find an operator $Y_{[n]}$, such that $X_{[n]}$ satisfies
$\delta$-exactness relation \e{deltaexactop}. The following two operators
$A_{[n]}$ and $B_{[n]}$ are well-defined by the induction assumption.
\bea
A_{[n]}&:=&\sum_{k=1}^{n-1}\left[\delta_{[n-k]},Y_{[k]}\right]\~, 
\label{ahelpfulop} \\ 
B_{[n]}&:=&\sum_{k=1}^{n-1}\left[\delta_{[k]},X_{[n-k]}\right]
\~\equi{\e{deltaexactop}}\~\sum_{k=1}^{n-1}\left[\delta_{[k]},
\sum_{\ell=1}^{n-k}\left[\delta_{[n-k-\ell]},Y_{[\ell]}\right]\right]
\=\sum_{\ell=1}^{n-1}\sum_{k=1}^{n-\ell}
\left[\delta_{[k]},\left[\delta_{[n-k-\ell]},Y_{[\ell]}\right]\right] \cr
&=& \sum_{\ell=1}^{n-1}\left( \sum_{k=0}^{n-\ell}
\left[\delta_{[k]},\left[\delta_{[n-k-\ell]},Y_{[\ell]}\right]\right]
-\left[\delta_{[0]},\left[\delta_{[n-\ell]},Y_{[\ell]}\right]\right]\right)\cr
&=&\sum_{\ell=1}^{n-1}\left[ \Hf \left[\delta,
\delta\right]_{[n-\ell]},Y_{[\ell]}\right] 
- \left[\delta_{[0]}, \sum_{\ell=1}^{n-1}
\left[\delta_{[n-\ell]},Y_{[\ell]}\right]\right]
\~\equi{\e{d2nilrelop}+\e{ahelpfulop}}\~-\left[\delta_{[0]},A_{[n]}\right]\~.
\label{bhelpfulop}
\eea
The $n$th closeness relation is
\beq
0\=\left[\delta, X\right]_{[n]} 
\=\sum_{k=0}^{n-1}\left[\delta_{[k]},X_{[n-k]}\right]
\=\left[\delta_{[0]},X_{[n]}\right]+B_{[n]}
\~\equi{\e{bhelpfulop}}\~\left[\delta_{[0]},X_{[n]}\!-\!A_{[n]}\right]
\~,\label{deltaclosedop}
\eeq
so $X_{[n]}\!-\!A_{[n]}$ is $\delta_{[0]}$-closed. Because $n\!\geq\!1$, and
because of the acyclicity \e{dmoneacyclicop}, there exists an operator 
$Y_{[n]}$ such that 
$X_{[n]}\!=\!\left[\delta_{[0]},Y_{[n]}\right]\!+\!A_{[n]}
\!=\!\sum_{k=0}^{n-1}\left[\delta_{[k]},Y_{[n-k]}\right]$,
which is just the sought--for $\delta$-exactness relation \e{deltaexactop}.
\proofbox

\end{document}